\documentclass[12pt]{article}
\usepackage{mystyle}
\begin{document} %\linenumbers

\begin{center}
{\Large {\bf A spatial causal analysis of wildland fire-contributed $\PM$ using numerical model output}}\\ \vspace{12pt}

{\large Alexandra E Larsen\footnote{Department of Statistics, North Carolina State University, Raleigh, NC}\footnote{Oak Ridge Institute for Science and Education at the United States Environmental Protection Agency, National Health and Environmental Effects Research Laboratory, Environmental Public Health Division, Research Triangle Park, NC}, Shu Yang$^1$, Brian J Reich$^1$ and Ana G Rappold\footnote{United States Environmental Protection Agency, National Health and Environmental Effects Research Laboratory, Environmental Public Health Division, Research Triangle Park, NC.  Disclaimer: This work does not necessarily represent EPA views or policy.}}\vspace{12pt}

\today
\end{center}

\begin{abstract}
Wildland fire smoke contains hazardous levels of fine particulate matter ($\PM$), a pollutant shown to adversely effect health. Estimating fire attributable $\PM$ concentrations is key to quantifying the impact on air quality and subsequent health burden. This is a challenging problem since only total $\PM$ is measured at monitoring stations and both fire-attributable $\PM$ and $\PM$ from all other sources are correlated in space and time. We propose a framework for estimating fire-contributed $\PM$ and $\PM$ from all other sources using a novel causal inference framework and bias-adjusted chemical model representations of $\PM$ under counterfactual scenarios. The chemical model representation of $\PM$ for this analysis is simulated using Community Multi-Scale Air Quality Modeling System (CMAQ), run with and without fire emissions across the contiguous U.S. for the 2008-2012 wildfire seasons. The CMAQ output is calibrated with observations from monitoring sites for the same spatial domain and time period. We use a Bayesian model that accounts for spatial variation to estimate the effect of wildland fires on $\PM$ and state assumptions under which the estimate has a valid causal interpretation. Our results include estimates of absolute, relative and cumulative contributions of wildfire smoke to $\PM$ for the contiguous U.S. Additionally, we compute the health burden associated with the $\PM$ attributable to wildfire smoke. 

\end{abstract}

\section{Introduction}\label{s:intro} 

Wildfires have become a leading contributor to unhealthy air quality in many communities.  Among the pollutants found in smoke, fine particulate matter (mixtures of particles smaller than 2.5 $\mu m$ in diameter or $\PM$),  associated with a number of respiratory and cardiovascular outcomes, is of the largest public health concern \citep{Dennekamp2011,Rappold2011,Johnston2012,Dennekamp2015,Haikerwal2015,Haikerwal2016,Wettstein2018}. The objective of this study is to estimate wildland fire-attributable fraction of ambient $\PM$ in order to quantify the related health burden.  We introduce a potential outcomes framework to estimate the causal effect of wildland fires on ambient $\PM$ in the presence of spatial correlation. The framework leverages numerical model simulations of air quality serving as biased representations of the potential outcomes. A Bayesian spatial downscaling model is used to learn the relationship between the spatially and temporally resolved numerical model output and the sparsely observed $\PM$ from air quality monitors, and to provide unbiased estimates of counterfactual outcomes, quantification of uncertainty, and predictions that are both spatially and temporally resolved. 

To quantify the magnitude of the health burden attributable to the smoke from fire events, we need to distinguish the $\PM$ composition mixture attributable to fire from the $\PM$ mixture due to all other sources.  Total ambient $\PM$ concentrations are recorded at the monitoring sites across the country, however, these observations do not provide insight into the potential composition of particles that would have formed had there been no wildfires.   The mixture of particles measured on any given day depends on multiple sources of emissions and conditions of combustion by which particles were produced. Once released, particles and gases coalesce and interact with those already present in the atmosphere through non-additive chemical and physical processes.  Formation of $\PM$ is additionally confounded by external factors including fire weather conditions, vegetation, burned areas and areas unable to burn again, as well as anthropogenic and other natural emissions \citep{McKenzie,Stavros2014a}. Finally, in the presence of fire, non-fire emissions themselves can be altered through feedbacks. Together, these factors lead to  complex dependencies of $\PM$ concentrations  across space and time. 

To distinguish fire-contributed $\PM$ from total ambient concentrations we utilize numerical model representations  of air quality. The model simulates chemical reactions and transport of particle-mixtures in the atmosphere using deterministic representations of chemical processes under a set of input emissions and external forcings. By removing the forcing for wildfire emissions, these models produce air quality simulations from the counterfactual  scenario, i.e. $\PM$ composition that would have formed had there not been wildfires. In this study, we use the Community Multiscale Air Quality (CMAQ) numerical model to simulate air quality under observed and counterfactual forcings. The difference in $\PM$ under CMAQ representations of air quality with and without wildfire emissions is considered to be a modeled representation of fire-contributed $\PM$. 

Numerical models have been used to simulate counterfactual environmental conditions in other contexts, most notably to  investigate future unobserved or distant past climate trajectories \citep{Allen2003,Hegerl2011a,NationalAcademiesofSciences2016,Katzfuss2017,Knutson2017}. These studies,  referred to as  detection and attribution (D\&A) studies,  use global climate models to detect changes by varying an exogenous forcing while holding all else constant and to attribute the change to the specific forcings. These studies have been linked to causal counterfactual theory in \cite{Hannart2015} in which authors demonstrate the utility of deriving the probability of necessary and sufficient causality in formulating causal claims \citep{Hannart2015}. However, when outcomes are not directly observable (e.g. future or paleo climates), causal inference is limited due to lack of accounting for error and uncertainty \citep{Hannart2015}.  

Even when the outcomes are observable, such as in the case of air quality, numerical models are subject to systematic bias arising from misspecification of inputs or processes  governing model behavior. To calibrate the CMAQ $\PM$ output in our study, we develop a Bayesian statistical downscaling method that relates data at a lower observed resolution to a spatially resolved, higher resolution of CMAQ model and allows for spatial prediction at all locations. The calibration model is similar to the spatio-temporal downscaling method introduced by \cite{Berrocal2010e} in that it uses spatially-varying coefficients to estimate the relationship between sparse observations and numeric model output where data is available \citep{Gelfand2003,Schmidt2003,Gelfand2004,Berrocal2010e}. The method is computationally efficient and has high predictive performance relative to other downscaling methods \citep{Cressie1993,Chiles2012,Fuentes2005}.

The second challenge to estimating wildfire attributed $ \PM$ concentrations within a causal inference paradigm is the spatial interference between the observed $\PM$ at sites according to whether or not the site (observation unit) is impacted by wildfires (treatment). Indirect or spill-over effect across spatial locations violates the stable unit treatment value assumption (SUTVA), which is central to the potential outcomes framework for causal inference \citep{Rubin1978}. Estimating valid causal effects in the presence of interference has previously been addressed in the context of vaccines and infectious diseases \citep{Hong2006,Rosenbaum2007,Hudgens2008,Tchetgen2012}. 
 
Most commonly, interference has been addressed using the less stringent partial interference assumption, which assumes interference amongst units in the same group, but not between groups \citep{Halloran1991,Sobel2006,Hudgens2008}. The problem of interference among units has also been addressed in estimating the causal effect of air pollution regulations on health burden \citep{Zigler2012b,Dominici2014,Zigler2017}. \cite{Zigler2012b} introduced the first application of spatial models to predict unobserved potential outcomes  and develop causal effect estimate of air pollution regulations. To address the interference among units of observation, the authors rely on principal stratification and an assumption of partial interference. However, the correlation between units observed under opposite treatments is unidentifiable under their framework.

The main methodological contribution of this paper is to present a counterfactual framework that utilizes bias-corrected numerical model output to produce valid causal inference in the presence of spatial spillover effects. The proposed framework estimates counterfactual outcomes for each day and location under two treatment regimes: the observed regime with wildfires and the unobservable regime without wildfires. We specify a Bayesian model to fuse the numerical model output with monitor data. We assume that conditional on numerical model output, observations in the areas not affected by smoke are representative of the counterfactual regime without wildfires. This allows us to bias-correct the CMAQ output from the counterfactual regimes with observed data, which has been the limitation of previous studies. Through numerical model simulations under both regimes we are able to estimate correlation between the units observed under different regimes. We clarify the assumptions required for the estimates to have a causal interpretation, and show if these assumptions hold then the proposed method accounts for spillover effects and that all model parameters are identified.  

We apply our method to estimating the effect of wildfires on ambient $\PM$ during 2008-2012 wildfire seasons in the contiguous U.S.  We use these estimates to conduct a health burden analysis. Our Bayesian model provides full uncertainty quantification about the causal effects and resulting health burden assessment.  While we apply the method to the example of wildfire-contributed $\PM$, the approach is relevant to many applications. 

\section{Description of the $\PM$ data}\label{s:data}

\begin{figure}[H]
	\centering
	\caption{\textbf{Summary of the monitor data}. The locations and daily $\PM$ concentrations (\unit) observed at EPA monitoring stations averaged over the 2008 to 2012 wildfire seasons. The breakpoints correspond to the $25^{th}$, $50^{th}$ and $75^{th}$ percentiles of $\PM$.}\label{fig:AQS}		
	\includegraphics[width=0.8\textwidth]{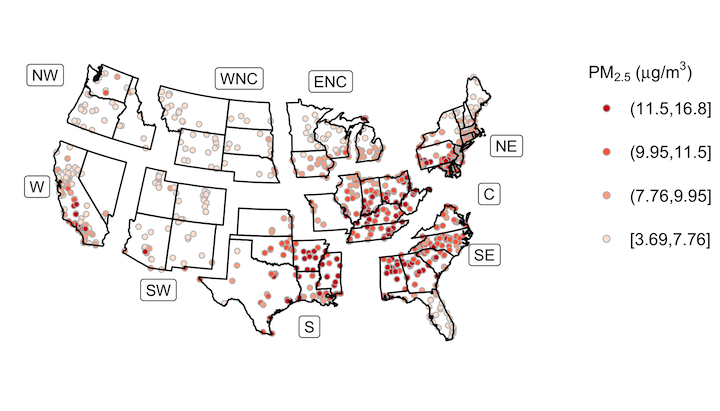}
\end{figure}

The analysis of fire-contributed $\PM$ is conducted over the 2008 to 2012 wildfire seasons (May 1 - October 31) in the contiguous U.S. There are two sources of $\PM$ data: monitor data from the Environmental Protection Agency's (EPA) Air Quality System (AQS) and simulated $\PM$ from the CMAQ model. Both data sources cover the contiguous U.S., but because of the large size we partition the data into regions with similar climates and conduct the analysis separately by region \citep{USEPA}. In the Supplemental Materials (Section 2), we conduct a sensitivity analysis to demonstrate that model results are robust to blocking by region. The nine regions are displayed in Figure \ref{fig:AQS}: West (W), Northwest (NW), West North Central (WNC), East North Central (ENC), Northeast (NE), Central (C), Southeast (SE), South (S) and Southwest (SW).

\subsection{AQS monitor data}\label{s:AQS}
We use $\PM$ data from Federal Reference Method monitoring sites in the EPA AQS monitoring network. There are more monitors in urban areas than rural, since monitors in the AQS network are distributed according to population density \citep{USEnvironmentalProtectionAgency2015}. At each site, daily average concentrations of $\PM$ are measured every one, three or six days. Figure \ref{fig:AQS} shows the monitor locations and the $\PM$ observed at each monitor in the network averaged over the study time period. The observed average concentrations range from 3.69 to 16.8 \unit. The highest concentrations of $\PM$ are in California and the Southeast, and the lowest are in the West North Central and Southwest regions.  

\subsection{CMAQ model output}\label{s:CMAQ}

CMAQ is a deterministic model of air quality which represents the most important processes related to atmospheric chemistry using cutting-edge scientific knowledge. The model utilizes emissions from a wide range of sources and transport by winds to predict concentrations of ambient composition and deposition due to precipitation. CMAQ characterizes production and loss of hundreds of particle and gas phase pollutants \citep{cmaq2019}. In the case of wildfire emissions, hourly information on fire location and size are determined using satellite information as well as on the ground reports. Wildland fire emissions are estimated based on the type, load, and conditions of vegetation at the detected burning site and uses vegetation-specific emission factors \citep{airfire2019}. The largest known sources of uncertainty arise due to misspecification in characterizing variability in weather patterns and anthropogenic emissions. In the case of wildfires, additional uncertainty arises from misclassification of plume rise, fire weather conditions and other factors.

The CMAQ-simulated $\PM$ data is the average $\PM$ concentrations for each day in the 2008 to 2012 wildfire seasons on a $12\times 12$ km grid over the contiguous U.S.;  see \cite{Rappold2017} for details. The model is run with and without the forcing for wildland fire emissions. The run without fire emissions is a CMAQ estimate of $\PM$ in the counterfactual scenario where no wildland fires are possible. The difference in $\PM$ concentrations between the two runs is a CMAQ estimate of fire-contributed $\PM$. CMAQ captures emissions, topology, weather conditions, fate and transport of air pollution among other factors. However, there are many possible remaining determinants or knowledge gaps that lead to either error and bias, which motivates our statistical approach.

\begin{figure}[H]
	\centering
	\caption{\textbf{Summary of the CMAQ-estimated $\PM$ concentrations}. The daily CMAQ $\PM$ concentrations (\unit) averaged over the 2008 to 2012 fire seasons on a $12 \times 12$ km grid across the contiguous U.S. by region. Panel (a) displays average $\PM$ from the CMAQ run without fire emissions; Panel (b) displays average $\PM$ from the CMAQ run with fire emissions; and panel (c) shows the difference between these runs reported as the percentage of the total.}\label{fig:cmaq_maps}
	\begin{subfigure}[b]{.49\textwidth}
	\centering
		\includegraphics[width=\textwidth]{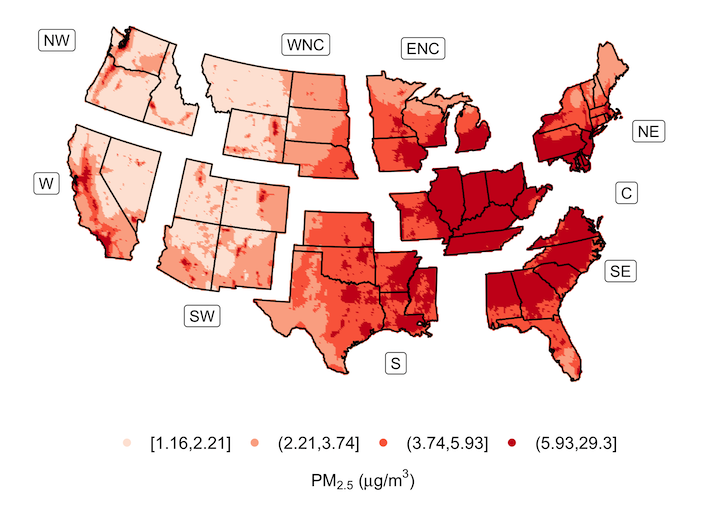}
		\caption{$\PM$ without fires (\unit)}
	\end{subfigure}
	\begin{subfigure}[b]{.49\textwidth}
	\centering
		\includegraphics[width=\textwidth]{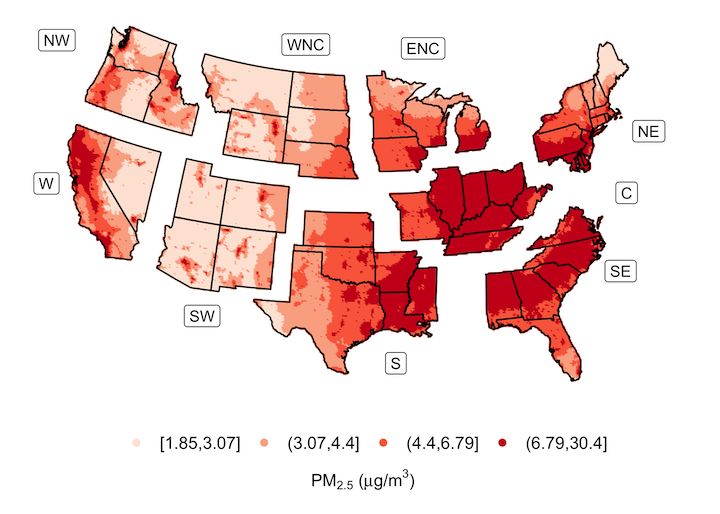}
		\caption{Total $\PM$ (\unit)}
	\end{subfigure}
	\vspace{\baselineskip}
	\begin{subfigure}[b]{\textwidth}
		\centering
		\includegraphics[width=\textwidth]{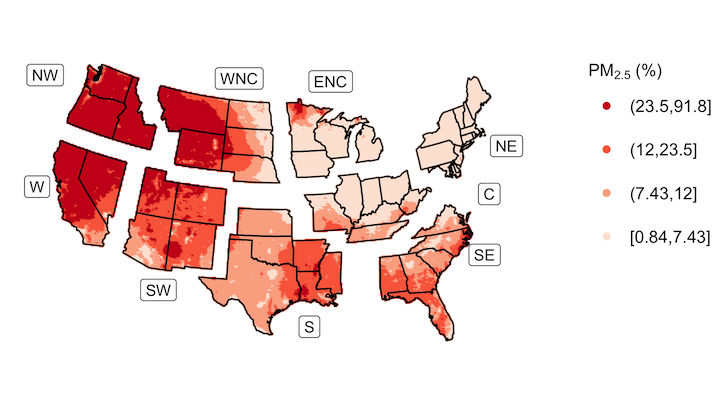}
		\caption{Fire-Contributed $\PM$ as \% of Total $\PM$}
	\end{subfigure}
\end{figure}

\begin{figure}[H]
	\centering
	    \caption{\textbf{Times series plot at one site.} Total $\PM$ from CMAQ run with and without the forcing for fire emissions, and total $\PM$ measured at an AQS monitoring site in Northern California ($-121.8^\circ, 39.8^\circ$) during the 2008 wildfire season.}\label{fig:oneSite_obs}
\includegraphics[width=0.75\textwidth]{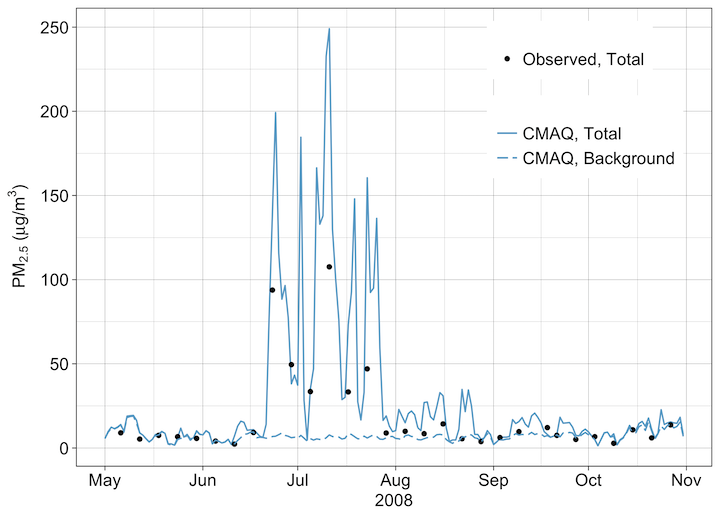}
\end{figure}

Figure \ref{fig:cmaq_maps} displays the $\PM$ modeled by CMAQ averaged over the 2008 to 2012 wildfire seasons. The western half of the U.S. has predominantly low concentrations of $\PM$ (1.16-2.21 \unit) when fire emissions are excluded, but higher concentrations when fire emissions are included (up to 6.78-30.4 \unit). This trend is particularly notable in the West and Northwest regions, where wildfire frequency is high and fire-contributed $\PM$ comprises 23.5-91.8\% of the total $\PM$ in parts of these regions (i.e. central and northern California, eastern Oregon and Washington, and central Idaho). In the South and Southeastern regions, contribution of both wildland and prescribed fires is evident. Figure \ref{fig:oneSite_obs} provides a time series of observed $\PM$ concentrations and CMAQ estimates at a site in Northern California. On days where wildfire activity is present, CMAQ tends to produce higher estimates of total $\PM$.  

\section{Methods\label{s:methods}}

\subsection{Notation}

We first establish some notation for the data. Let the observed monitor data be $Y_{t}(\bs)$ for day $t$ and location $\bs\in{\cal R}^2$ in spatial domain $\calD$. We denote CMAQ output from the no-fire run as $\hat{\theta}_{t}(\bs)$ and the difference between the fire and no-fire runs as $\hat{\delta}_{t}(\bs)$. We denote other environmental factors that are related to both fire activity and $\PM$ (confounders such as non-fire natural emissions, anthropogenic sources, wind, land type, etc.) as $X_{t}(\bs)$ and binary fire presence as $A_{t}(\bs)$, so that $A_{t}(\bs)=1$ if there is a fire burning at $\bs$ on day $t$ and $A_{t}(\bs)=0$ otherwise. The smoke plumes associated with the fires determine which locations' air quality are affected by the fires, and so we define $C_t(\bs)=1$ if site $\bs$ is in a plume on day $t$ and $C_t(\bs)=0$ otherwise to capture spillover effects. The collection of data across space is denoted in bold, e.g. $\bA_{t}=\{A_{t}(\bs):\bs\in\calD\}$ for the fire indicators.

\subsection{Fire regimes and potential outcomes framework\label{s:counterfactuals} }

To estimate fire-attributable $\PM$, i.e., the amount of $\PM$ that would not have occurred were it not for wildland fires, we use a potential outcomes framework \citep{Rubin1978}. To this end, we define regimes $R=1$ and $R=0$ over spatial domain $\calD$: the fire regime ($R=1$) under which wildfires occur in $\calD$, and the no-fire regime ($R=0$) under which fires do not occur anywhere in $\calD$. We also define the potential $\PM$ under regimes $R=0$ and $R=1$ as $Y_{t}(\bs,0)$ and $Y_{t}(\bs,1)$, respectively, and model each as 
\beqn \label{Y0Y1} 
	Y_t(\bs,0) &=& \theta_t(\bs), \\
	Y_t(\bs,1) &=& \theta_t(\bs) + \delta_t(\bs), \nonumber 
\eeqn
where $\theta_{t}(\bs)$ and $\delta_{t}(\bs)$ are stochastic processes representing non-fire and fire-attributed $\PM$, respectively. 

Under regime $R=1$ the potential outcomes are generated by sampling $\bX_t$,  $\bA_t|\bX_t$, $\bC_t|\bA_t,\bX_t$, and finally $\bY_t(1)|\bX_t,\bA_t,\bC_t$ in sequence.  Similarly, under regime $R=0$, the potential outcome are generated by  sampling $\bX_t$ and then $\bY_t(0)|\bX_t$ setting $\bA_t=\bC_t=0$. Therefore, the amount of $\PM$ caused by wildland fires is quantified by the average (over $\bX_t$, $\bA_t$ and $\bC_t$) difference in the potential outcomes \citep{Rubin1978,Holland1986,Hernan2008}: 
\[
	\Delta(\bs)=\mbox{E}[Y_{t}(\bs,1)-Y_{t}(\bs,0)]=\mbox{E}[\delta_{t}(\bs)].
\]
In our analysis, we average over time throughout the entire fire season and years of the study, although this framework could be applied to estimate the causal effect annually, seasonally or even daily.

\subsection{Assumptions\label{s:assumptions} }
The fundamental problem in causal inference is that not all potential outcomes are observed for each $\bs$ and $t$ \citep{Holland1986}. Therefore,  the potential outcome models and the causal effect, $\Delta(\bs)$, are not identifiable without assumptions, which we discuss below. 

We assume there exist bias-correction functions, $B_{0}$ and $B_{1}$, and we observe binary indicator $C_{t}(\bs)\in\{0,1\}$ where $\bs$ is affected by wildfire smoke if and only if $C_{t}(\bs)=1$ so that the following hold: 

\begin{assumption}[Potential Outcomes Model]\label{assumption:models}
	The counterfactual processes can be decomposed as, 
	\beqn\label{theta_delta_bias}
		\theta_t(\bs) = B_0(\hat\theta_t(\bs),\bs) + e_{0t}(\bs)\text{\ \ \ \ \  and \ \ \ \ \ } \delta_t(\bs)  = B_1(\hat\delta_t(\bs),\bs) + e_{1t}(\bs),\nonumber
	\eeqn 
	where $[e_{0t}(\bs),e_{1t}(\bs)]$ is a bivariate spatial process independent of $\bX_{t}$, $\bA_{t}$ and ${\bf C}_{t}$ given ${\hat \btheta}_t$ and ${\hat \bdelta}_t$. 
\end{assumption} 
\noindent 
The bias correction functions $B_0$ and $B_1$ can be flexible nonlinear functions (e.g., splines) and vary by spatial location and the discrepancy terms $e_{0t}$ and $e_{1t}$ account for model misspecification and are modeled as spatial processes \citep{kennedy2001bayesian}. To allow for learning where the model is under-performing relative to the truth, we model CMAQ bias function $B_j$ to be a flexible spatially varying surface. As such, this bias function enables us to gain insights into possible spatially-varying confounding by examining the residual variation.

Equation (\ref{Y0Y1}) and Assumption \ref{assumption:models} specify the full joint model between counterfactual outcomes $Y_{t}(\bs,0)$ and $Y_{t}(\bs,1)$ 
given $[{\hat \theta}_t(\bs),{\hat \delta}_t(\bs)]$.
Under Assumption \ref{assumption:models}, $[Y_t(\bs,0),Y_t(\bs,1)]$ are independent of $\bX_t$ given $[{\hat \theta}_t(\bs),{\hat \delta}_t(\bs)]$. Therefore, $[{\hat \theta}_t(\bs),{\hat \delta}_t(\bs)]$ can be called the prognostic score of \citet{hansen2008prognostic}, which is the prognostic analogue of the propensity score. Also, Assumption \ref{assumption:models} implies that all confounders with the regime realizations and potential outcomes are captured through CMAQ output, since Assumption \ref{assumption:models} implies $[\theta_t(\bs),\delta_t(\bs)]$, and thus $[Y_t(\bs,0),Y_t(\bs,1)]$, are independent of $\bC_t$ given $[{\hat \btheta}_t(\bs),{\hat \delta}_t(\bs)]$.  This is similar to the unconfounded network influence assumption  of \citet{kao2017causal} under the social network framework. Although Assumption \ref{assumption:models} is key for identification and dramatically simplifies the analysis compared to modeling the effect of $\bX_t$, it cannot be verified from the observed data. Because CMAQ uses  most important meteorological and environmental factors for fire activity, smoke transportation, and $\PM$ as well as state-of-the-art computer simulations (see Section \ref{s:CMAQ}), this assumption is plausible.

\begin{assumption}[Consistency]\label{assumption:consistency}
	Ignoring measurement errors, the observation at $\bs$ equals the potential outcome at $\bs$ under regime given by $C_{t}(\bs)$, 
	\[
		Y_{t}(\bs)=\begin{cases}
		Y_{t}(\bs,0), & \mbox{\ \ if }C_{t}(\bs)=0,\\
		Y_{t}(\bs,1), & \mbox{\ \ if }C_{t}(\bs)=1.
		\end{cases}
	\]
\end{assumption} 
\noindent Assumption \ref{assumption:consistency} links the potential outcomes with the observed outcomes. In particular, it allows for partial realizations of $Y_{t}(\bs,0)$ removing the need to actualize a situation under the counterfactual no-fire regime. For example, this assumption implies that a set of monitors far removed from fires and plumes on  day $t$ can be assumed to follow the potential outcomes distribution under the no fires regime and thus be used to identify parameters in this distribution such as those that determine $B_0$ and the spatial covariance of $e_{0t}(\bs)$. As long as an appropriate variable $C_t(\bs)$ can be identified from the observed data, this assumption is plausible.  In our analysis we use CMAQ output to determine $C_t(\bs)$. Namely, we let $C = 1(\hat\delta > \tau)$, where $\tau$ is a fixed threshold chosen through cross-validation and sensitivity analysis.

Theorem \ref{Thm:identification} gives the main identification result with the proof deferred to the Appendix.
\begin{theorem}\label{Thm:identification}Under Assumptions \ref{assumption:models} and \ref{assumption:consistency} and further assuming  that $\mathbf{C}_{t}$ is not degenerate, the 
parameters in the potential outcome models are identifiable via the distribution of the observed data, i.e. the distribution of $\bY_{t}$ given
$(\hat{\btheta}_{t},\hat{\bdelta}_{t},\mathbf{C}_{t} )$.
\end{theorem} 

\noindent While we never observe complete $\bY_t$ (i.e., for all $\bs$) under the no fires regime, Assumptions 1 and 2 along with the non-degeneracy of $\bC_t$ are sufficient for identification. By Theorem \ref{Thm:identification}, causal parameter estimation only requires inspecting the implied model for $Y_{t}(\bs)$ and confirming parameter identification. In Section \ref{s:BayesianModel} we specify parametric models for the bias correction functions $B_{0}$ and $B_{1}$ and the spatial process $\bfe_{t}(\bs)=[e_{0t}(\bs),e_{1t}(\bs)]^{T}$.
We then argue in Section \ref{s:estimability} that all parameters, including the correlation between counterfactuals, are identifiable in our spatial setting. This setup serves as a basis for using a Bayesian approach to estimating the causal effect, $\Delta(\bs)$.

Defining the intervention as the fire regime instead of individual fires ($A_t(\bs)$) is key for two reasons. First, this is parallel to how the numerical model simulates fire and no-fire $\PM$. Second, the amount of fire-contributed $\PM$ at any site in the spatial domain depends on the fire status at other sites, because the smoke from fires at neighboring sites is transported. This is called interference or spill-over and it is problematic because we could not reasonably claim that changes in $\PM$ at site $\bs$ were only due to fire presence or absence at site $\bs$, i.e., $Y_{t}(\bs,A_{t}(\bs))$ is not well-defined. There would be a different potential outcome for every possible $\bA_t$, resulting in $2^n$ potential outcomes per site for a spatial domain containing $n$ sites.  

\subsection{Bayesian hierarchical model}\label{s:BayesianModel}
Assumption \ref{assumption:models} and the addition of measurement error give the following model for the observed $\PM$:

\beqn
	Y_t(\bs) &=  \theta_t(\bs) + C_t(\bs) \delta_t(\bs) + \epsilon_t(\bs),
\eeqn

\noindent where $\epsilon_t(\bs) \distas{iid}\calN(0, \sigma^2)$ are measurement errors. To separate background $\PM$ from fire-contributed $\PM$, we assign priors to $\theta_t(\bs)$ and $\delta_t(\bs)$ based on bias-adjusted CMAQ runs, as per Assumption \ref{assumption:models}. We model the means of both processes as 

$$
	B_j(z,\bs) = \alpha_j(\bs) + \beta_j(\bs)z
$$
for $j=0, 1$, where $\alpha_j(\bs)$ is the additive bias and $\beta_j(\bs)$ is the multiplicative bias. The bias terms have Gaussian process priors with means $\mbox{E}[\alpha_j(\bs)]=\mu_{\alpha_j}$ and $\mbox{E}[\beta_j(\bs)]=\mu_{\beta_j}$ and covariances $\mbox{Cov}[\alpha_j(\bs),\alpha_j(\bs')] = \sigma^2_{\alpha_j}\exp(-||\bs-\bs'||/\phi_2)$ and $\mbox{Cov}[\beta_j(\bs),\beta_j(\bs')] = \sigma^2_{\beta_j}\exp(-||\bs-\bs'||/\phi_2)$. The prior distributions for all hyper-parameters are detailed in Appendix \ref{app:Bayesian}. 

The background and fire-contributed $\PM$ then have the following form:\begin{eqnarray}
  \theta_t(s) &=& \alpha_0(\bs)+\beta_0(\bs)\hat\theta_t(\bs) + e_{0t}(\bs),\\
  \delta_t(s) &=& \alpha_1(\bs)+\beta_1(\bs)\hat\delta_t(\bs) + e_{1t}(\bs),\nonumber
\end{eqnarray}
where $\bfe_{t}(\bs)=[e_{0t}(\bs),e_{1t}(\bs)]^{T}$ is a bivariate spatial process with mean $\mbox{E}[e_{jt}(\bs)]=0$ and separable exponential covariance $\mbox{Cov}[\bfe_t(\bs), \bfe_t(\bs')] = \bSigma\exp(-||\bs-\bs'||/\phi_1)$. The $2\times 2$ cross-covariance matrix $\bSigma$ has diagonal elements $\sigma_1^2$ and $\sigma_2^2$, and off-diagonal element $\sigma_{12} = \sigma_1 \sigma_2 \gamma$, so $\gamma$ gives the correlation between counterfactual outcomes. 

\subsection{Estimability}\label{s:estimability}
In this section, we argue that all parameters in the joint model specified above are estimable. 

Consider the parameters in the mean,
\beq\label{observations}
	E[Y_t(\bs)] = \mu_t(\bs) = \alpha_0(\bs) + \beta_0(\bs){\hat \theta}_t(\bs) + \alpha_1(\bs)C_t(\bs) + \beta_1(\bs)[C_t(\bs){\hat \delta}_t(\bs)].
\eeq
Assuming the four covariates $(1, \hat\theta_t(\bs), C_t(\bs), C_t(\bs)\hat\delta_t(\bs))$ are not linearly dependent at $\bs$, then the four parameters $\alpha_0(\bs)$, $\alpha_1(\bs)$, $\beta_0(\bs)$ and $\beta_1(\bs)$ are estimable. For example, ordinary least squares would provide an unbiased and consistent estimator (as the number of days increases).   This result clearly relies on Assumption 1 or it would not be possible to identify both $\alpha_0(\bs)$ and $\alpha_1(\bs
)$.  

Under the model $Y_t(\bs) = \mu_t(\bs) + e_{0t}(\bs) + C_t(\bs)e_{1t}(\bs) + \epsilon_t(\bs)$, the covariance is
\beq\label{eq:covY}
\mbox{Cov}[Y_t(\bs), Y_t(\bs')|{\hat \theta}_t(\bs),{\hat \delta}_t(\bs)] = 
  \begin{cases} 
   \sigma_1^2\exp(-h/\phi_1) & \text{if } C_t(\bs)=C_t(\bs')=0 \\
   \sigma_1^2(1+\frac{\sigma_2}{\sigma_1}\gamma)\exp(-h/\phi_1) & \text{if } C_t(\bs)\ne C_t(\bs')\\
   (\sigma_1^2 + 2\sigma_1\sigma_2\gamma + \sigma_2^2)\exp(-h/\phi_1) & \text{if } C_t(\bs)=C_t(\bs')=1, 
  \end{cases}
\eeq
where $h=||\bs-\bs'||$. By examining the spatial correlation between pairs of points separately according to their values of $C_t(\bs)$, the parameters $\sigma_1^2$, $\sigma_2^2$, $\gamma$ and $\phi_1$ are estimable.  For example, simple variogram-based methods could be used to estimate these parameters. More importantly, under the full Bayesian model, $\mbox{Cov}[Y_t(\bs, 0), Y_t(\bs,1) | Y_t(\bs)] =\sigma_1^2(1+\frac{\sigma_2}{\sigma_1}\gamma)$ is estimable, although $Y_t(\bs, 0)$ and $Y_t(\bs, 1)$ are never jointly observed at one location.  In our analysis we use Bayesian modeling to jointly estimate the mean and covariance parameters.

\subsection{Posterior inference}\label{PosteriorInferece}

The causal effect at $\bs$ is approximated as,

$$
	\Delta(\bs) \approx \frac{1}{T} \sum_{t=1}^T  C_t(\bs)\delta_t(\bs).
$$
Our fully Bayesian analysis produces the entire posterior distribution of the causal effect, including the posterior mean ${\bar \Delta}(\bs)=\frac{1}{T} \sum_{t=1}^T  C_t(\bs){\bar \delta}_t(\bs)$, where $\bar\delta_t(\bs)$ is the posterior mean of $\delta_t(\bs)$. The estimate, $\bar\delta_t(\bs)$, includes both bias-corrected CMAQ, $B_0(\hat\theta_t(\bs),
\bs)$ and $B_1(\hat\delta_t(\bs),\bs)$, and observed data, $Y_t(\bs)$, allowing us to account for any daily variation in fire-attributable $\PM$ not captured by CMAQ. Finally, $\bar\delta_t(\bs)$ is based on estimable parameters defined in the previous section, making $\bar\Delta(\bs)$ an estimable quantity as well.

We multiply $\bar\delta_t(\bs)$ by $C_t(\bs)$ because given Assumption \ref{assumption:consistency}, this relates the observations to the potential outcomes, thereby imparting the causal interpretation on $\bar\Delta(\bs)$. Multiplying by $C_t(\bs)$ also allows the model to only identify $\delta_t(\bs)$ as a causal quantity if $C_t(\bs) = 1$, which is important as we are not interested in the $\PM$ if there were fires affecting $\bs$ every day, but the causal estimate if the fires we observed were removed.

Assuming conditional independence of $C_t(\bs)$ and $\delta_t(\bs)$ given $Y_t(\bs), \hat\theta_t(\bs), \hat\delta_t(\bs)$
over time, $\bar\Delta(\bs)$ satisfies%can be written as 
$$\begin{aligned} 
	E[\bar\Delta(\bs)] 
    	&= E[C_t(\bs)\bar\delta_t(\bs)]\\
        &= E[C_t(\bs)E[\delta_t(\bs) | Y_t(\bs), \hat\theta_t(\bs), \hat\delta_t(\bs)]]\\ 
        &= E[C_t(\bs)\delta_t(\bs)].
\end{aligned}$$
Hence it is reasonable to use $\bar\Delta(\bs)$ to approximate the causal effect.

\subsection{Computation}
To approximate the posterior of the causal effect $\Delta(\bs)$, we implement the spatial Bayesian analysis using a Markov chain Monte Carlo sampling. The missing values of observed $\PM$ are imputed and every model parameter is iteratively updated by the algorithm, conditional on all other parameters. The spatial range parameters, $\phi_1$ and $\phi_2$ are estimated empirically using variograms. All other model parameters have conditionally-conjugate priors and are accordingly updated with Gibbs steps where each step samples from their respective full conditional distributions (see Appendix \ref{app:FullConditionals} for derivations of the full conditional distributions). We use Gaussian Kriging to estimate smooth spatial surfaces across each study region for both the posterior means and standard deviations of each model parameter. We Kriged each estimate to the centroids of the $12\times 12$ km CMAQ grid. Our MCMC has a burn-in period of length 5,000, after which we collect samples every 100 iterations until a total of 30,000 iterations have been completed. To verify that the MCMC algorithm converged, we computed the effective sample size of the causal effect estimate, $\Delta(\bs)$, for each $\bs$. We also monitored convergence using visual inspection of trace plots for several representative parameters. Summary statistics and figures of the effective sample sizes and trace plots are included in the Supplemental Materials, Section 1.

\section{Fire-contributed $\PM$ estimates}

We let $C_t(\bs) = I[\hat\delta_t(\bs) > \tau]$, where ${\hat \delta}_t(\bs)$ is the CMAQ estimate of fire-attributed $\PM$ and $\tau$ is a fixed threshold. To select the threshold, we ran several models for a range of values of $\tau$ and used five-fold cross-validation to evaluate each model's ability to predict total $\PM$. We found little variation between the prediction metrics between each model. For example, mean-squared error (MSE) ranged from $12.58\mu g/m^3$ ($\tau=1\mu g/m^3$) to $12.71\mu g/m^3$ ($\tau=5\mu g/m^3$) (Supplemental Materials, Section 4, Table 1). We also examined variation in the causal effect when estimated with different values of $\tau$ and found the differences to be negligible except if $\tau$ is selected to be extreme (e.g. 0 or 10 \unit) (Supplemental Materials, Section 4, Figure 5). Based on these findings, we concluded that the model is robust to moderate choices for the threshold and we let $\tau = 1\mu g/m^3$ for the remaining analysis.

\begin{figure}[H]
	\centering
	\caption{\textbf{Maps of the bias terms}. Posterior means and standard deviations (SD) of the multiplicative bias for CMAQ's fire-contributed $\PM$, $\beta_1(\bs)$. The breakpoints correspond to the $2^{nd}$, $25^{th}$, $50^{th}$, $75^{th}$ and $98^{th}$ percentiles.}\label{fig:bias1_map}
	\begin{subfigure}[b]{.49\textwidth}
		\includegraphics[width=\textwidth]{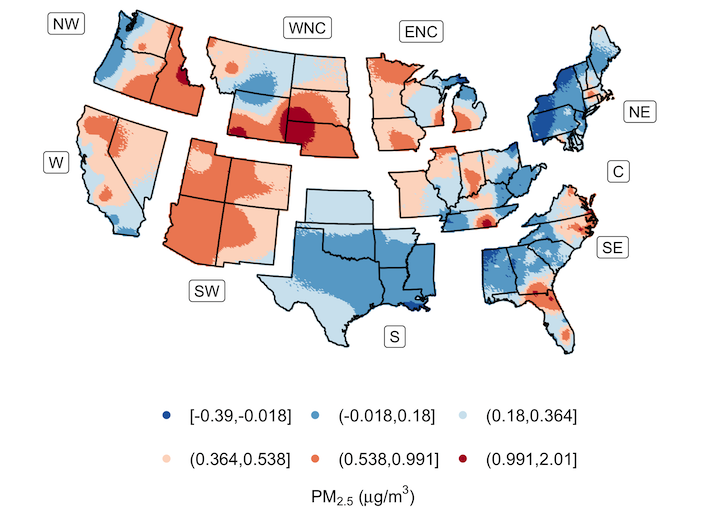}
		\caption{Posterior Mean of $\beta_1(\bs)$ (\unit)}\label{fig:bias1_mean}
	\end{subfigure}
	\begin{subfigure}[b]{.49\textwidth}
		\includegraphics[width=\textwidth]{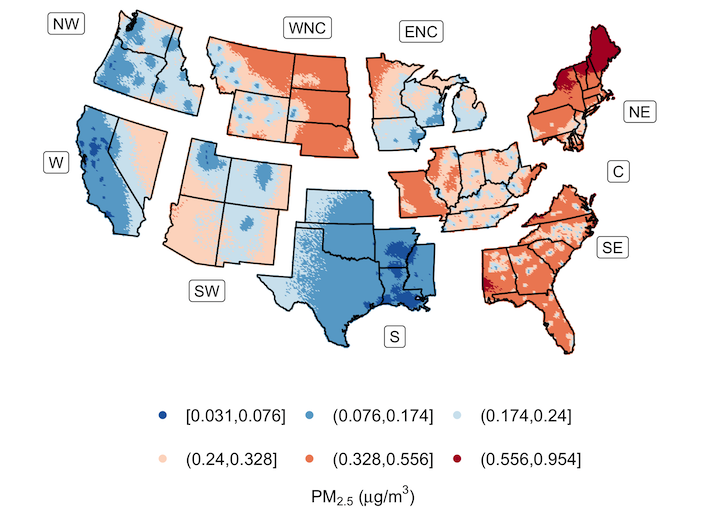}
		\caption{Posterior SD of $\beta_1(\bs)$ (\unit)}\label{fig:bias1_se}
	\end{subfigure}
\end{figure}

Figure \ref{fig:bias1_map} shows the posterior mean and standard deviation for the multiplicative bias parameter for the fire-contribution process, $\beta_1(\bs)$. Similar maps for the other bias parameters are included in Appendix \ref{app:effect-estimates}. The highest  $2^{nd}$ percent of $\beta_1(\bs)$ values reached ($0.991$, $2.01$ \unit), meaning that the strongest estimated association between CMAQ estimates and the monitor data occurs in the Northwest and West North Central (WNC) regions, along with parts of the East North Central region, the Southwest and parts of the Southeast region (Figure \ref{fig:bias1_map}). The lowest values ($-0.39$, $-0.018$ \unit) in the northern part of the East North Central region, the South and parts of the Northeast region. These have fewer wildfires (Figure \ref{fig:cmaq_maps}) and thus it is more difficult for CMAQ to estimate the relationship between model-estimated contribution and observed $\PM$. This is neither surprising nor problematic because these regions rarely experience fire smoke.

\begin{figure}[H]
	\caption{\textbf{Correlation and 95\% credible intervals.} The posterior means and 95\% credible intervals of $\gamma$, the correlation between $\theta_t(\bs)$ and $\delta_t(\bs)$.}\label{fig:corr}
	\centering
	\includegraphics[width=.75\textwidth]{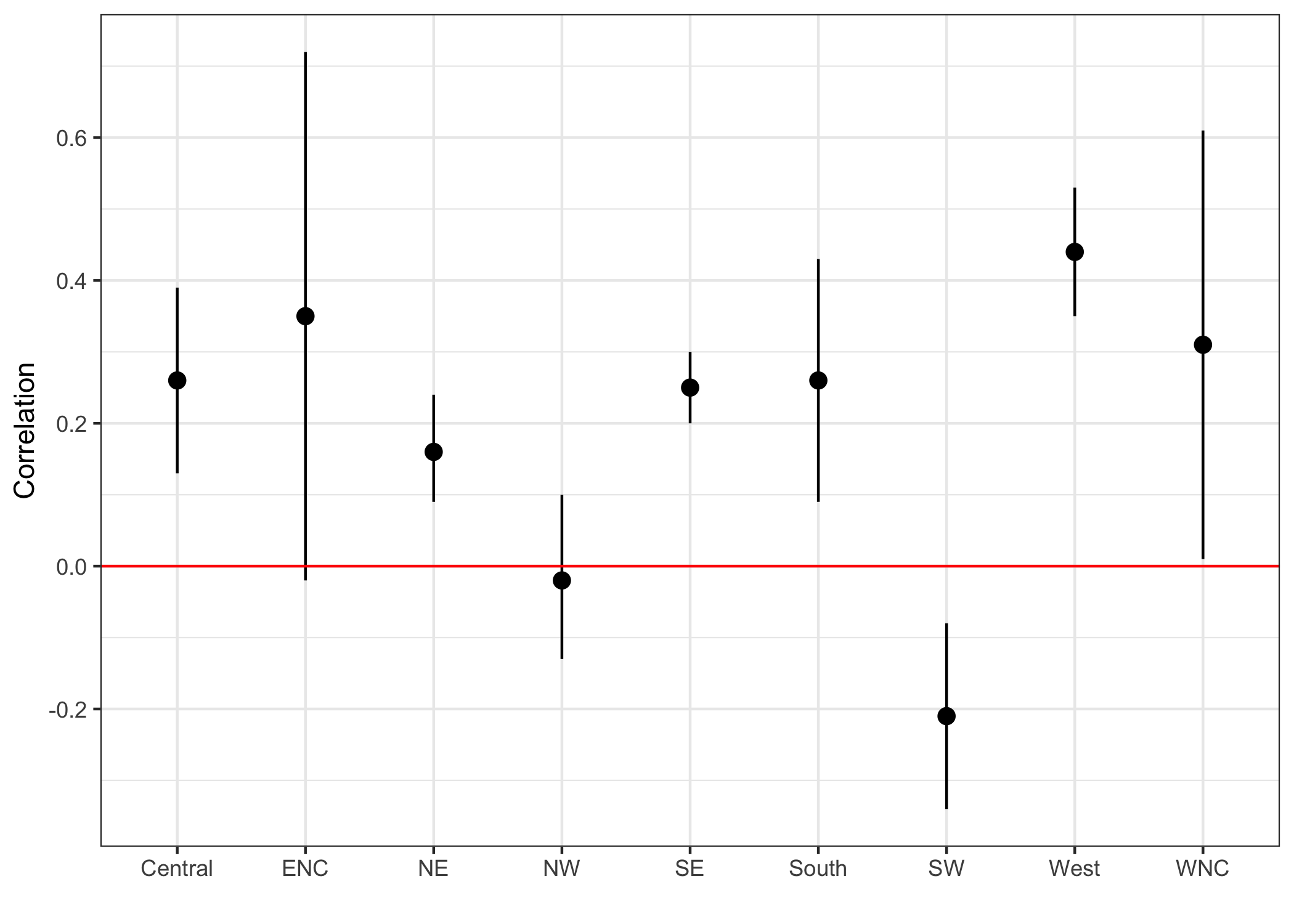}
\end{figure}

The posterior mean of the correlation between the counterfactual processes is summarized in Figure \ref{fig:corr}. Observing a positive correlation in a given region is indicative of fire smoke occurring in areas where non-fire $\PM$ emissions are present. A negative correlation indicates the converse. The highest estimated correlation is in the West region $(0.44 \pm 0.05)$, followed by the WNC region $(0.31 \pm 0.15)$ and then the South, Central and Southeast regions $(0.26 \pm 0.08, 0.26 \pm 0.06, 0.25 \pm 0.02)$. The Northeast region exhibits low correlation $(0.16 \pm 0.04)$. The correlation estimate for the Southwest region $(-0.21 \pm 0.06)$ is negative, and the only areas for which the correlation was plausibly zero were the ENC and the Northwest regions. To further illustrate the spatial correlation between observations, we also provide plots of Equation \ref{eq:covY} evaluated at the posterior mean of the model parameters for each region and combination of $C_t(\bs)$ in the Supplemental Materials (Section 3, Figure 4).

\begin{figure}[H]
	\centering
	\caption{\textbf{Causal effect estimate}. The maps show the posterior mean and standard deviation of the causal effects, $\Delta(\bs)$. Values are averaged over the 2008 to 2012 wildfire seasons at each site.}\label{fig:ce_map}
	\begin{subfigure}[b]{.49\textwidth}
		\includegraphics[width=\textwidth]{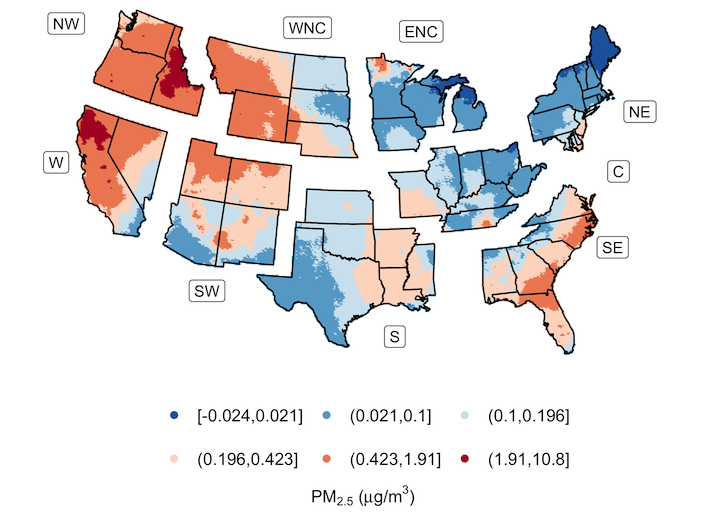}
		\caption{Posterior Mean of $\Delta(\bs)$ (\unit)}\label{fig:ce_map_mean}
	\end{subfigure}
	\begin{subfigure}[b]{.49\textwidth}
		\includegraphics[width=\textwidth]{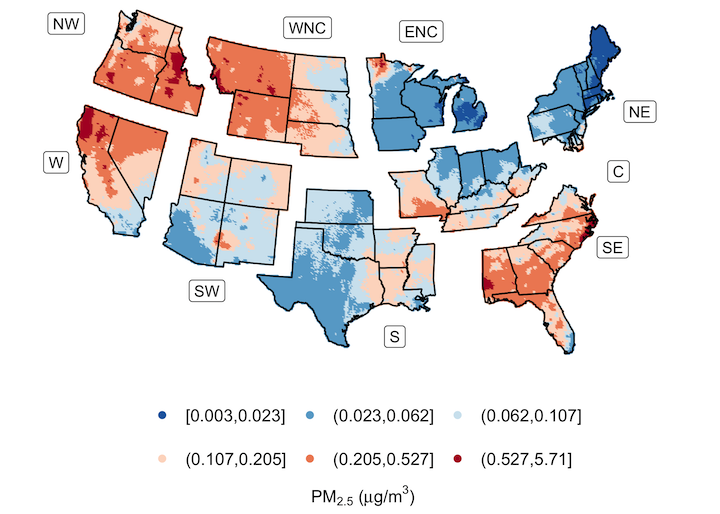}
		\caption{Posterior SD of $\Delta(\bs)$ (\unit)}\label{fig:ce_map_sd}
	\end{subfigure}
	\vskip\baselineskip
	\begin{subfigure}[b]{\textwidth}
		\centering
		\includegraphics[width=\textwidth]{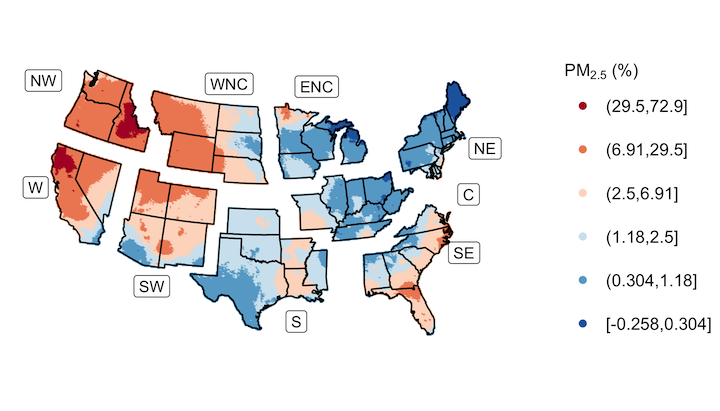}
		\caption{Fire-Contributed $\PM$ as \% of Total $\PM$ }\label{fig:ce_map_percent}
	\end{subfigure}
\end{figure}

Figure \ref{fig:ce_map} displays the causal effect estimates (Panel a), posterior standard deviation (Panel b) and the causal effect as percent of total estimated $\PM$ (Panel c). The largest estimates occur in the West, Northwest and WNC regions where wildfires are most prevalent. In these areas, between 29.5\% and 72.9\% of $\PM$ is attributable to wildfire smoke (Figure \ref{fig:ce_map}c). Moderate effects are estimated in areas of the South and Southeast, where prescribed burning is prevalent. The causal estimates in the East North Central region are in both the top and bottom two percent of fire-contributed $\PM$. This area is typically only effected by long-range smoke transport from the western U.S. or from Canadian wildland fires further north.  Large areas of the Northeast region have estimates near zero (some locations have very small negative values, likely due to statistical uncertainty).  

\begin{figure}[H]
	\centering
   	\caption{\textbf{Times series plot of estimates for one site}. Background ($\hat\theta_t(\bs)$) and total $\PM$ ($\hat\theta_t(\bs)+\hat\delta_t(\bs)$) from CMAQ, the posterior mean from the Bayesian model for background ($\theta_t(\bs)$) and total $\PM$ ($\theta_t(\bs) + C_t(\bs)\Delta_t(\bs)$), and the station measurements for an AQS site in Northern California ($-121.8^\circ, 39.8^\circ$) during the 2008 wildfire season.}\label{fig:oneSite_unobs}
 \includegraphics[width=0.75\textwidth]{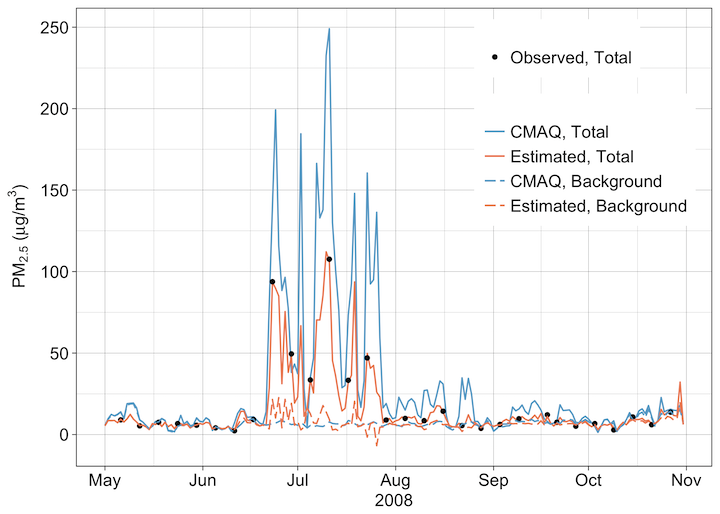}
\end{figure}

Figure \ref{fig:ce_map} shows Bayesian model estimates of background and total $\PM$, CMAQ-simulated background and total $\PM$, as well as observed $\PM$ during the 2008 wildfire season. 
Although the spatial pattern of the causal estimates resembles the CMAQ estimates, there are notable differences in the range of the estimates. Figure \ref{fig:oneSite_unobs} illustrates these differences at the same site on Northern California shown in Figure \ref{fig:oneSite_obs}. The estimates from the Bayesian causal model tend to fit closely to the observed values of $\PM$ from the monitor rather than to the CMAQ-simulated total $\PM$ and that the CMAQ model estimates are, on average, much higher. 

\begin{figure}[H]
	\centering
	\caption{\textbf{Causal estimates at monitoring stations}. Fire-contributed $\PM$ from the Bayesian model ($\Delta(\bs)$) versus the CMAQ model ($\hat\delta(\bs)$) at the AQS monitoring sites. Vertical error bars denote 95\% credible intervals. The dashed lines represent $x=y$ and $y=0$.}\label{fig:scatter_obs}
	\begin{subfigure}[b]{0.3\textwidth}
		\includegraphics[width=\textwidth]{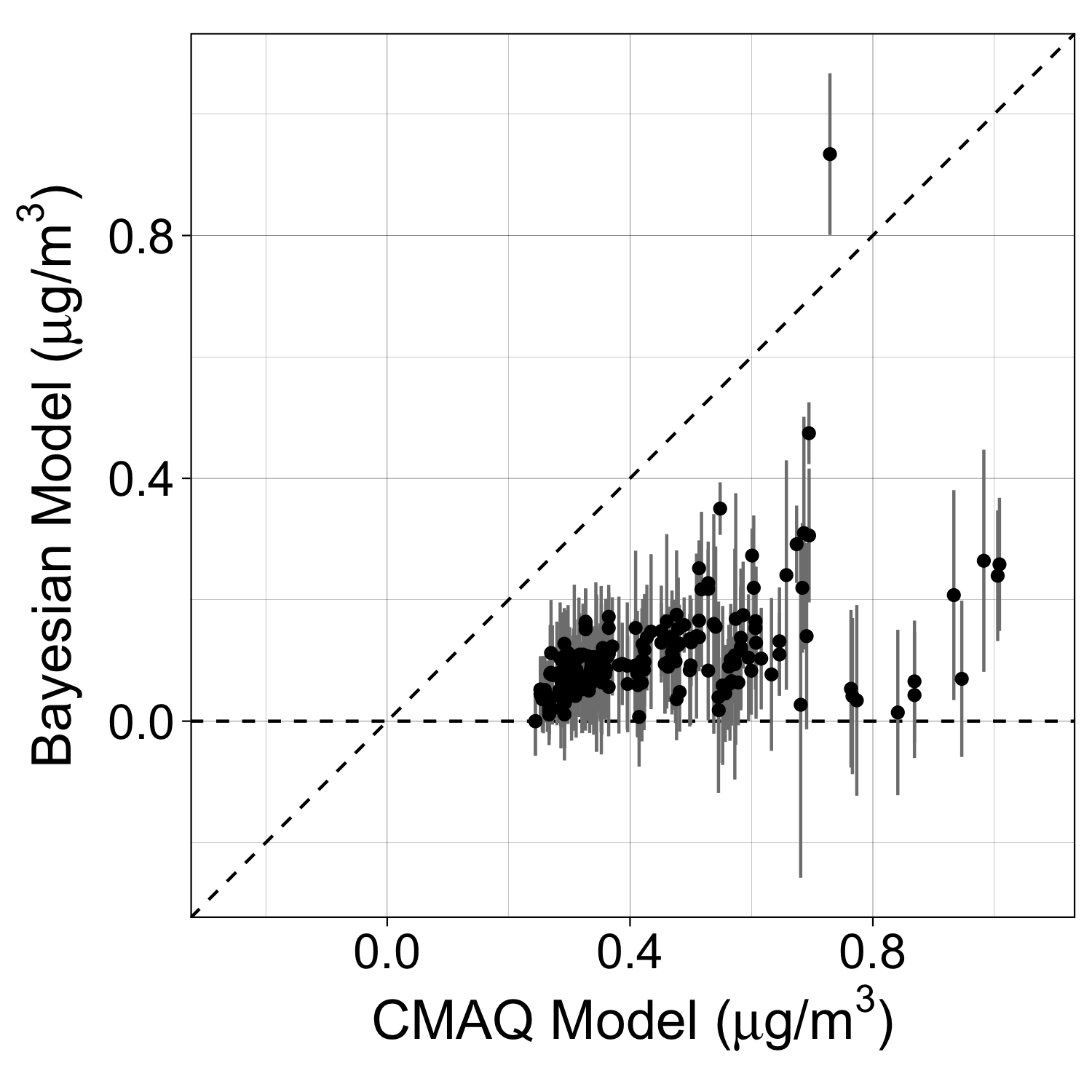}
		\caption{Central}
	\end{subfigure}	
	\quad
	\begin{subfigure}[b]{0.3\textwidth}
		\includegraphics[width=\textwidth]{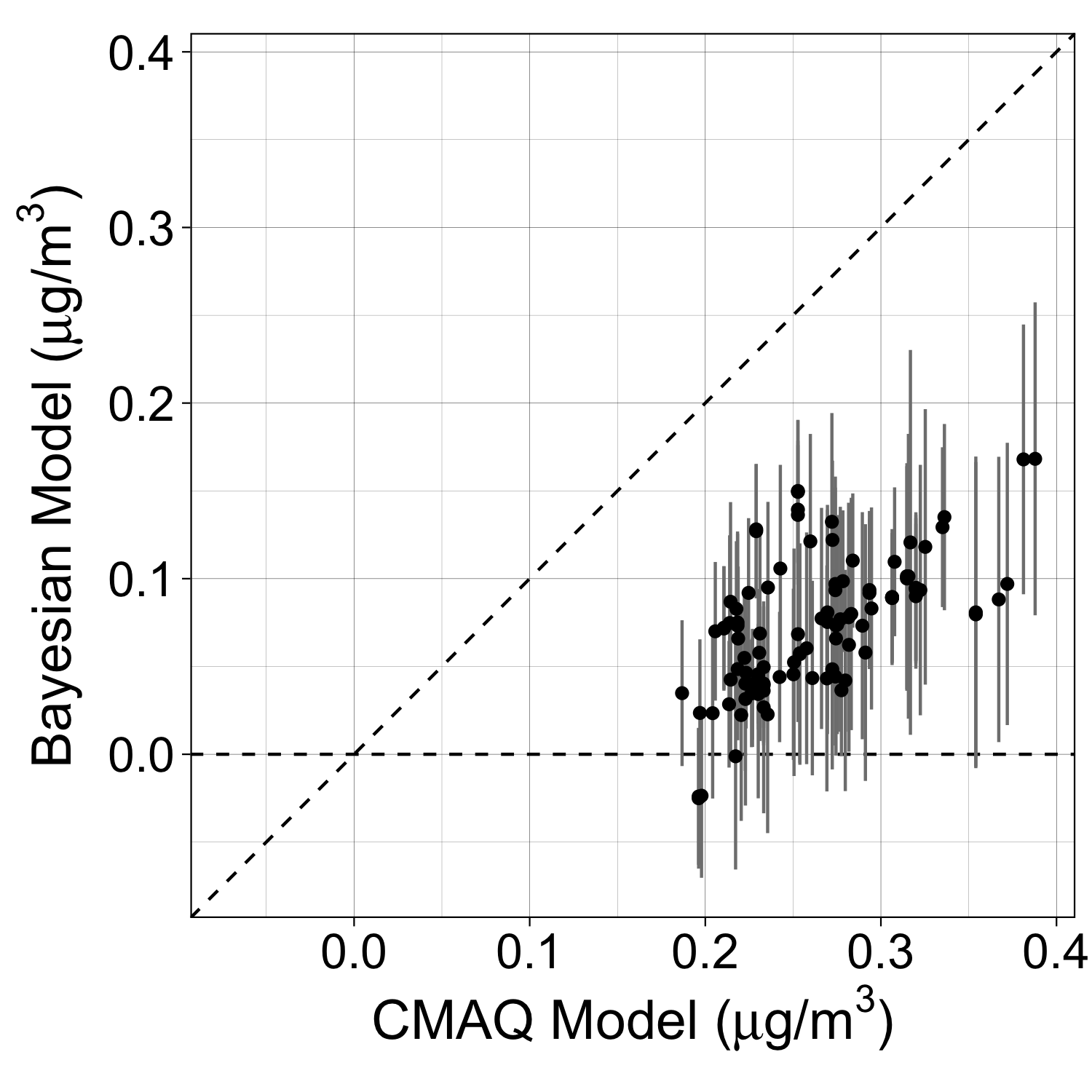}
		\caption{East North Central}
	\end{subfigure}
	\quad
	\begin{subfigure}[b]{0.3\textwidth}
		\includegraphics[width=\textwidth]{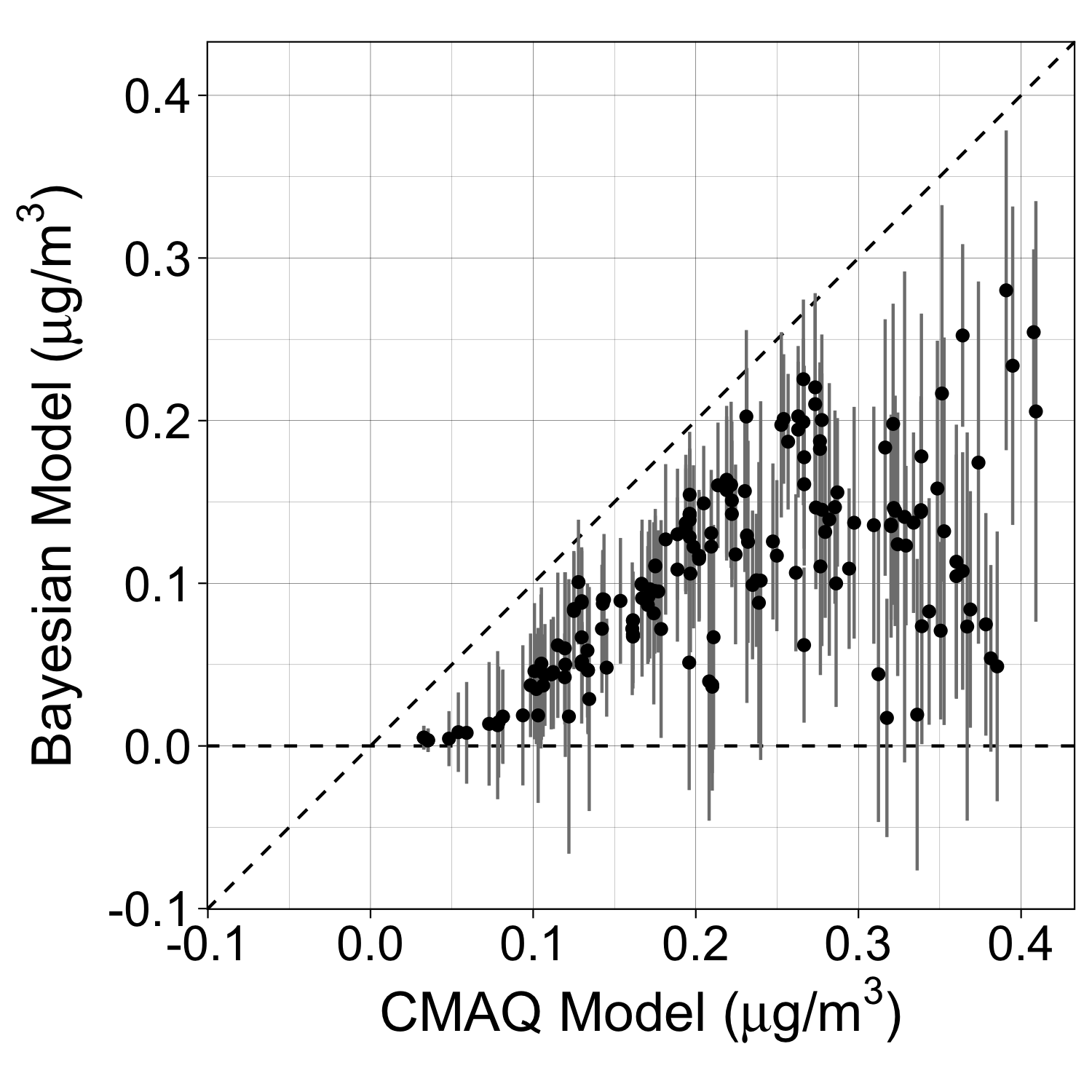}
		\caption{Northeast}
	\end{subfigure}
	\vskip\baselineskip
	\begin{subfigure}[b]{0.3\textwidth}
		\includegraphics[width=\textwidth]{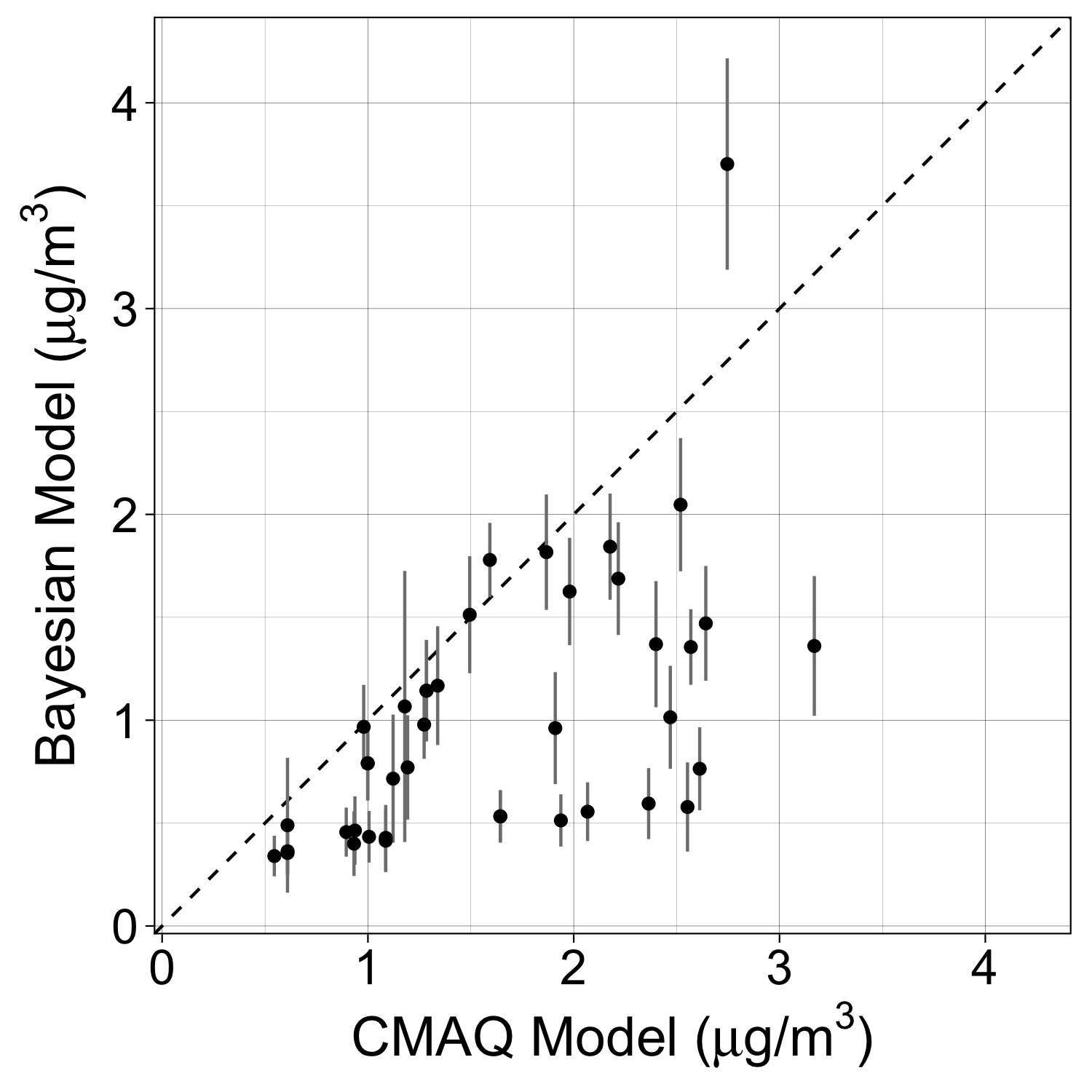}
		\caption{Northwest}
	\end{subfigure}
	\quad
	\begin{subfigure}[b]{0.3\textwidth}
		\includegraphics[width=\textwidth]{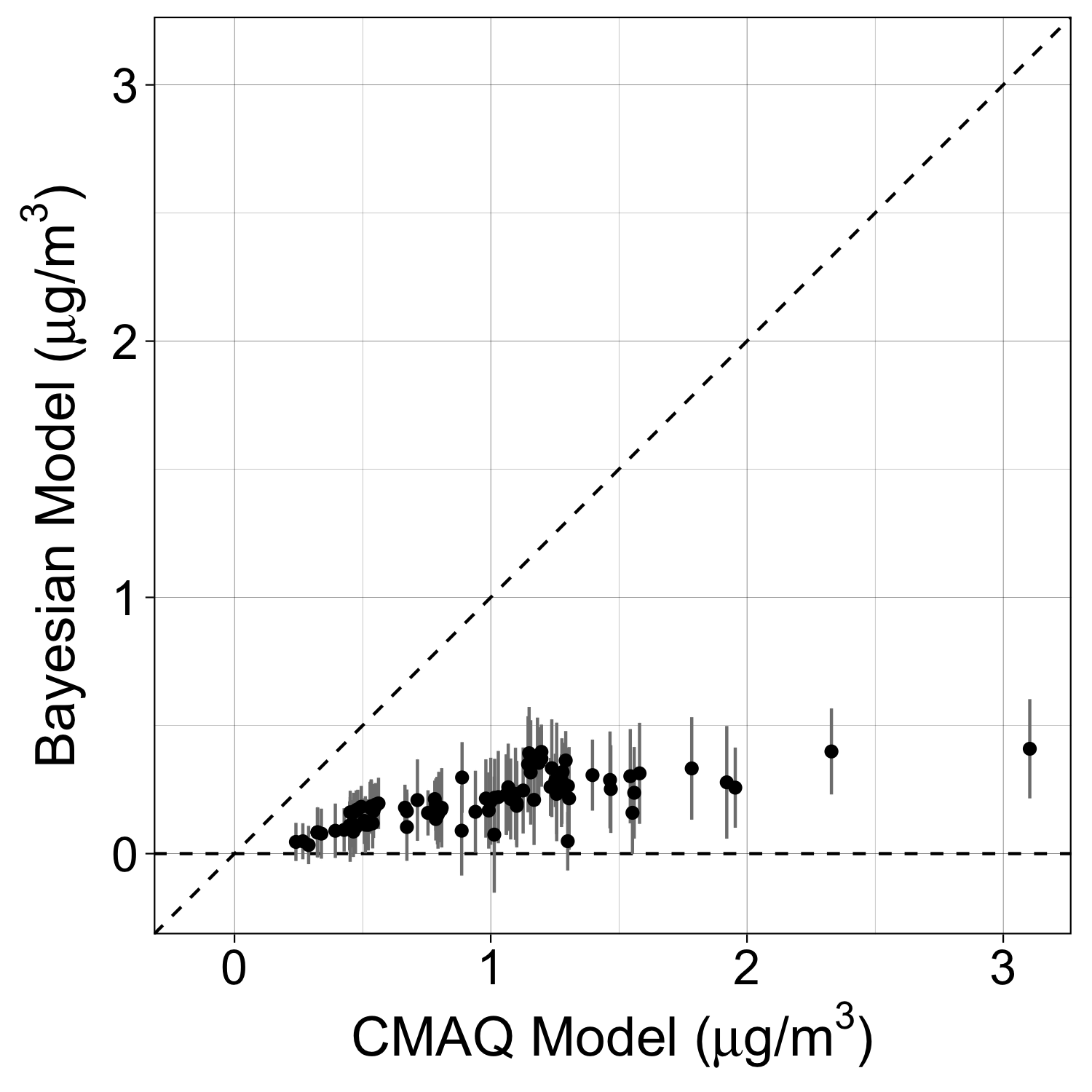}
		\caption{South}
	\end{subfigure}
	\quad
	\begin{subfigure}[b]{0.3\textwidth}
		\includegraphics[width=\textwidth]{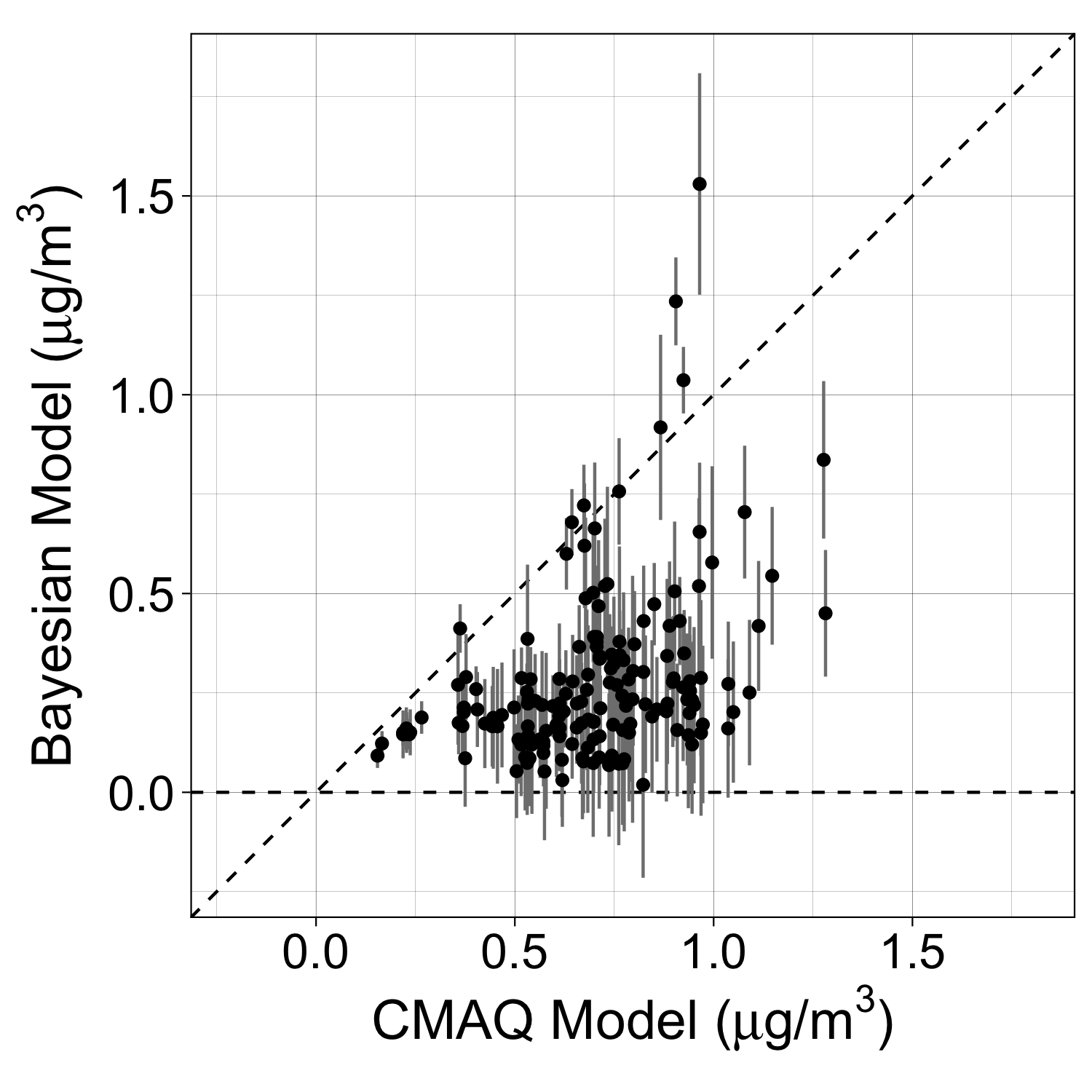}
		\caption{Southeast}
	\end{subfigure}
	\vskip\baselineskip
	\begin{subfigure}[b]{0.3\textwidth}
		\includegraphics[width=\textwidth]{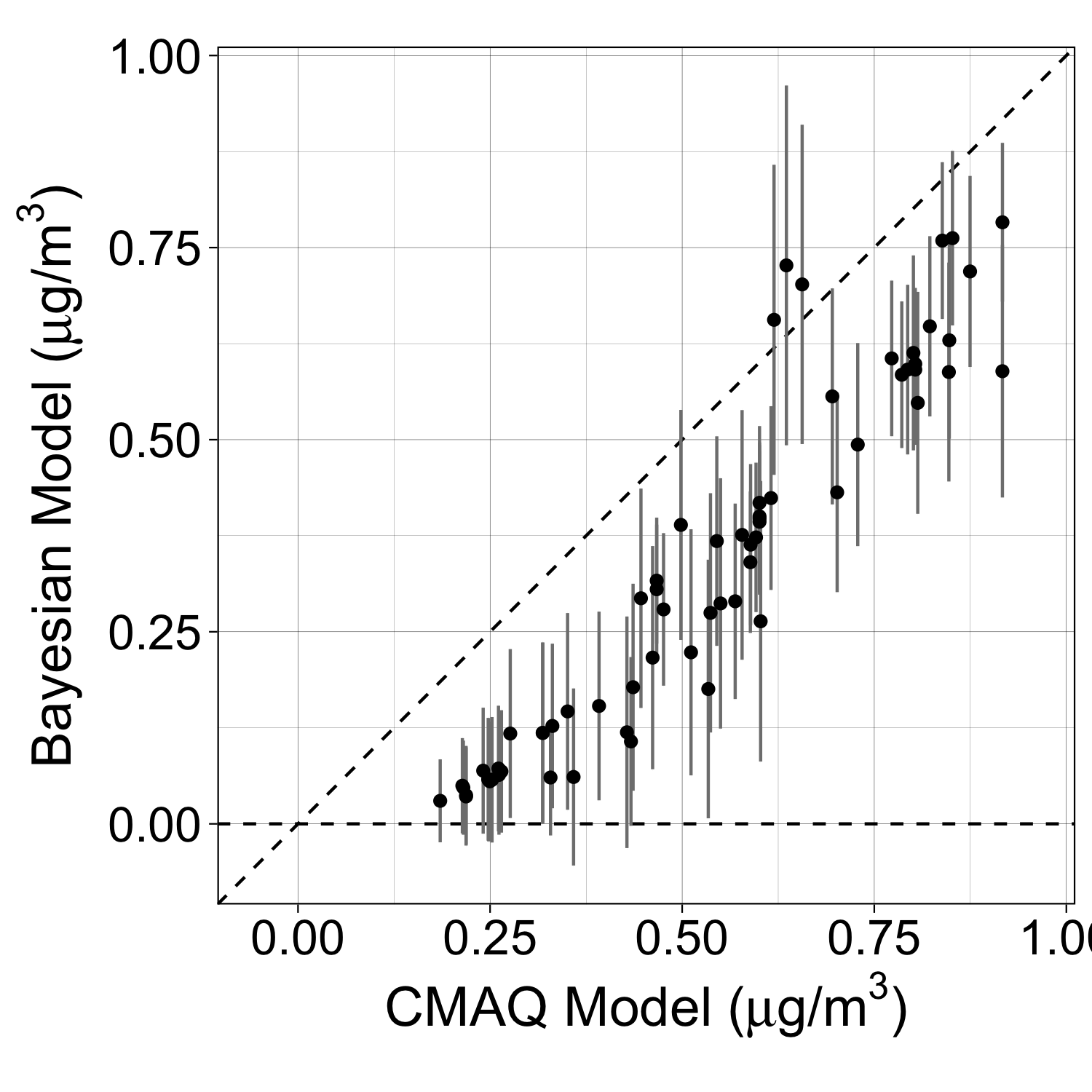}
		\caption{Southwest}
	\end{subfigure}
	\quad
	\begin{subfigure}[b]{0.3\textwidth}
		\includegraphics[width=\textwidth]{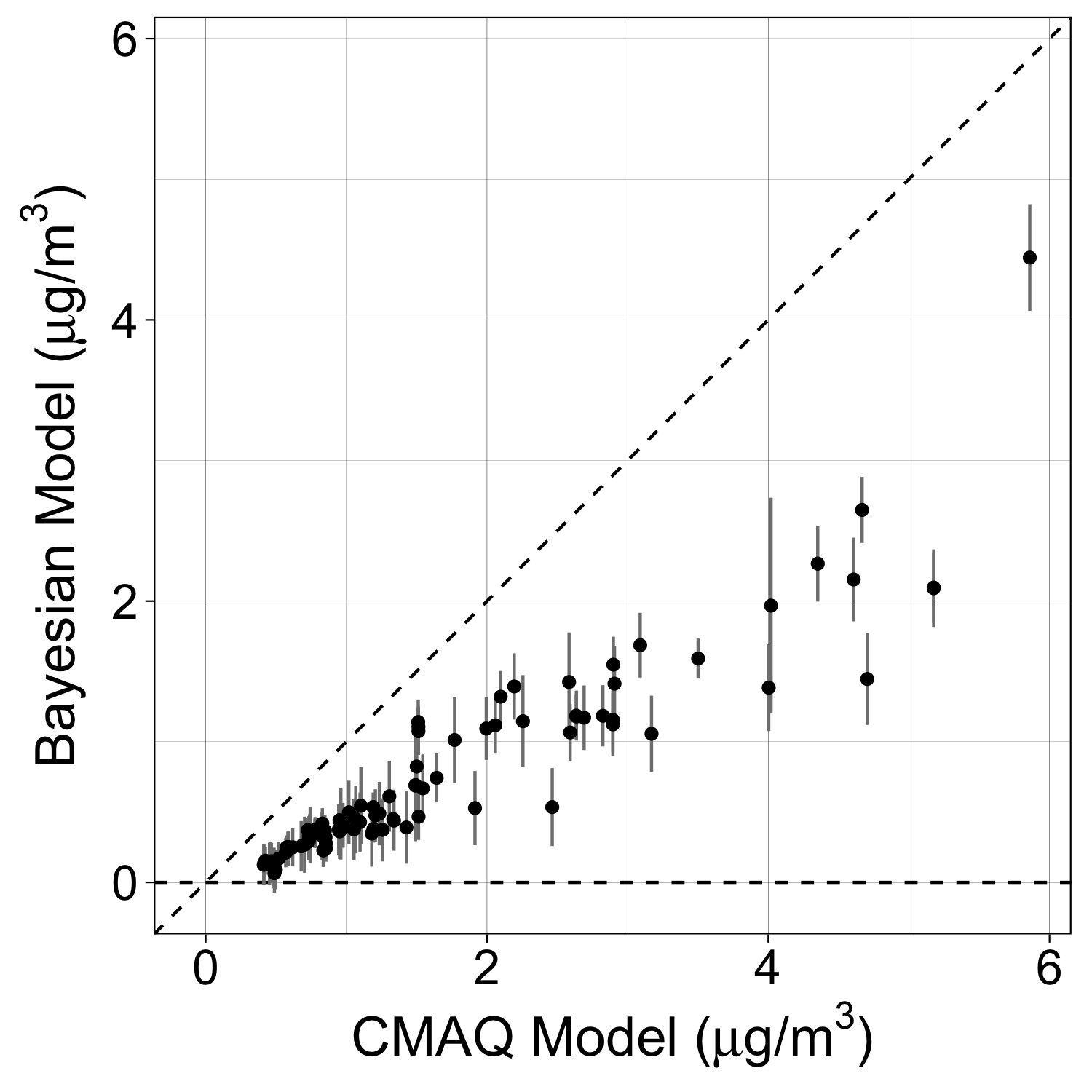}
		\caption{West}
	\end{subfigure}
	\quad
	\begin{subfigure}[b]{0.3\textwidth}
		\includegraphics[width=\textwidth]{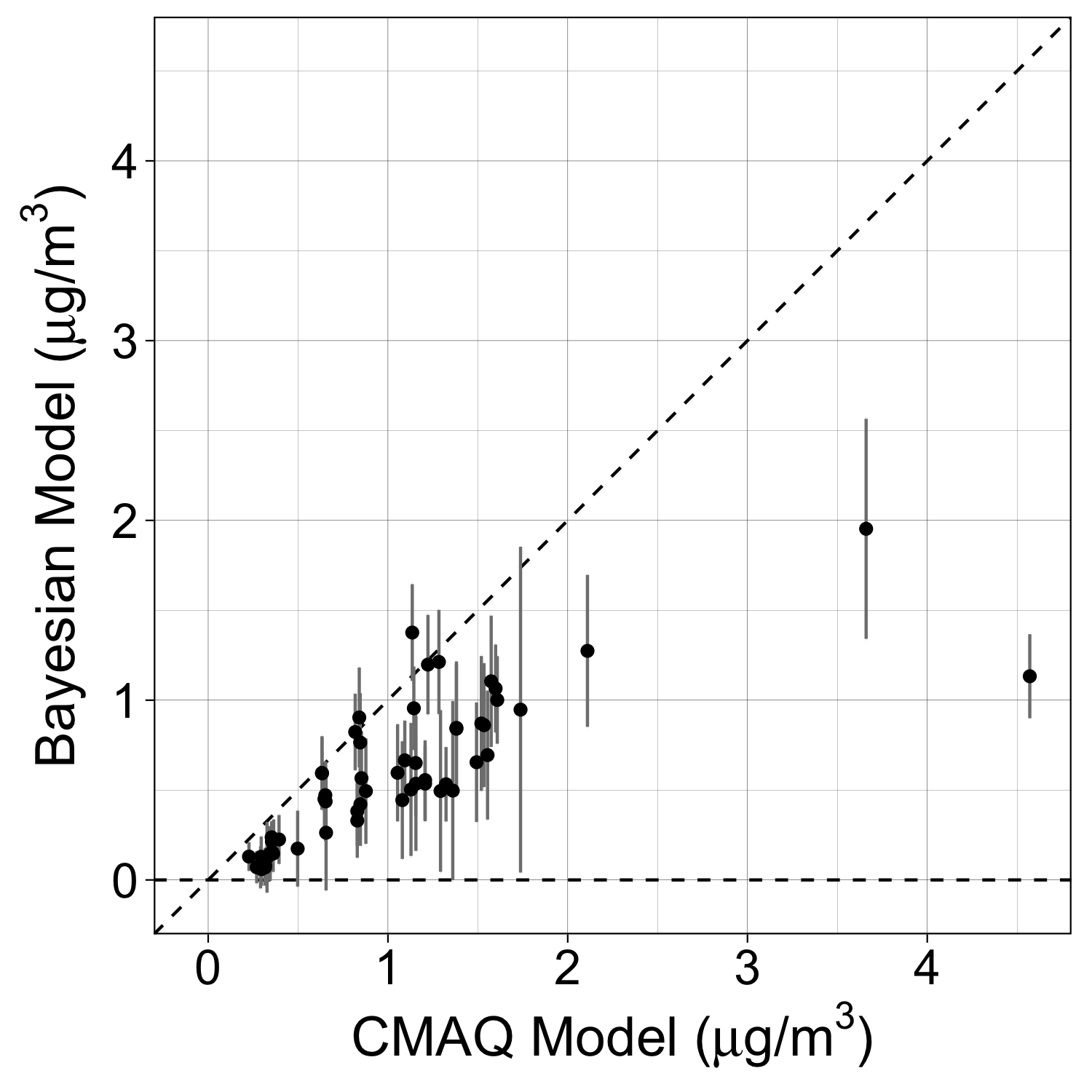}
		\caption{West North Central}		
	\end{subfigure}
\end{figure}

\begin{figure}[H]
	\centering
	\caption{\textbf{Causal estimates at prediction sites}. Fire-contributed $\PM$ from the Bayesian model versus the CMAQ model kriged to the $12\times 12$ km CMAQ grid. Vertical error bars denote 95\% credible intervals. The dashed lines represent $x=y$ and $y=0$.}\label{fig:scatter_pred}
	\begin{subfigure}[b]{0.3\textwidth}
		\includegraphics[width=\textwidth]{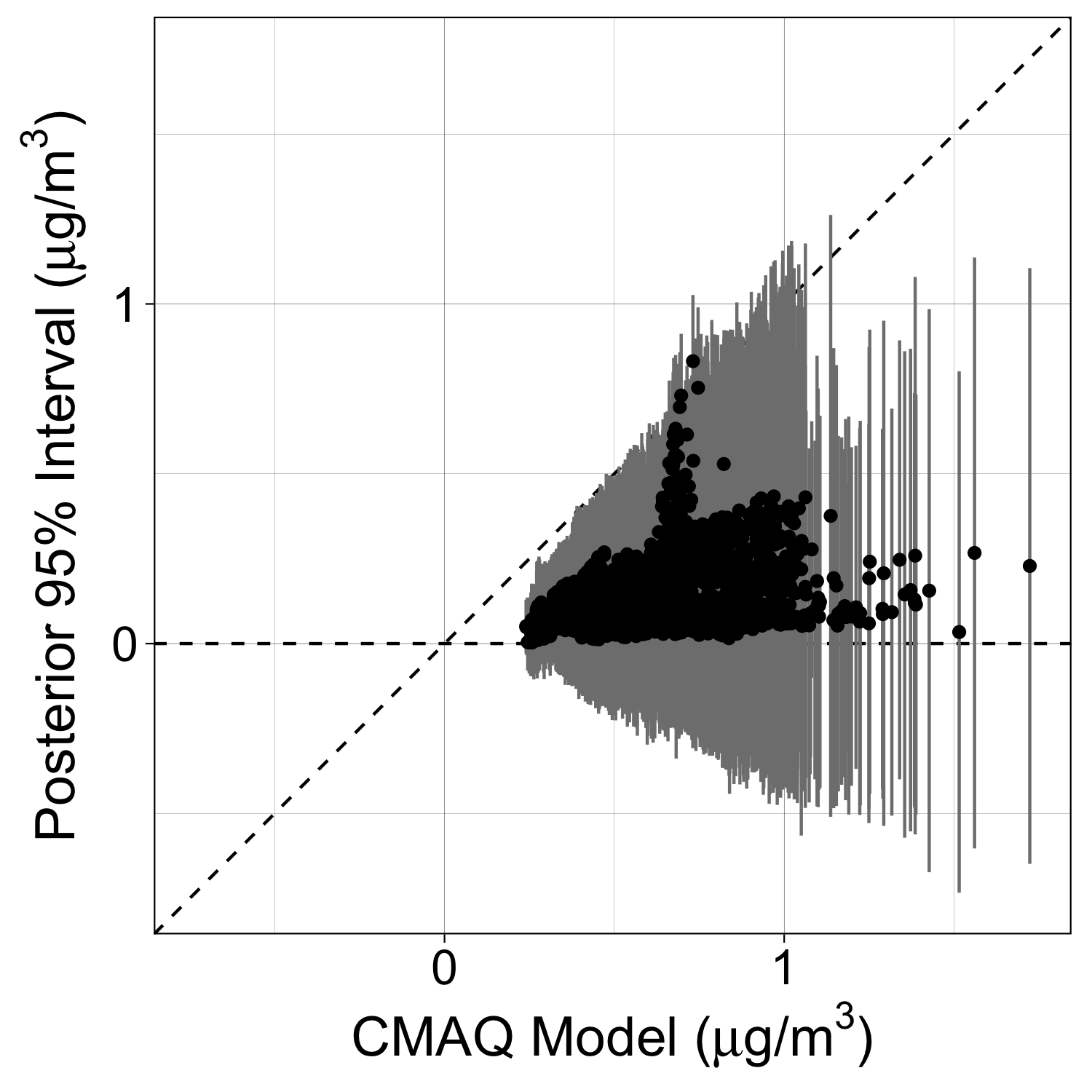}
		\caption{Central}
	\end{subfigure}	
	\quad
	\begin{subfigure}[b]{0.3\textwidth}
		\includegraphics[width=\textwidth]{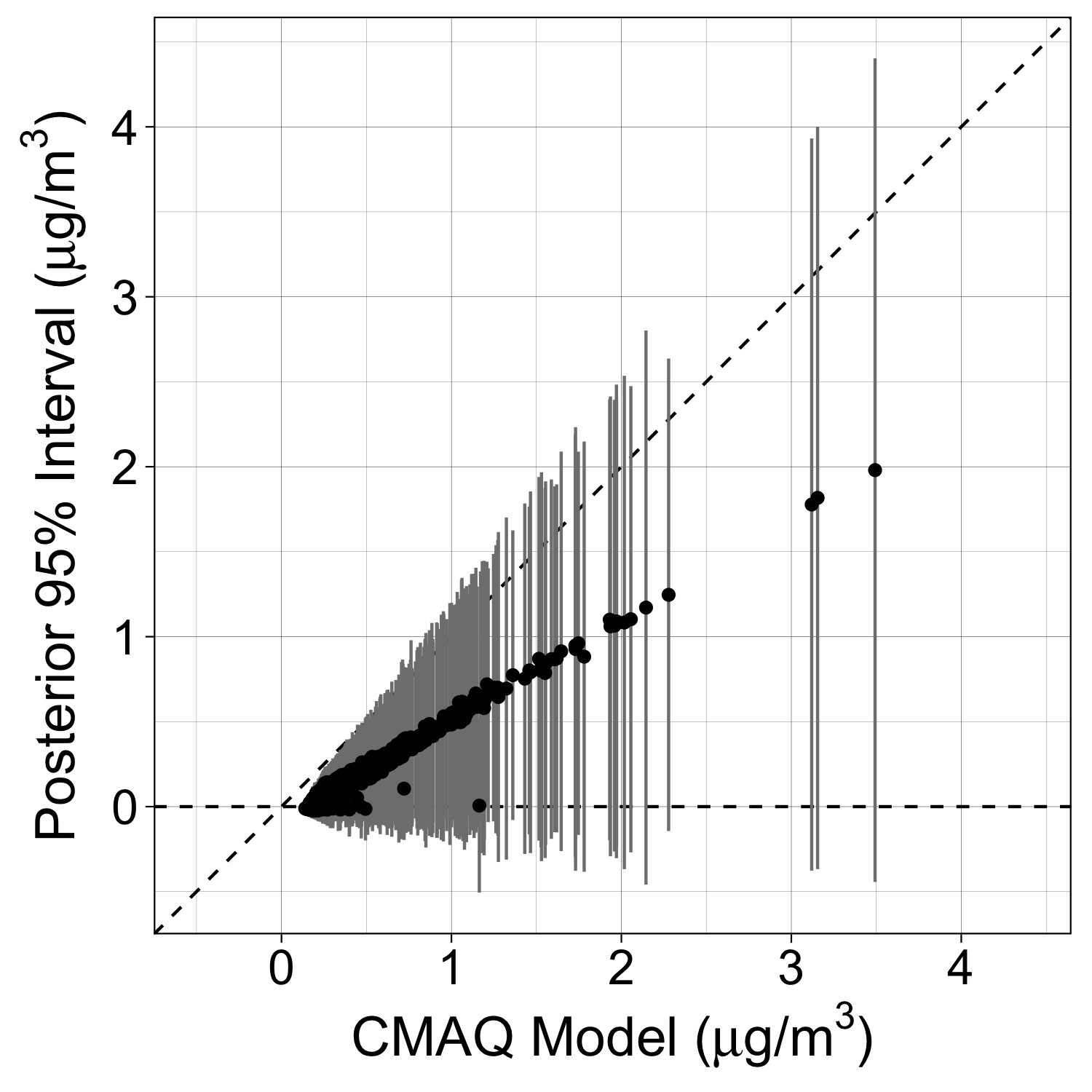}
		\caption{East North Central}
	\end{subfigure}
	\quad
	\begin{subfigure}[b]{0.3\textwidth}
		\includegraphics[width=\textwidth]{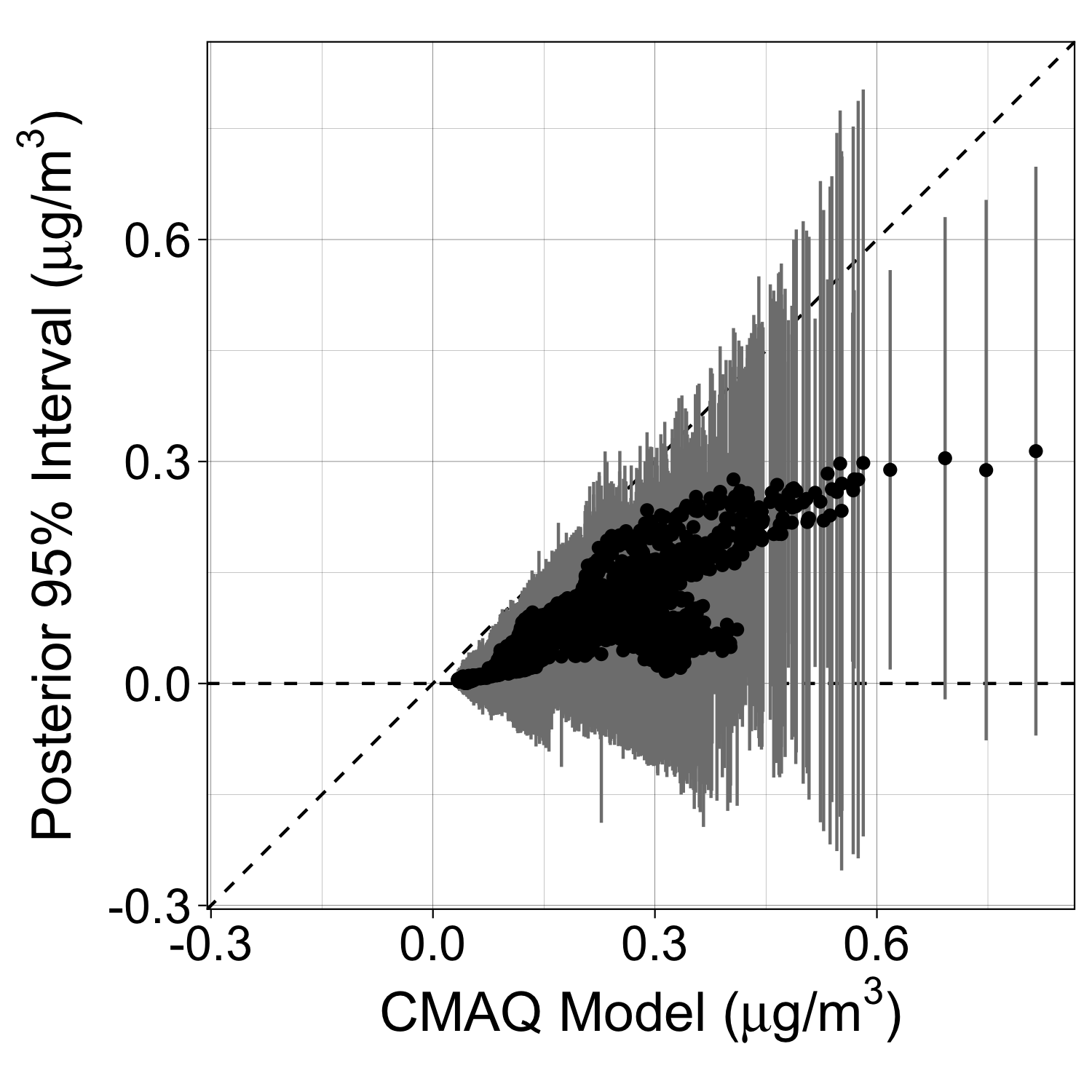}
		\caption{Northeast}
	\end{subfigure}
	\vskip\baselineskip
	\begin{subfigure}[b]{0.3\textwidth}
		\includegraphics[width=\textwidth]{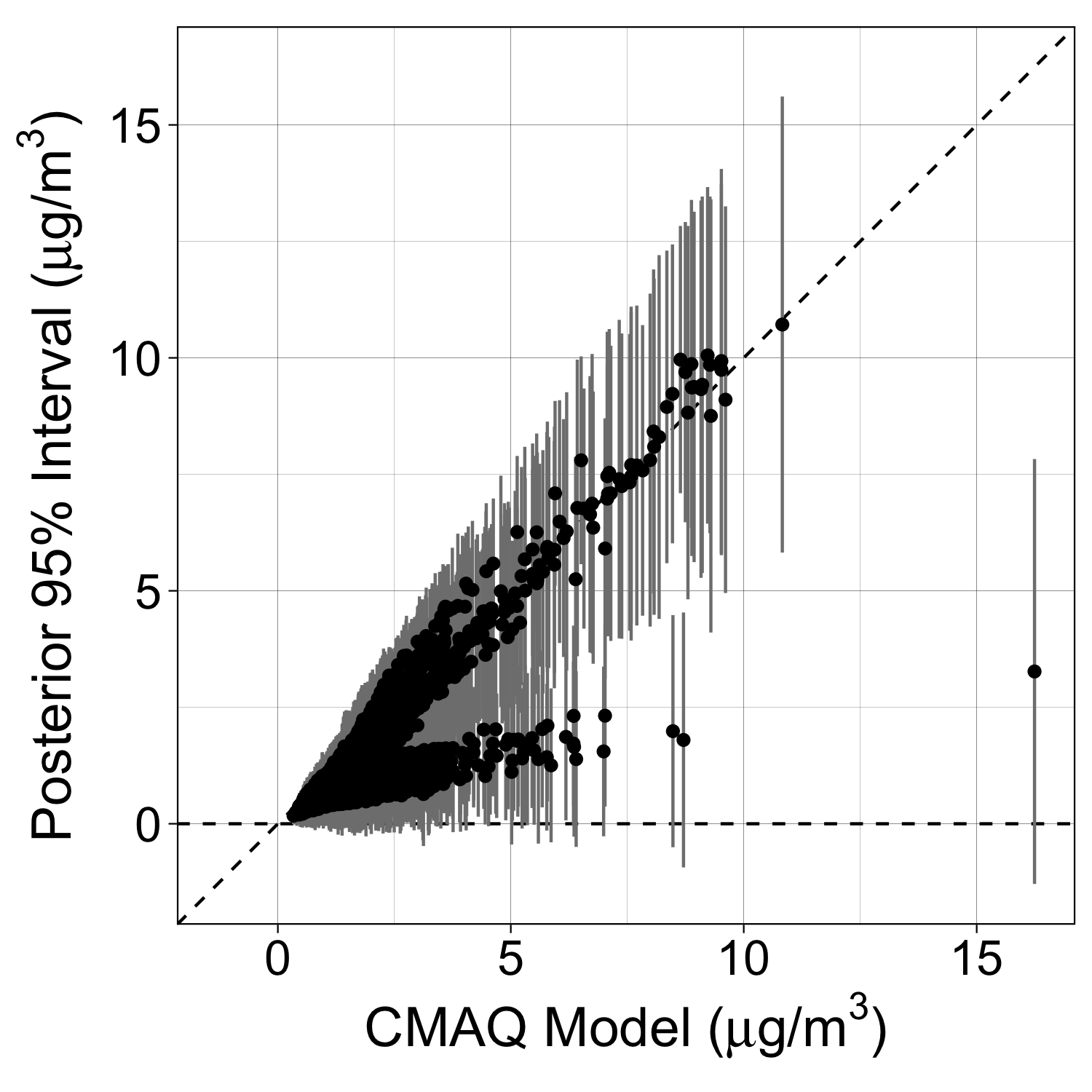}
		\caption{Northwest}
	\end{subfigure}
	\quad
	\begin{subfigure}[b]{0.3\textwidth}
		\includegraphics[width=\textwidth]{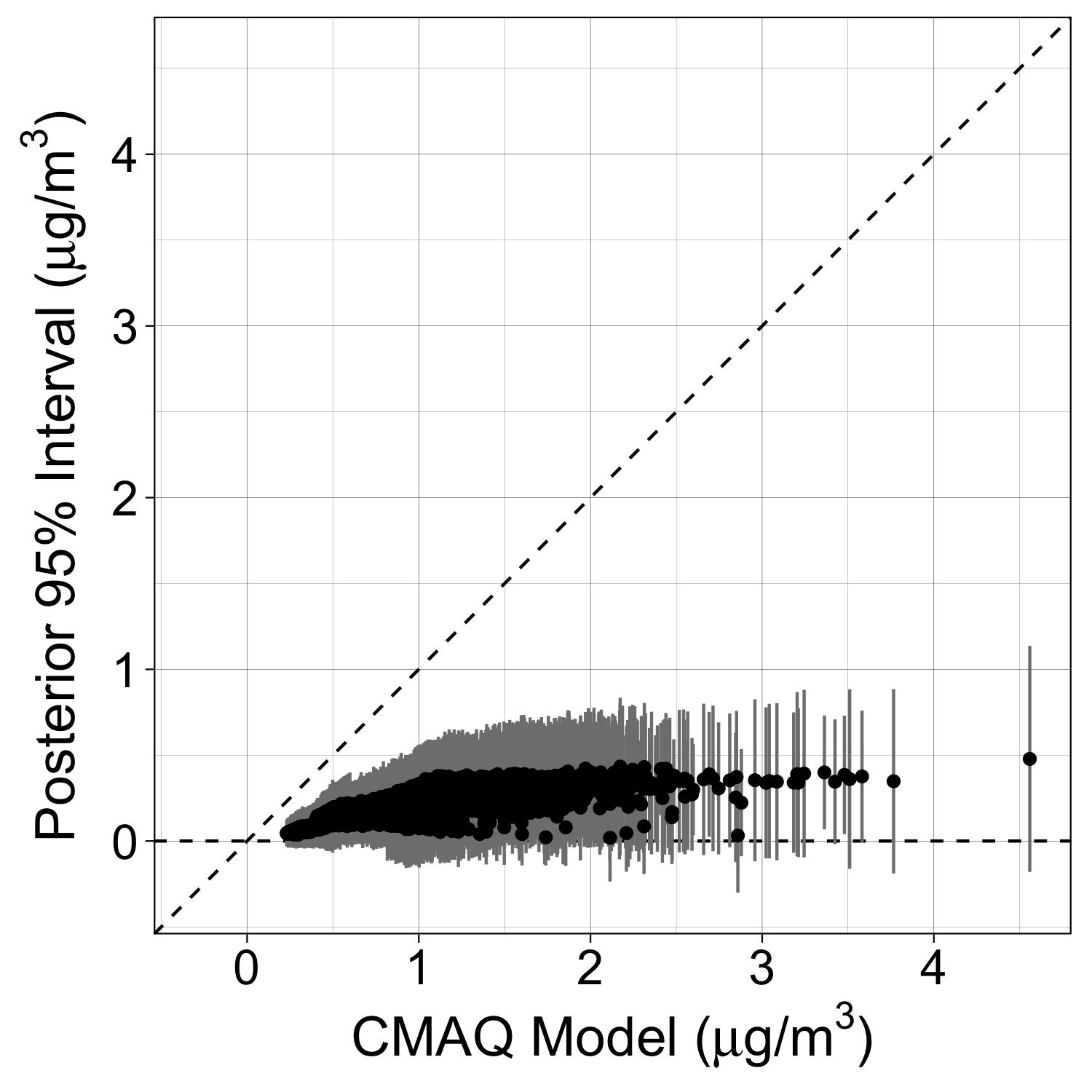}
		\caption{South}
	\end{subfigure}
	\quad
	\begin{subfigure}[b]{0.3\textwidth}
		\includegraphics[width=\textwidth]{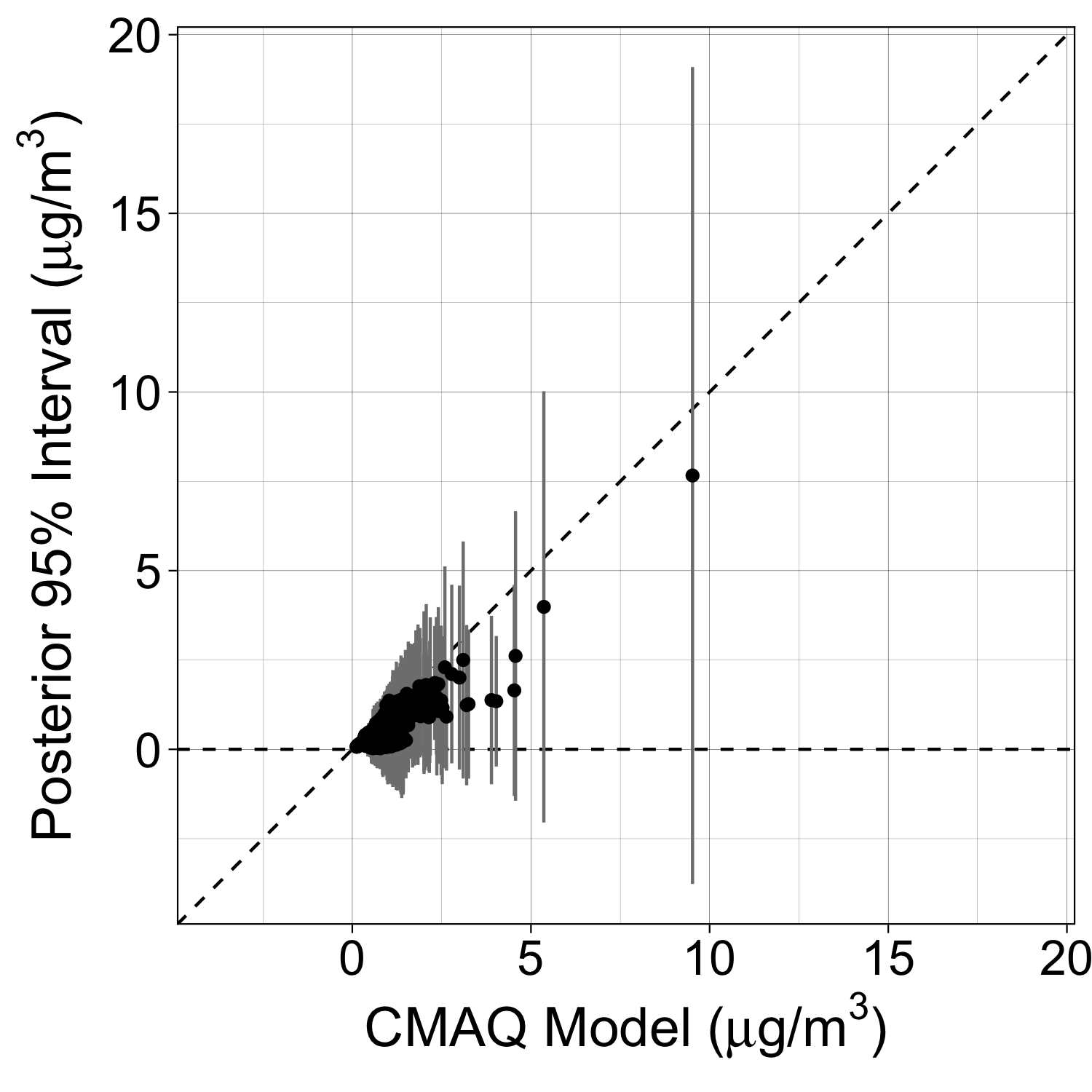}
		\caption{Southeast}
	\end{subfigure}
	\vskip\baselineskip
	\begin{subfigure}[b]{0.3\textwidth}
		\includegraphics[width=\textwidth]{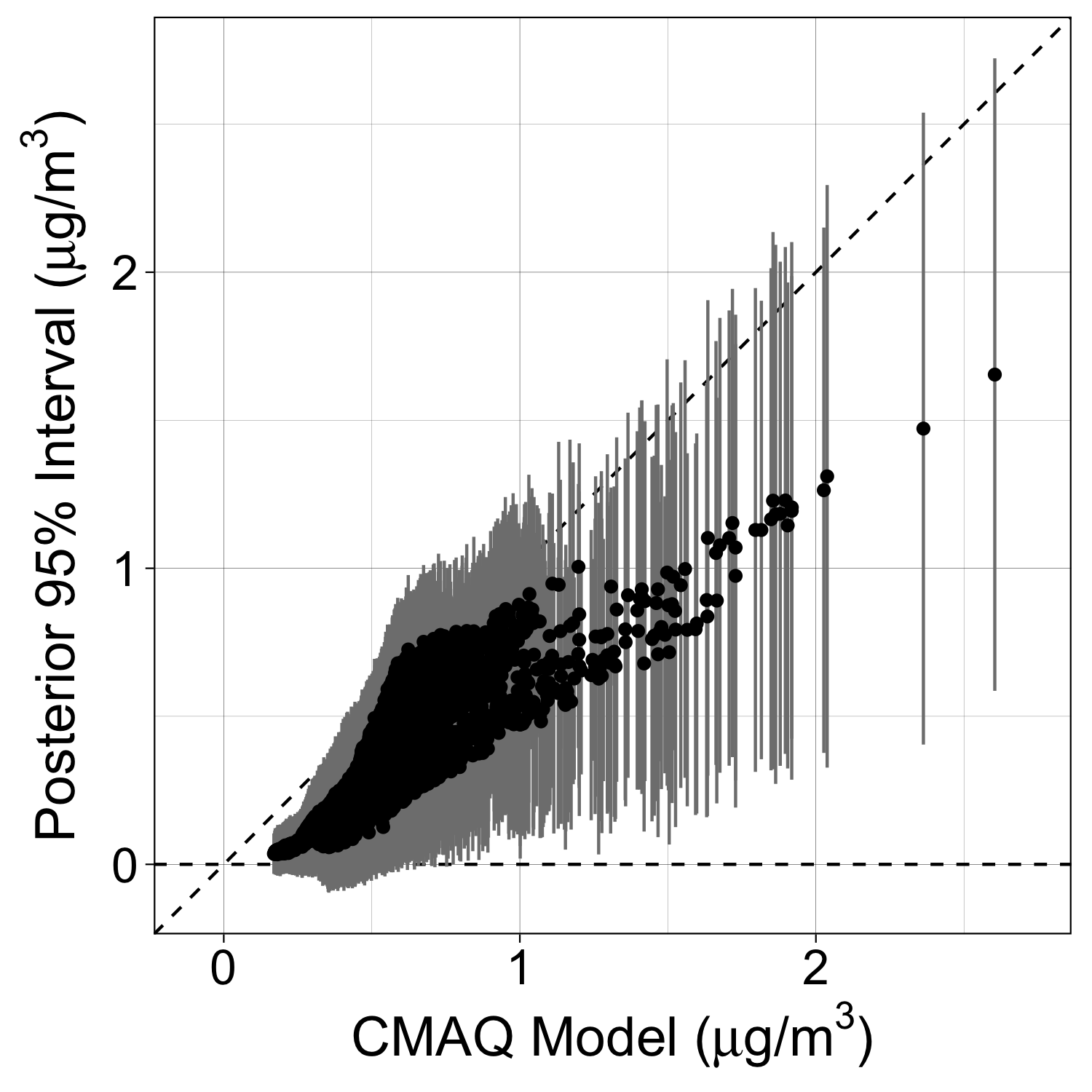}
		\caption{Southwest}
	\end{subfigure}
	\quad
	\begin{subfigure}[b]{0.3\textwidth}
		\includegraphics[width=\textwidth]{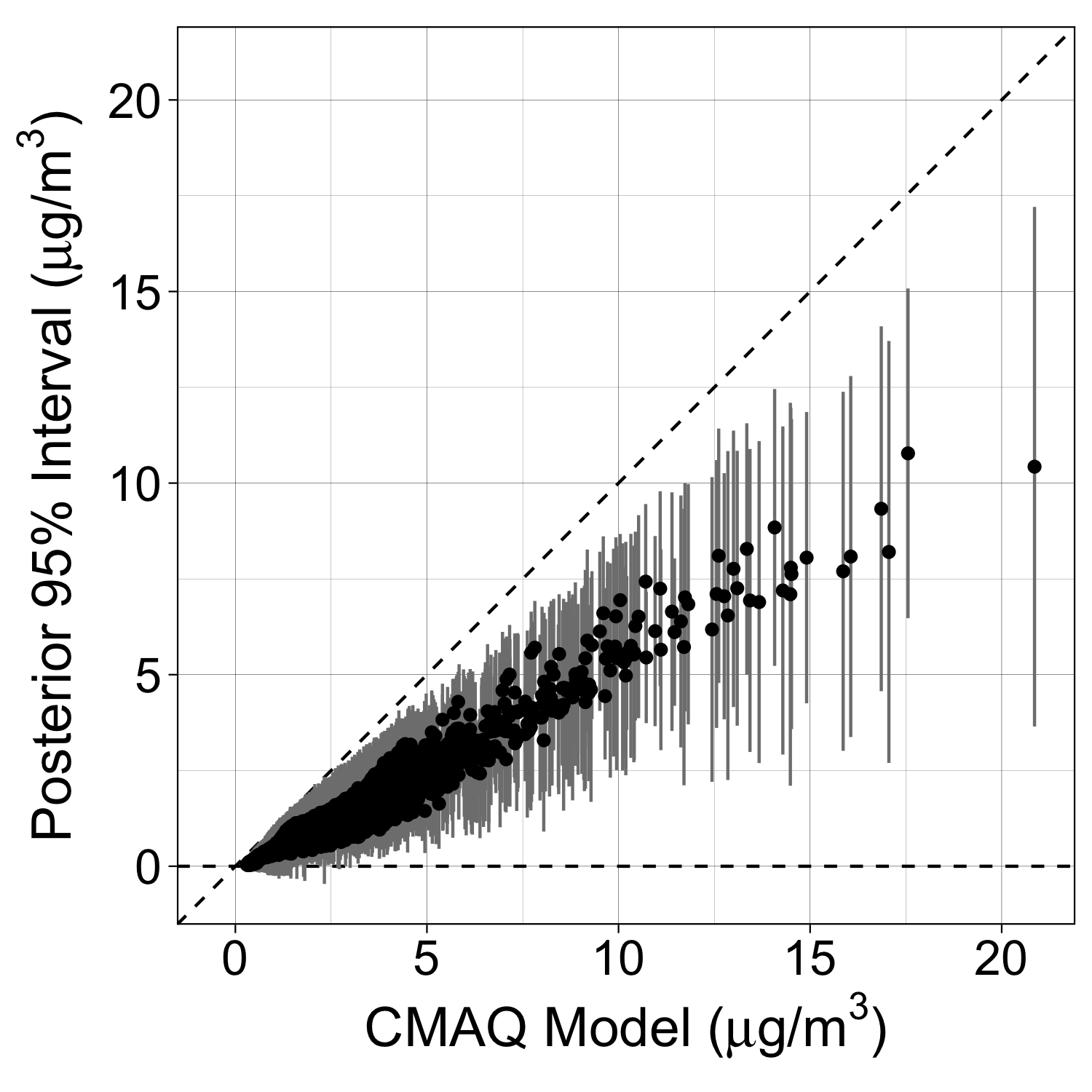}
		\caption{West}
	\end{subfigure}
	\quad
	\begin{subfigure}[b]{0.3\textwidth}
		\includegraphics[width=\textwidth]{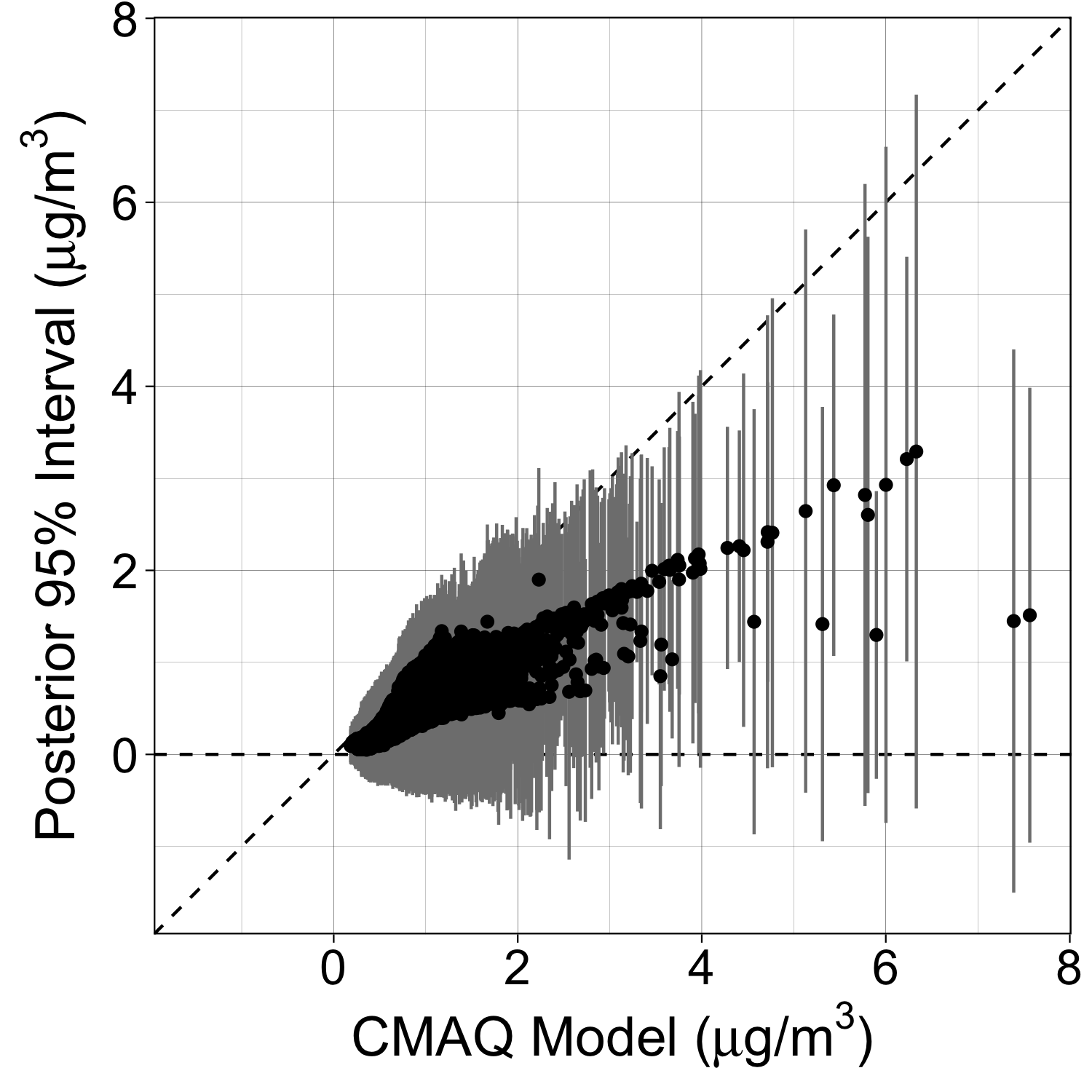}
		\caption{West North Central}		
	\end{subfigure}
\end{figure}

We also compare the estimates from the Bayesian causal model to those from CMAQ for all monitoring sites (Figure \ref{fig:scatter_obs}) and the prediction sites (Figure \ref{fig:scatter_pred}). As in Figure \ref{fig:oneSite_unobs}, the Bayesian model generally produces lower estimates of fire-contributed $\PM$ than CMAQ at all regions, both at monitoring sites (Figure \ref{fig:scatter_obs}) and prediction sites (Figure \ref{fig:scatter_pred}). The 95\% credible intervals are longer at the prediction locations than at the monitoring sites, which is to be expected in an interpolation analysis. Additionally, only in regions where fires are prevalent (e.g., West, Northwest, WNC) do we see causal effect estimates significantly different from zero. 

\section{Health burden analysis}
\begin{figure}[H]
	\centering
	\caption{\textbf{Distribution of cumulative health burden by county}. For each county, we aggregated the number of hospitalizations for respiratory illness across all age groups related to fire-contributed $\PM$ (the Bayesian estimate). The map displays the number of hospital admissions estimated over the 2008-2012 wildfire seasons.}\label{fig:map_burden}
	\includegraphics[width=\textwidth]{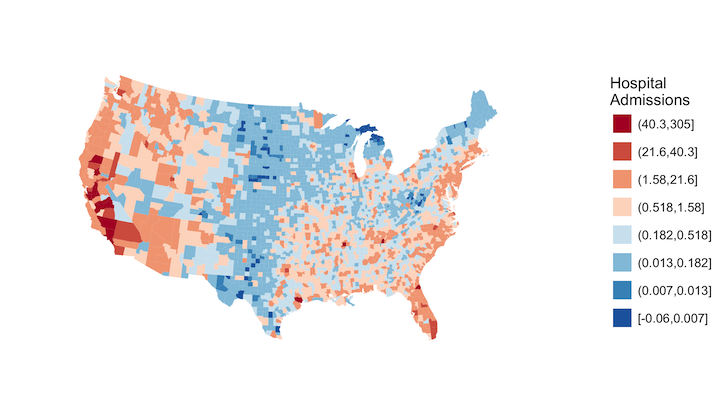}
\end{figure}

We use a log-linear concentration-response function to describe the relationship between $\PM$ and the number of hospitalizations due to respiratory illness. This analysis is conducted at the county level as well as by age group, $a$.  We define $\Delta_c$ as the integrated causal effect $\Delta(\bs)$ for $\bs$ in county $c$. The health impact function relating fire-contributed $\PM$ to changes in the incidence rate of hospitalizations due to respiratory illness is 

$$
	R_{ac} = r^0_a n_c(e^{r_a \Delta_c} - 1)
$$ 
\noindent where $n_c$ is the population of county $c$ based on the July 2010 U.S. Census and $r_a^0$ is the incidence rate of hospitalizations for respiratory illness by county and age group \citep{Benmap2017}. Using $r_a$, we calculate cumulative daily burden over all days in the study (May-October, 2008-2012) by county and age group (Figure \ref{fig:map_burden}) \citep{Delfino2009}. Cumulative $R_{ac}$ over all counties in each region is summarized in Table \ref{tab:avg_cum_burden} based on both the Bayesian and the CMAQ estimate of fire-contributed $\PM$. We note that these estimates have a causal interpretation only if the estimates in \cite{Delfino2009} have a causal interpretation.  While \cite{Delfino2009} account for many known confounders for fire-contributed $\PM$ and respiratory illness and the \cite{ISAPM2009} declares that the adverse effects of short-term $\PM$ exposure on respiratory outcomes is likely to be causal (using the Hill criteria), this remains an important caveat.

The Bayesian estimate yields more conservative estimates of the impact of fire-contributed $\PM$ on hospital admission rates for respiratory illness. The highest estimated burden is observed in the West region, notably in Southern California with upwards of 300 hospitalizations estimated cumulatively over the 2008 to 2012 fire seasons in some counties (Figure \ref{fig:map_burden}). In Table \ref{tab:avg_cum_burden}, the highest estimated burden for any region is in the West with 1513.9 hospitalizations over the 2008-2012 fire seasons using the Bayesian estimate of the causal effect. If the CMAQ estimate for the causal effect is used, the cumulative burden in the West is estimated to be 3500.4 hospitalizations per day. Most counties in the rest of the country exhibit lower burden with less than 5 hospitalizations per county over the 2008-2012 fire seasons (Figure \ref{fig:map_burden}).

\newgeometry{margin=1cm}
\begin{landscape}
\begin{table}[H]
\centering
\caption{\textbf{Number of hospitalizations in each region.} Cumulative number of hospitalizations for respiratory illness due to wildland fires over the 2008-2012 fire seasons in each region calculated using the Bayesian and the CMAQ estimate of the causal effect, $\Delta_t(\bs)$, by region. 95\% confidence intervals are provided.}
\label{tab:avg_cum_burden}
\begin{tabular}{ccccccc}
\cline{1-7}
\multirow{2}{*}{Region} & \multirow{2}{*}{$\Delta$} & \multicolumn{5}{c}{Age Group (years)} \\ \cline{3-7} 
        &                           &  0-1 & 2-34 & 35-64 & 65-99 & 0-99   \\ \cline{1-7} 
\multirow{2}{*}{Central}
	& Bayesian           & 150.5 (33.9, 270.6) & 60.1 (-19.5, 146.6)  & 159.8 (33.5, 290.7)  &  283.7 (104.8, 460.1) &  654.1 (152.7,1168.0) \\
	& CMAQ               & 612.8 (137.7, 1103.6)  &   242.3 (-77.9, 592.1)  &   663.5 (139.1, 1208.7)  &  1161.1 (428.4, 1884.9)  &  2679.8  (627.4, 4789.3)     \\ \cline{3-7}
\multirow{2}{*}{ENC}   
	& Bayesian           &  26.7(6.0, 48.2)   &   10.4 (-3.4, 25.4) &    28.5 (6.0, 52.0) & 56.3 (20.8, 91.4) &  121.9 (29.3, 217.1) \\
         & CMAQ              & 95.4 (21.4, 172.3) & 37.1 (-12.0, 91.1) &  103.7 (21.7, 189.1) & 207.7 (76.6, 337.5) &  443.9 (107.7, 790.0) \\ \cline{3-7}
\multirow{2}{*}{South}     
	& Bayesian           & 134.9 (30.4, 242.1)  &   47.8 (-15.4, 116.2)  &   127.3 (26.7, 231.5)  & 248.9 (92.0, 403.4) & 558.9 (133.8, 993.1) \\
        & CMAQ               & 604.0 (135.1, 1093.7) &  211.0 (-67.2, 518.7) &   562.1 (117.5, 1027.1) & 1102.0  (405.4, 1794.2) &  2479.1 (590.8, 4433.7) \\ \cline{3-7}
\multirow{2}{*}{Southeast} 
	& Bayesian           & 279.3 (62.4, 506.0) & 118.7 (-37.1, 290.9)  &   324.8 (67.9, 593.3)  &   565.6 (208.2, 920.5)  & 1288.3  (301.4, 2310.6) \\
        & CMAQ               & 642.5 (144.1, 1160.6) &  272.7 (-86.8, 667.4) & 746.6 (156.4, 1361.3) & 1284.4 (473.4, 2087.2) &  2946.1 (687.1, 5276.5) \\ \cline{3-7}
\multirow{2}{*}{Southwest} 
	& Bayesian           & 113.6 (25.5, 205.1) &   33.8 (-13.1, 85.4) & 49.1 (10.3, 89.4) &   85.6 (31.6, 139.1)   & 282.1  (54.2, 519.2)  \\
        & CMAQ               & 183.6 (41.2, 330.8) & 57.3 (-21.7, 144.0) &   89.3 (18.7, 162.7) & 157.0 (58.0, 254.9) &  487.3 (96.2, 892.4) \\ \cline{3-7}
\multirow{2}{*}{Northeast} 
	& Bayesian           & 116.5 (26.1, 210.1) & 51.1 (-18.0, 127.0) & 118.4 (24.8, 215.7) & 231.6 (85.4, 376.2) & 517.6 (118.4, 929.1) \\
         & CMAQ              & 209.9 (47.3, 377.1) & 93.2 (-32.2, 229.6) & 231.0 (48.5, 420.2) & 456.2 (168.5, 739.6) & 990.3 (232.1, 1766.5) \\ \cline{3-7}
\multirow{2}{*}{Northwest} 
	& Bayesian           & 101.8 (22.6, 185.2) & 40.5 (-11.6, 98.6) & 116.0 (24.2, 212.5) & 217.0  (79.7, 354.2) & 475.3 (114.9, 850.6) \\
        & CMAQ               & 184.1 (40.4, 340.9) & 72.6 (-20.0, 179.5) & 213.3 (44.1, 394.3) & 401.5 (146.1, 661.3) & 871.5 (210.6, 1575.9) \\ \cline{3-7}
\multirow{2}{*}{West}      
	& Bayesian           & 391.9 (86.3, 722.1) & 142.5 (-51.0, 365.2) & 312.8 (64.7, 578.4) & 666.7 (242.7, 1097.7) & 1513.9 (342.7, 2763.3) \\
        & CMAQ               & 906.3 (195.9, 1712.1) & 330.5 (-116.0, 873.2) & 714.8 (145.9, 1342.5) & 1548.9 (556.3, 2589.3) & 3500.4 (782.1, 6517.0)   \\ \cline{3-7}
\multirow{2}{*}{WNC}       
	& Bayesian           & 50.4 (11.3, 91.2) & 33.5 (-8.9, 80.2) & 24.8 (5.2, 45.3) & 30.6 (11.3, 49.7) & 139.3 (18.9, 266.4) \\
        & CMAQ               & 93.8 (20.7, 171.8) & 61.0 (-15.9, 148.4) & 45.3 (9.4, 83.2) & 57.9 (21.3, 94.3) & 258.0 (35.5, 497.7) \\ \cline{1-7} 
\end{tabular}
\end{table}
\end{landscape}
\restoregeometry

\section{Discussion}
We present a novel potential outcomes framework that leverages numerical model output to estimate fire-contributed $\PM$ while taking spatial correlation into account and modeling interference between sites. Using a Bayesian spatial downscaling model and monitoring data, we bias-correct CMAQ-estimated counterfactual outcomes for $\PM$ under fire and no-fire regimes, and model correlation between potential outcomes. Assuming consistency between the potential outcomes and the observations based on a CMAQ-derived treatment indicator, and that confounding is accounted for conditional on CMAQ data, we show  that the resulting estimate of fire-contributed $\PM$ has a valid causal interpretation.

We provide a spatially-resolved estimate for fire-contributed $\PM$ and uncertainty across the contiguous U.S. We found that the causal estimate of wildland fires on $\PM$ reached the highest levels in the West, Northwest and Southeast regions. The western parts of the U.S. are impacted by large wildfires and frequent prescribed and agricultural burns are observed in the Southeast. The number of estimated hospitalizations due to exposure to fire-contributed $\PM$ also reached a maximum in these regions, particularly in central California. Our estimates are lower than those produced by CMAQ.  This particular application can be used by health professionals and environmental managers to better understand the health burden associated with fire events in their communities.  Equipped with health burden estimates and uncertainty, they would be able to better anticipate the number of patients to expect and to plan accordingly. In this analysis, we estimated the number of cumulative respiratory hospitalizations per county; it is possible to compute other outcomes related to $\PM$ exposure including all-cause mortality, cardiovascular outcomes, etc.

The study has limitations and strengths. We take a model-based approach that relies on relatively simple separable stationary Gaussian processes. Given the size of the data for this particular study, this approach is warranted, but could be revisited if used for smaller spatial regions. The approach is however generalizable to related research questions concerning how fire-contributed $\PM$ depends on the specific features of wildland fires such as their location, strength, etc. or attribution of $\PM$ to a single fire in which case CMAQ model would be run with corresponding forcings.  These questions are critical in the environmental management context when it has to be shown that a specific fire caused exceedances of regulatory air quality standards.  The proposed causal inference framework can also be generalized to wider range of attribution studies where potential outcomes can be represented using numerical modeling approaches e.g. in climate science, forestry, materials science, etc. In each case, the potential outcomes would differ by the factor of attribution whose impact is the objective of inference. Under the given assumptions and with bias correction we show that the resulting inference has a valid causal interpretation. 

\bibliographystyle{myplainnat}
\bibliography{myProject_2}

\begin{thebibliography}{45}
\providecommand{\natexlab}[1]{#1}
\providecommand{\url}[1]{\texttt{#1}}
\expandafter\ifx\csname urlstyle\endcsname\relax
  \providecommand{\doi}[1]{doi: #1}\else
  \providecommand{\doi}{doi: \begingroup \urlstyle{rm}\Url}\fi

\bibitem[Nat(2016)]{NationalAcademiesofSciences2016}
{Attribution of Extreme Weather Events in the Context of Climate Change}.
\newblock Technical report, National Academies of Sciences, Engineering, and
  Medicine, Washington, D.C., (2016).

\bibitem[Allen and Stott(2003)]{Allen2003}
Allen, M.~R. and Stott, P.~A. (2003).
\newblock {Estimating signal amplitudes in optimal fingerprinting, Part I:
  Theory}.
\newblock \emph{Climate Dynamics}, {\bf {\bf 21}\penalty0  (5-6)}, \penalty0
  477--491.

\bibitem[BenMAP(2017)]{Benmap2017}
BenMAP.
\newblock Technical report.
\newblock {Environmental Benefits Mapping and Analysis Program, Community
  Edition, User's Manual Appendices}, pages 34--44, (2017).

\bibitem[Berrocal et~al.(2010)Berrocal, Gelfand, and Holland]{Berrocal2010e}
Berrocal, V.~J., Gelfand, A.~E., and Holland, D.~M. (2010).
\newblock {A spatio-temporal downscaler for output from numerical models}.
\newblock \emph{Journal of Agricultural, Biological, and Environmental
  Statistics}, {\bf {\bf 15}\penalty0  (2)}, \penalty0 176--197.

\bibitem[Chil{\`{e}}s and Delfiner(2012)]{Chiles2012}
Chil{\`{e}}s, J.-P. and Delfiner, P., (2012).
\newblock \emph{{Geostatistics: Modeling Spatial Uncertainty}}.
\newblock Wiley Series in Probability and Statistics. John Wiley {\&} Sons,
  Inc., Hoboken, NJ, USA.

\bibitem[Cressie(1993)]{Cressie1993}
Cressie, N. A.~C., (1993).
\newblock \emph{{Statistics for Spatial Data}}.
\newblock Wiley Series in Probability and Statistics. John Wiley {\&} Sons,
  Inc., Hoboken, NJ, USA.

\bibitem[Delfino et~al.(2009)Delfino, Brummel, Wu, Stern, Ostro, Lipsett,
  Winer, Street, Zhang, Tjoa, and Gillen]{Delfino2009}
Delfino, R.~J., Brummel, S., Wu, J., Stern, H., Ostro, B., Lipsett, M., Winer,
  A., Street, D.~H., Zhang, L., Tjoa, T., and Gillen, D.~L. (2009).
\newblock {The relationship of respiratory and cardiovascular hospital
  admissions to the southern California wildfires of 2003.}
\newblock \emph{Occupational and Environmental Medicine}, {\bf {\bf
  66}\penalty0  (3)}, \penalty0 189--197.

\bibitem[Dennekamp and Abramson(2011)]{Dennekamp2011}
Dennekamp, M. and Abramson, M.~J. (2011).
\newblock {The effects of bushfire smoke on respiratory health}.
\newblock \emph{Respirology}, {\bf {\bf 16}\penalty0  (2)}, \penalty0 198--209.

\bibitem[Dennekamp et~al.(2015)Dennekamp, Straney, Erbas, Abramson, Keywood,
  Smith, Sim, Glass, {Del Monaco}, Haikerwal, and Tonkin]{Dennekamp2015}
Dennekamp, M., Straney, L.~D., Erbas, B., Abramson, M.~J., Keywood, M., Smith,
  K., Sim, M.~R., Glass, D.~C., {Del Monaco}, A., Haikerwal, A., and Tonkin,
  A.~M. (2015).
\newblock {Forest fire smoke exposures and out-of-hospital cardiac arrests in
  Melbourne, Australia: A case-crossover study}.
\newblock \emph{Environmental Health Perspectives}, {\bf {\bf 123}\penalty0
  (10)}, \penalty0 959--964.

\bibitem[Dominici et~al.(2014)Dominici, Greenstone, and Sunstein]{Dominici2014}
Dominici, F., Greenstone, M., and Sunstein, C.~R. (2014).
\newblock {Particulate Matter Matters}.
\newblock \emph{Science}, {\bf 344}, \penalty0 257--259.

\bibitem[Fuentes and Raftery(2005)]{Fuentes2005}
Fuentes, M. and Raftery, A.~E. (2005).
\newblock {Model Evaluation and Spatial Interpolation by Bayesian Combination
  of Observations with Outputs from Numerical Models}.
\newblock \emph{Biometrics}, {\bf {\bf 61}\penalty0  (1)}, \penalty0 36--45.

\bibitem[Gelfand et~al.(2003)Gelfand, Kim, Sirmans, and Banerjee]{Gelfand2003}
Gelfand, A.~E., Kim, H.-J., Sirmans, C.~F., and Banerjee, S. (2003).
\newblock {Spatial Modeling with Spatially Varying Coefficient Processes}.
\newblock \emph{Source Journal of the American Statistical Association}, {\bf
  {\bf 98}\penalty0  (462)}, \penalty0 387--396.

\bibitem[Gelfand et~al.(2004)Gelfand, Schmidt, Banerjee, and
  Sirrnans]{Gelfand2004}
Gelfand, A.~E., Schmidt, A.~M., Banerjee, S., and Sirrnans, C.~F. (2004).
\newblock {Nonstationary Multivariate Process Modeling through Spatially
  Varying Coregionalization}.
\newblock \emph{Test}, {\bf {\bf 13}\penalty0  (2)}, \penalty0 263--312.

\bibitem[Haikerwal et~al.(2015)Haikerwal, Akram, {Del Monaco}, Smith, Sim,
  Meyer, Tonkin, Abramson, and Dennekamp]{Haikerwal2015}
Haikerwal, A., Akram, M., {Del Monaco}, A., Smith, K., Sim, M.~R., Meyer, M.,
  Tonkin, A.~M., Abramson, M.~J., and Dennekamp, M. (2015).
\newblock {Impact of Fine Particulate Matter (PM2.5) Exposure During Wildfires
  on Cardiovascular Health Outcomes.}
\newblock \emph{Journal of the American Heart Association}, {\bf {\bf
  4}\penalty0  (7)}, \penalty0 e001653.

\bibitem[Haikerwal et~al.(2016)Haikerwal, Akram, Sim, Meyer, Abramson, and
  Dennekamp]{Haikerwal2016}
Haikerwal, A., Akram, M., Sim, M.~R., Meyer, M., Abramson, M.~J., and
  Dennekamp, M. (2016).
\newblock {Fine particulate matter (PM2.5) exposure during a prolonged wildfire
  period and emergency department visits for asthma}.
\newblock \emph{Respirology}, {\bf {\bf 21}\penalty0  (1)}, \penalty0 88--94.

\bibitem[Halloran and Struchiner(1991)]{Halloran1991}
Halloran, M.~E. and Struchiner, C.~J. (1991).
\newblock {Study designs for dependent happenings.}
\newblock \emph{Epidemiology}, {\bf {\bf 2}\penalty0  (5)}, \penalty0 331--8.

\bibitem[Hannart et~al.(2015)Hannart, Pearl, Otto, Naveau, and
  Ghil]{Hannart2015}
Hannart, A., Pearl, J., Otto, F. E.~L., Naveau, P., and Ghil, M. (2015).
\newblock {Causal counterfactual theory for the attribution of weather and
  climate-related events}.
\newblock \emph{Bulletin of the American Meteorological Society}, {\bf 97},
  \penalty0 99--110.

\bibitem[Hansen(2008)]{hansen2008prognostic}
Hansen, B.~B. (2008).
\newblock The prognostic analogue of the propensity score.
\newblock \emph{Biometrika}, {\bf 95}, \penalty0 481--488.

\bibitem[Hegerl and Zwiers(2011)]{Hegerl2011a}
Hegerl, G. and Zwiers, F. (2011).
\newblock {Use of models in detection and attribution of climate change}.
\newblock \emph{Wiley Interdisciplinary Reviews: Climate Change}, {\bf {\bf
  2}\penalty0  (4)}, \penalty0 570--591.

\bibitem[Hern{\'{a}}n et~al.(2008)Hern{\'{a}}n, Alonso, Logan, Grodstein,
  Michels, Willett, Manson, and Robins]{Hernan2008}
Hern{\'{a}}n, M.~A., Alonso, A., Logan, R., Grodstein, F., Michels, K.~B.,
  Willett, W.~C., Manson, J.~E., and Robins, J.~M. (2008).
\newblock {Observational Studies Analyzed Like Randomized Experiments}.
\newblock \emph{Epidemiology}, {\bf {\bf 19}\penalty0  (6)}, \penalty0
  766--779.

\bibitem[Holland(1986)]{Holland1986}
Holland, P.~W. (1986).
\newblock {Statistics and Causal Inference: Rejoinder}.
\newblock \emph{Journal of the American Statistical Association}, {\bf {\bf
  81}\penalty0  (396)}, \penalty0 968--970.

\bibitem[Hong and Raudenbush(2006)]{Hong2006}
Hong, G. and Raudenbush, S.~W. (2006).
\newblock {Evaluating Kindergarten Retention Policy}.
\newblock \emph{Journal of the American Statistical Association}, {\bf {\bf
  101}\penalty0  (475)}, \penalty0 901--910.

\bibitem[Hudgens and Halloran(2008)]{Hudgens2008}
Hudgens, M.~G. and Halloran, M.~E. (2008).
\newblock {Toward Causal Inference With Interference}.
\newblock \emph{Journal of the American Statistical Association}, {\bf
  103(482)}, \penalty0 832--842.

\bibitem[Johnston et~al.(2012)Johnston, Henderson, Chen, Randerson, Marlier,
  DeFries, Kinney, Bowman, and Brauer]{Johnston2012}
Johnston, F.~H., Henderson, S.~B., Chen, Y., Randerson, J.~T., Marlier, M.,
  DeFries, R.~S., Kinney, P., Bowman, D. M. J.~S., and Brauer, M. (2012).
\newblock {Estimated global mortality attributable to smoke from landscape
  fires}.
\newblock \emph{Environmental Health Perspectives}, {\bf {\bf 120}\penalty0
  (5)}, \penalty0 695--701.

\bibitem[Kao(2017)]{kao2017causal}
Kao, E.~K. (2017).
\newblock Causal inference under network interference: A framework for
  experiments on social networks.
\newblock \emph{arXiv preprint arXiv:1708.08522}.

\bibitem[Katzfuss et~al.(2017)Katzfuss, Hammerling, and Smith]{Katzfuss2017}
Katzfuss, M., Hammerling, D., and Smith, R.~L. (2017).
\newblock {A Bayesian hierarchical model for climate change detection and
  attribution}.
\newblock \emph{Geophysical Research Letters}, {\bf {\bf 44}\penalty0  (11)},
  \penalty0 5720--5728.

\bibitem[Kennedy and O'Hagan(2001)]{kennedy2001bayesian}
Kennedy, M.~C. and O'Hagan, A. (2001).
\newblock Bayesian calibration of computer models.
\newblock \emph{Journal of the Royal Statistical Society: Series B (Statistical
  Methodology)}, {\bf {\bf 63}\penalty0  (3)}, \penalty0 425--464.

\bibitem[Knutson et~al.(2017)Knutson, Kossin, Mears, Perlwitz, and
  Wehner]{Knutson2017}
Knutson, T., Kossin, J., Mears, C., Perlwitz, J., and Wehner, M., (2017).
\newblock {Detection and attribution of climate change}.
\newblock In \emph{Climate Science Special Report: Fourth National Climate
  Assessment, Volume I}, pages 114--132. U.S. Global Change Research Program,
  Washington, DC, USA.

\bibitem[McKenzie et~al.(2014)McKenzie, Shankar, Keane, Stavros, Heilman, Fox,
  and Riebau]{McKenzie}
McKenzie, D., Shankar, U., Keane, R.~E., Stavros, E.~N., Heilman, W.~E., Fox,
  D.~G., and Riebau, A.~C. (2014).
\newblock {Earth's Future Smoke consequences of new wildfire regimes driven by
  climate change}.
\newblock \emph{Earth's Future}, {\bf {\bf 2}\penalty0  (2)}, \penalty0 35--59.

\bibitem[Rappold et~al.(2011)Rappold, Stone, Cascio, Neas, Kilaru, Carraway,
  Szykman, Ising, Cleve, Meredith, Vaughan-Batten, Deyneka, and
  Devlin]{Rappold2011}
Rappold, A.~G., Stone, S.~L., Cascio, W.~E., Neas, L.~M., Kilaru, V.~J.,
  Carraway, M.~S., Szykman, J.~J., Ising, A., Cleve, W.~E., Meredith, J.~T.,
  Vaughan-Batten, H., Deyneka, L., and Devlin, R.~B. (2011).
\newblock {Peat bog wildfire smoke exposure in rural North Carolina is
  associated with cardiopulmonary emergency department visits assessed through
  syndromic surveillance}.
\newblock \emph{Environmental Health Perspectives}, {\bf {\bf 119}\penalty0
  (10)}, \penalty0 1415--1420.

\bibitem[Rappold et~al.(2017)Rappold, Reyes, Pouliot, Cascio, and
  Diaz-Sanchez]{Rappold2017}
Rappold, A.~G., Reyes, J., Pouliot, G., Cascio, W.~E., and Diaz-Sanchez, D.
  (2017).
\newblock {Community Vulnerability to Health Impacts of Wildland Fire Smoke
  Exposure}.
\newblock \emph{Environmental Science {\&} Technology}, {\bf {\bf 51}\penalty0
  (12)}, \penalty0 6674--6682.

\bibitem[Rosenbaum(2007)]{Rosenbaum2007}
Rosenbaum, P.~R. (2007).
\newblock {Interference Between Units in Randomized Experiments}.
\newblock \emph{Journal of the American Statistical Association}, {\bf {\bf
  102}\penalty0  (477)}, \penalty0 191--200.

\bibitem[Rubin(1978)]{Rubin1978}
Rubin, D.~B. (1978).
\newblock {Bayesian Inference for Causal Effects: The Role of Randomization}.
\newblock \emph{The Annals of Statistics}, {\bf {\bf 6}\penalty0  (1)},
  \penalty0 34--58.

\bibitem[Schmidt and Gelfand(2003)]{Schmidt2003}
Schmidt, A.~M. and Gelfand, A.~E. (2003).
\newblock {A Bayesian coregionalization approach for multivariate pollutant
  data}.
\newblock \emph{Journal of Geophysical Research: Atmospheres}, {\bf {\bf
  108}\penalty0  (D24)}.

\bibitem[Sobel(2006)]{Sobel2006}
Sobel, M.~E. (2006).
\newblock {What Do Randomized Studies of Housing Mobility Demonstrate?}
\newblock \emph{Journal of the American Statistical Association}, {\bf {\bf
  101}\penalty0  (476)}, \penalty0 1398--1407.

\bibitem[Stavros et~al.(2014)Stavros, Mckenzie, and Larkin]{Stavros2014a}
Stavros, E.~N., Mckenzie, D., and Larkin, N. (2014).
\newblock {The climate-wildfire-air quality system: Interactions and feedbacks
  across spatial and temporal scales}.
\newblock \emph{Wiley Interdisciplinary Reviews: Climate Change}, {\bf {\bf
  5}\penalty0  (6)}, \penalty0 719--733.

\bibitem[Tchetgen and VanderWeele(2012)]{Tchetgen2012}
Tchetgen, E. J.~T. and VanderWeele, T.~J. (2012).
\newblock {On causal inference in the presence of interference}.
\newblock \emph{Statistical Methods in Medical Research}, {\bf {\bf
  21}\penalty0  (1)}, \penalty0 55--75.

\bibitem[{US Environmental Protection
  Agency}(2015)]{USEnvironmentalProtectionAgency2015}
{US Environmental Protection Agency}.
\newblock {AirData Download Files Documentation}, (2015).

\bibitem[{U.S. EPA}(2009)]{ISAPM2009}
{U.S. EPA}. (2009).
\newblock Integrated science assessment for particulate matter.
\newblock \emph{US Environmental Protection Agency: Washington DC, USA}.

\bibitem[{US EPA}(2018)]{USEPA}
{US EPA}. (2018).
\newblock {Particulate Matter (PM2.5) Trends}.

\bibitem[{U.S. EPA}(2019)]{cmaq2019}
{U.S. EPA}.
\newblock {Overview of science processes in CMAQ}, (2019).
\newblock URL \url{https://www.epa.gov/cmaq/overview-science-processes-cmaq}.

\bibitem[{U.S. Forest Service}(2019)]{airfire2019}
{U.S. Forest Service}.
\newblock {AirFire research team}, (2019).
\newblock URL
  \url{https://sites.google.com/firenet.gov/wfaqrp-external/smoke-modeling}.

\bibitem[Wettstein et~al.(2018)Wettstein, Hoshiko, Fahimi, Harrison, Cascio,
  and Rappold]{Wettstein2018}
Wettstein, Z.~S., Hoshiko, S., Fahimi, J., Harrison, R.~J., Cascio, W.~E., and
  Rappold, A.~G. (2018).
\newblock {Cardiovascular and Cerebrovascular Emergency Department Visits
  Associated With Wildfire Smoke Exposure in California in 2015.}
\newblock \emph{Journal of the American Heart Association}, {\bf {\bf
  7}\penalty0  (8)}, \penalty0 e007492.

\bibitem[Zigler et~al.(2012)Zigler, Dominici, and Wang]{Zigler2012b}
Zigler, C.~M., Dominici, F., and Wang, Y. (2012).
\newblock {Estimating causal effects of air quality regulations using principal
  stratification for spatially correlated multivariate intermediate outcomes}.
\newblock \emph{Biostatistics}, {\bf {\bf 13}\penalty0  (2)}, \penalty0
  289--302.

\bibitem[Zigler et~al.(2017)Zigler, Choirat, and Dominici]{Zigler2017}
Zigler, C.~M., Choirat, C., and Dominici, F. (2017).
\newblock {Impact of National Ambient Air Quality Standards nonattainment
  designations on particulate pollution and health}.
\newblock \emph{Epidemiology}, {\bf {\bf 29}\penalty0  (2)}, \penalty0
  165--174.

\end{thebibliography}

\section*{Acknowledgements} \noindent This work was partially supported by grants from the National Institutes of Health (R01ES027892, R01DE024984-01A1), the National Science Foundation (DMS-1513579 and DMS-1811245), the National Cancer Institute (P01CA142538), the Department of the Interior (14-1-04-9) and Oak Ridge Associated Universities.

\begin{appendices}

\section{Proof of Theorem \ref{Thm:identification}}\label{app:proof1}

\begin{proof} 
To relate the potential outcome processes to the induced model of the observed outcome process, consider $Y_{t}^{\text{miss}}(\bs)$ as the observation of the potential outcome that is missing under each regime, i.e.,
\[
\begin{aligned}Y_{t}(\bs,0) & =\begin{cases}
Y_{t}(\bs) & \text{ if \ensuremath{C_{t}(\bs)=0}}\\
Y_{t}^{\text{miss}}(\bs) & \text{ if \ensuremath{C_{t}(\bs)=1}}
\end{cases}\mbox{\ \ \ \ and \ \ \ \ }Y_{t}(\bs,1) & =\begin{cases}
Y_{t}^{\text{miss}}(\bs) & \text{ if \ensuremath{C_{t}(\bs)=0}}\\
Y_{t}(\bs) & \text{ if \ensuremath{C_{t}(\bs)=1}.}
\end{cases}\end{aligned}
\]
Hence, the joint distribution of the potential outcomes, $Y_{t}(\bs,0)$
and $Y_{t}(\bs,1)$, is the joint distribution of the observed and
missing observations $Y_{t}(\bs)$ and $Y_{t}^{\text{miss}}(\bs)$. 

Denoting $\Theta$ as all parameters in the potential outcomes model,
the likelihood function of $\Theta$ is 
\begin{equation}
\begin{aligned} & \prod_{t=1}^{T}\int f(\bY_{t},\bY_{t}^{\text{miss}}|\hat{\btheta}_{t},\hat{\bdelta}_{t},\Theta)d\bY_{t}^{\text{miss}}\\
= & \prod_{t=1}^{T}\int f(\bY_{t},\bY_{t}^{\text{miss}}|\hat{\btheta}_{t},\hat{\bdelta}_{t},\mathbf{C}_{t},\Theta)d\bY_{t}^{\text{miss}}\\
= & \prod_{t=1}^{T}\left[\int f(\bY_{t}^{\text{miss}}|\bY_{t},\hat{\btheta}_{t},\hat{\bdelta}_{t},\mathbf{C}_{t},\Theta)d\bY_{t}^{\text{miss}}\right]f(\bY_{t}|\hat{\btheta}_{t},\hat{\bdelta}_{t},\mathbf{C}_{t},\Theta)\\
= & \prod_{t=1}^{T}f(\bY_{t}|\hat{\btheta}_{t},\hat{\bdelta}_{t},\mathbf{C}_{t},\Theta),
\end{aligned}
\label{eq:identification}
\end{equation}
where the second line follows by Assumption \ref{assumption:models}.
By (\ref{eq:identification}), $\Theta$ depends only on the observed
processes, which completes the proof.
\end{proof} 

\section{Bayesian Analysis and Computing}\label{app:Bayesian}
We conduct our analysis using a Bayesian framework, placing prior distributions on all model parameters. At the first level, we have observations $\bY_t = (Y_t(\bs_1), \hdots, Y_t(\bs_n))^T$ denoting measured $\PM$ measured on day $t$, $t = 1, \hdots, m$, for $n$ sites, $\bs_1, \hdots, \bs_n$, modeled as multivariate normal:

$$
	\bY_1, \hdots, \bY_m \overset{indp}{\sim} MVN\left(\btheta_t + \bC_t  \bsy{\delta_t}, \sigma^2 I_n\right),
$$

\noindent where $\bsy{\theta_t}$ and $\bsy{\delta_t}$ are $n$-vectors of the true background and fire-contributed $\PM$ processes on day $t$, respectively, $\bC_t$ is an $n\times n$ diagonal matrix with binary entries, $C_t(\bs)$, that indicate fire impacts at all sites on day $t$ and $\sigma^2$ is error variance. We place a non-informative inverse-gamma prior on $\sigma^2$: $\sigma^2 \sim IG\left(0.1, 0.1\right)$. 

We assume the priors for the true $\PM$ spatial processes follow a bivariate Gaussian process for $t=1,\hdots,m$:

\beqn\label{bivariate}
	\begin{pmatrix} 
	\bsy{\theta_t}  \\ 
	\bsy{\delta_t}
	\end{pmatrix} \sim \calGP 
		\begin{pmatrix}
			\begin{bmatrix}
				B_0(\bsy{\hat\theta_t}) \\
				B_1(\bsy{\hat\delta_t})
			\end{bmatrix},
			\begin{bmatrix}
				\sigma_1^2 & \sigma_{12}\\
				\sigma_{12} & \sigma_2^2
			\end{bmatrix}
			\otimes
			C(\phi_1)
		\end{pmatrix}
\eeqn

\noindent where $\hat\theta_t$ and $\hat\delta_t$ are the numerical model output on (background) and wildfire-contributed $\PM$ for all sites on day $t$, respectively. The mean functions are $B_0(\bsy{\theta_t}) = \balpha_0 + \bbeta_0 \bsy{\hat\theta_t}$ and $B_1(\bsy{\delta_t}) = \balpha_1 + \bbeta_1 \bsy{\hat\delta_t}$, where $\balpha_0$ and $\balpha_1$ are $n$-vectors of additive bias for sites $\bs_1, \hdots, \bs_n$, and $\bbeta_0$ and $\bbeta_1$ are $n$-vectors of multiplicative bias for sites $\bs_1, \hdots, \bs_n$. The $n\times n$ matrix, $C(\phi_1)$, is an exponential decay correlation matrix such that $C_{i, j} = \exp(-\|\bs_i - \bs_j \| / \phi_1)$ for two sites, $\bs_i$ and $\bs_j$. Also, the covariance parameter can be written as $\sigma_{12} = \sigma_1\sigma_2\gamma$.

We re-parameterize the model in (\ref{bivariate}) as a linear model of co-regionalization \citep{Gelfand2004} for ease of computation. Define $s_1^2 = \sigma_1^2$, $\rho = \frac{\sigma_{12}}{\sigma_1^2}$ and $s_2^2 = \sigma_2^2 - \frac{\sigma_{12}^2}{\sigma_1^2}$. Then we have

$$\begin{aligned}
	\bsy{\theta_t} &\sim \calGP \left( B_0( \bsy{\hat\theta_t}), s_1^2 C(\phi_1) \right) \\
	\bsy{\delta_t} | \bsy{\theta_t} & \sim \calGP \left(B_1(\bsy{\hat\delta_t}) + \rho(\bsy{\theta_t} - B_0(\bsy{\hat\theta_t})), s_2^2 C(\phi_1) \right) 
\end{aligned}$$

\noindent We assign noninformative priors to the covariance parameters: $s_1^2 \sim IG(0.1, 0.1)$, $s_2^2 \sim IG(0.1, 0.1)$, $\rho \sim N(0, 100)$. To get posterior means of $\sigma_1^2$, $\sigma_2^2$, $\sigma_{12}$ and the correlation parameter, $\gamma$,

$$
	\hat\sigma_1^2 = \hat s_1^2, \hspace{6pt} \hat \sigma_{12} = \hat\rho \hat s_1^2, \hspace{6pt} \hat\sigma_2^2 = \hat\rho^2 \hat s_1^2 + \hat s_2^2
$$

\noindent and since $\gamma = \frac{\sigma_{12}}{\sigma_1\sigma_2}$,

$$
	\hat\gamma = \frac{\hat\rho \hat s_1}{\sqrt{\hat\rho^2 \hat s_1^2 + \hat s_2^2}}.\\
$$

We let the priors of the bias terms vary spatially with constant mean:

$$\begin{aligned}
	\balpha_0 &\sim MVN\left(\mu_{\alpha_0}\bsy{1_n}, \sigma_{\alpha_0}^2C(\phi_2)\right) \\
	\bbeta_0   &\sim MVN\left(\mu_{\beta_0}\bsy{1_n}, \sigma_{\beta_0}^2C(\phi_2)\right) \\
	\balpha_1 &\sim MVN\left(\mu_{\alpha_1}\bsy{1_n}, \sigma_{\alpha_1}^2C(\phi_2)\right) \\
	\bbeta_1   &\sim MVN\left(\mu_{\beta_1}\bsy{1_n}, \sigma_{\beta_1}^2C(\phi_2)\right).\\
\end{aligned}$$\\

We assign non-informative priors to the hyper-parameters:

$$\begin{aligned}
	\mu_{\alpha_0}, \mu_{\beta_0}, \mu_{\alpha_1}, \mu_{\beta_1} &\sim N(0, 100^2) \\
	\sigma_{\alpha_0}^2, \sigma_{\beta_0}^2, \sigma_{\alpha_1}^2, \sigma_{\beta_1}^2 &\sim IG(0.1, 0.1) \\
	\log(\phi_2) &\sim N(0, 500).
\end{aligned}$$

\subsection{Derivation of Full Conditionals}\label{app:FullConditionals}

Most of the parameter models have conditionally conjugate priors and are thus updated using Gibbs sampling. Below we give the full conditional distribution for these parameters:

% Theta ----------------------------------------------------------
\begin{flalign*}
	&\bsy{\theta_t} | \bY_t \sim \mathcal{N} 
		\left((\Sigma_y^{-1} + \Sigma_\theta^{-1} + \rho^2\Sigma_\delta^{-1})^{-1}
			(\Sigma_y^{-1}(\bsy{Y_t} - C_t\bsy{\delta_t}) + 
			\Sigma_\theta^{-1}B_0(\bsy{\hat\theta_t}) +  
			\rho\Sigma_\delta^{-1}(\bdelta_t - B_1(\bsy{\hat\delta_t}) + \rho B_0(\bsy{\hat\theta_t}))), \right. \\ 
		&\left. \qquad \qquad \qquad \qquad \qquad (\Sigma_y^{-1} + \Sigma_\theta^{-1} + \rho^2\Sigma_\delta^{-1})^{-1}\right) &
\end{flalign*}
% Delta ---------------------------------------------------------
\begin{flalign*}
	&\bsy{\delta}_t | \bY_t \sim \mathcal{N} 
		\left((C_t^T\Sigma_y^{-1}C _t+ \Sigma_\delta^{-1})^{-1}
		(C_t^T\Sigma_y^{-1}(\bsy{Y}_t - \bsy{\theta}_t) + 
		\Sigma_\delta^{-1}(B_1(\bsy{\hat\delta_t}) + \rho(\btheta_t - B_0(\bsy{\hat\theta_t})))), \right. \\ 
	&\left. \qquad \qquad \qquad \qquad \qquad (C_t^T\Sigma_y^{-1}C_t + \Sigma_\delta^{-1})^{-1}\right) &
\end{flalign*}
% Sigma2 -----------------------------------------------------
\begin{flalign*}
	& \sigma^2 | \bY_t \sim \mathcal{IG}
		\left(\frac{nm}{2} + 0.1, 
		\left(\frac{1}{2}\sum_{t=1}^m(\bsy{Y}_t - (\btheta_t+C_t\bdelta_t))^T(\bsy{Y}_t - (\btheta_t+C_t\bdelta_t)) \right) +  0.1\right) &
\end{flalign*}
% A0 ------------------------------------------------------------
\begin{flalign*}
	& \balpha_0 | \bY_t, \btheta_t, \bdelta_t \sim \mathcal{N} 
		\left(\left(m\Sigma_\theta^{-1} + m\rho^2\Sigma_\delta^{-1} + \Sigma_{a_0}^{-1}\right)^{-1}
		\left(\sum_{t=1}^m \Sigma_\theta^{-1}\left(\bsy{\theta}_t - \bsy{\hat\theta_t}\bbeta_0 \right) +
		\rho\Sigma_\delta^{-1}\left( \rho(\btheta_t - \bsy{\hat\theta_t}\bbeta_0) - \right. \right. \right. \\ 		
	& \left. \left. \left. \qquad \qquad \qquad \qquad \qquad
		 (\bdelta_t - B_1(\bsy{\hat\delta_t})) \right) + 
		\mu_{a_0}\Sigma_{a_0}^{-1}\bsy{1}_n\right),
	 	\left(m\Sigma_\theta^{-1} + m\rho^2\Sigma_\delta^{-1} + \Sigma_{a_0}^{-1}\right)^{-1} \right) &
\end{flalign*}
% B0 ------------------------------------------------------------
\begin{flalign*}
	 & \bbeta_0 | \bY_t, \btheta_t, \bdelta_t \sim \mathcal{N} 
	 	\left(\left(\sum_{t=1}^m\hat\theta_t^T( \Sigma_\theta^{-1} +\rho^2\Sigma_\delta^{-1} )\hat\theta_t + 
		\Sigma_{\beta_0}^{-1}\right)^{-1}
		\left(\sum_{t=1}^m\hat\theta_t^T\Sigma_\theta^{-1}(\btheta_t - \balpha_0) + 
		\rho\bsy{\hat\theta_t}\Sigma_\delta^{-1}\left(\rho(\btheta_t - \balpha_0) - \right.  \right. \right. \\
	 & \left. \left. \left. \qquad \qquad \qquad \qquad \qquad 
		(\bdelta_t - B_1(\bsy{\hat\delta_t})) \right) +
		\mu_{\beta_0}\Sigma_{\beta_0}^{-1}\bsy{1}_n\right),
		\left(\sum_{t=1}^m\hat\theta_t^T( \Sigma_\theta^{-1} +\rho^2\Sigma_\delta^{-1} )\hat\theta_t + \Sigma_{\beta_0}^{-1}\right)^{-1} \right) &
\end{flalign*}
% A1 -------------------------------------------------------------
\begin{flalign*}	
	& \balpha_1 | \bY_t, \btheta_t, \bdelta_t \sim \mathcal{N} 
		\left((m\Sigma_\delta^{-1} + \Sigma_{\alpha_1}^{-1})^{-1}
		\left(\sum_{t=1}^m\Sigma_\delta^{-1} \left(\bdelta_t - (\hat\delta_t \bbeta_1 + \rho(\btheta_t - B_0(\bsy{\hat\theta_t})))\right) + \mu_{\alpha_1}\Sigma_{\alpha_1}^{-1}\bsy{1}_n\right), \right. \\
	& \left. \qquad \qquad \qquad \qquad \qquad 
		(m\Sigma_\delta^{-1} + \Sigma_{\alpha_1}^{-1})^{-1} \right) &
\end{flalign*}
% B1 -------------------------------------------------------------
\begin{flalign*}
	& \bbeta_1 | \bY_t, \btheta_t, \bdelta_t \sim \mathcal{N} 
		\left(\left(\sum_{t=1}^m\hat\delta_t^T\Sigma_\delta^{-1}\hat\delta_t + \Sigma_{\beta_1}^{-1}\right)^{-1}
		\left(\sum_{t=1}^m\hat\delta_t^T\Sigma_\delta^{-1}(\bdelta_t - \balpha_1 - \rho(\btheta_t - B_0(\bsy{\hat\theta_t}))) + \mu_{\beta_1}\Sigma_{\beta_1}^{-1}\bsy{1}_n\right), \right. \\
	& \left. \qquad \qquad \qquad \qquad \qquad  
		\left(\sum_{t=1}^m\hat\delta_t^T\Sigma_\delta^{-1}\hat\delta_t + \Sigma_{\beta_1}^{-1}\right)^{-1} \right) &
\end{flalign*}
% Rho -----------------------------------------------------------
\begin{flalign*}
	& \rho | \bY_t, \btheta_t, \bdelta_t \sim \mathcal{N} 
		\left(\left(\frac{1}{100} + \sum_{t=1}^m(\btheta_t - B_0(\bsy{\hat\theta_t}))^T \Sigma_\delta^{-1} (\btheta_t - B_0(\bsy{\hat\theta_t}))\right)^{-1}
		\sum_{t=1}^m (\bdelta_t - B_1(\bsy{\hat\delta_t}))^T \Sigma_\delta^{-1}(\bdelta_t - B_1(\bsy{\hat\delta_t})),\right. \\
	& \left. \qquad \qquad \qquad \qquad \qquad  
		 \left(\frac{1}{100} + \sum_{t=1}^m(\btheta_t - B_0(\bsy{\hat\theta_t}))^T \Sigma_\delta^{-1} (\btheta_t - B_0(\bsy{\hat\theta_t}))\right)^{-1}\right) &
\end{flalign*}
% s_1^2 ---------------------------------------------------------
\begin{flalign*}
	& s_1^2 | \bY_t, \btheta_t \sim \mathcal{IG} 
		\left(\frac{nm}{2} + 0.1, 
		\left(\frac{1}{2}\sum_{t=1}^m(\btheta_t - B_0(\bsy{\hat\theta_t}))^TC(\phi_1)^{-1}(\btheta_t - B_0(\bsy{\hat\theta_t})) + 0.1 \right)\right) &
\end{flalign*}
% s_2^2 ---------------------------------------------------------
%Define $\mu_\delta = (B_1(\bsy{\hat\delta_t}) + \rho \left(\btheta_t - B_0(\bsy{\hat\theta_t}) \right))$. Then,
\begin{flalign*}
	& s_2^2 | \bY_t, \bdelta_t \sim \mathcal{IG} 
		\left(\frac{nm}{2} + 0.1, 
		\left(\frac{1}{2}\sum_{t=1}^m(\bdelta_t - (B_1(\bsy{\hat\delta_t}) + \rho \left(\btheta_t - B_0(\bsy{\hat\theta_t}) \right)))^TC(\phi_1)^{-1}(\bdelta_t - (B_1(\bsy{\hat\delta_t}) + \right. \right. \\
	& \left. \left. \qquad \qquad \qquad \qquad \qquad  
		\rho \left(\btheta_t - B_0(\bsy{\hat\theta_t}) \right))) + 0.1 \right)\right) &
\end{flalign*}
% mu_alpha ---------------------------------------------------
\begin{flalign*}
	&\mu_{\alpha_i} | \bY_t \sim \mathcal{N} 
		\left(\left(\bsy{1}_n^T\Sigma_{\alpha_i}^{-1}\bsy{1}_n + \frac{1}{100^2}\right)^{-1}
			\bsy{1}_n^T\Sigma_{\alpha_i}^{-1}\alpha_i, 
		\left(\bsy{1}_n^T\Sigma_{\alpha_i}^{-1} \bsy{1}_n + \frac{1}{100^2}\right)^{-1}\right) &
\end{flalign*}
% mu_beta ----------------------------------------------------
\begin{flalign*}
	& \mu_{\beta_i} | \bY_t \sim \mathcal{N}
		\left(\left(\bsy{1}_n^T\Sigma_{\beta_i}^{-1} \bsy{1}_n + \frac{1}{100^2}\right)^{-1}
			\bsy{1}_n^T\Sigma_{\beta_i}^{-1} \bbeta_i, 
		\left(\bsy{1}_n^T\Sigma_{\beta_i}^{-1}\bsy{1}_n + \frac{1}{100^2}\right)^{-1}\right) &
\end{flalign*}
% sigma2_alpha --------------------------------------------
\begin{flalign*}
	& \sigma_{\alpha_i}^2 | \bY_t \sim \mathcal{IG}
		\left(\frac{n}{2} + 0.1, 
		\frac{1}{2}(\balpha_i - \mu_{\alpha_i} \bsy{1}_n)^TC(\phi_2)^{-1}(\balpha_i - \mu_{\alpha_i} \bsy{1}_n) + 0.1\right) &
\end{flalign*}
% sigma2_beta ---------------------------------------------
\begin{flalign*}
	& \sigma_{\beta_i}^2 | \bY_t \sim \mathcal{IG} 
		\left(\frac{n}{2} + 0.1, 
		\frac{1}{2}(\bbeta_i - \mu_{\beta_i} \bsy{1}_n)^TC(\phi_2)^{-1}(\bbeta_i - \mu_{\beta_i}\bsy{1}_n) + 0.1\right) &
\end{flalign*}

\noindent The range parameters, $\phi_1$ and $\phi_2$ are estimated empirically using variograms. 

\subsection{Effect Estimates} \label{app:effect-estimates}

\begin{figure}[H]
	\centering
	\caption{\textbf{Bias estimates for CMAQ's background $\PM$}. The posterior mean   standard deviation (SD) for additive bias (a,b) and multiplicative bias (c, d).}\label{fig:bias0_maps}
	\begin{subfigure}[b]{.49\textwidth}
		\includegraphics[width=\textwidth]{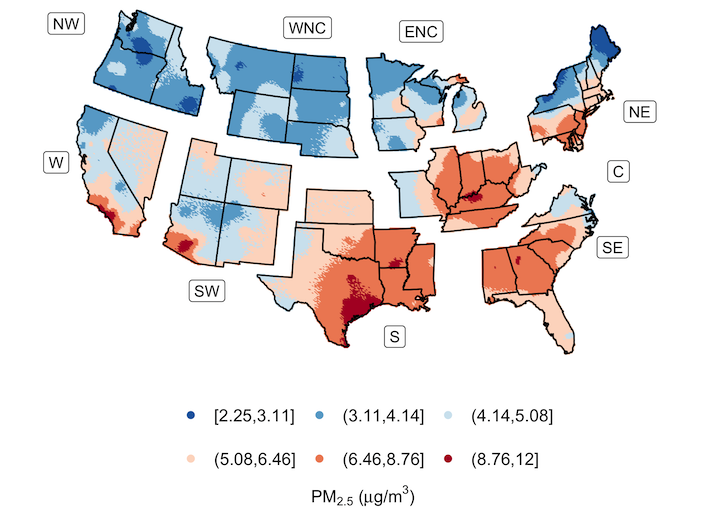}
		\caption{Posterior Mean of $\alpha_0(\bs)$ (\unit)}
	\end{subfigure}    
	\begin{subfigure}[b]{.49\textwidth}
		\includegraphics[width=\textwidth]{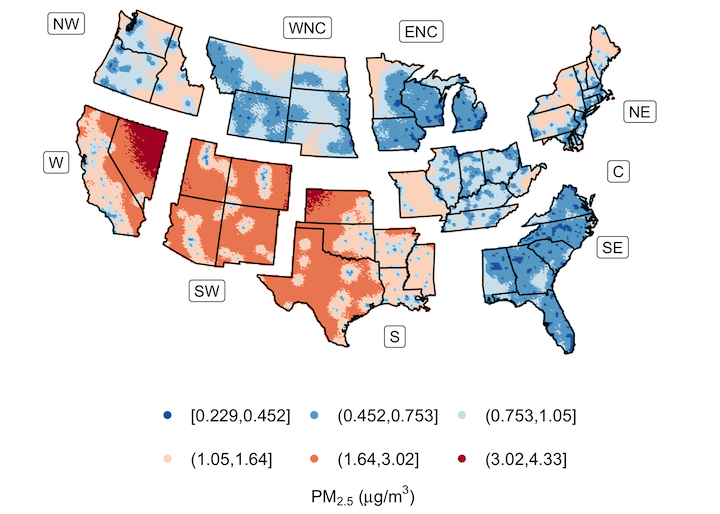}
		\caption{Posterior SD of $\alpha_0(\bs)$ (\unit)}
	\end{subfigure}
	\vskip\baselineskip    
	\begin{subfigure}[b]{.49\textwidth}
		\includegraphics[width=\textwidth]{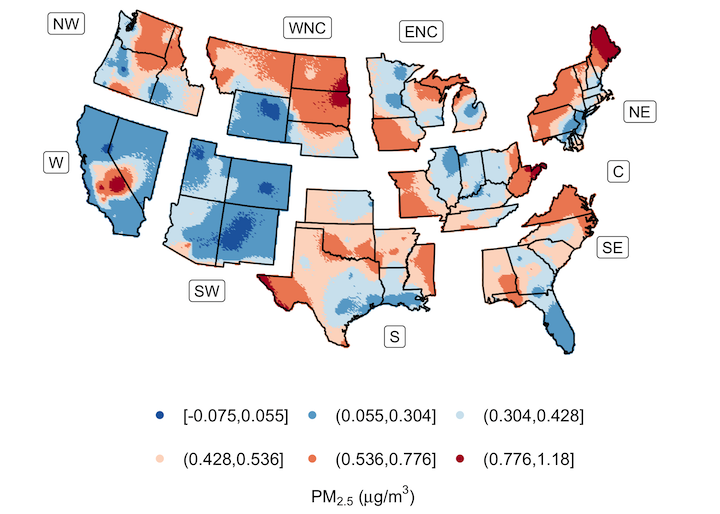}
		\caption{Posterior Mean of $\beta_0(\bs)$ (\unit)}
	\end{subfigure}    
	\begin{subfigure}[b]{.49\textwidth}
		\includegraphics[width=\textwidth]{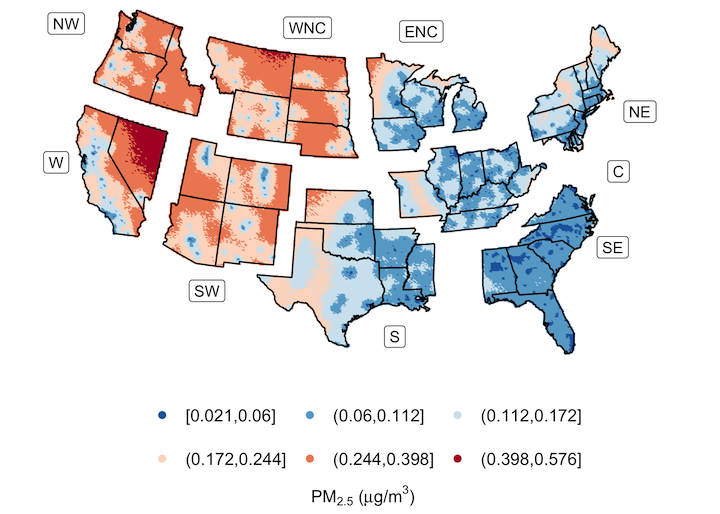}
		\caption{Posterior SD of $\beta_0(\bs)$ (\unit)}
	\end{subfigure}
\end{figure}

\begin{figure}[H]
	\centering
	\caption{\textbf{Bias estimates for CMAQ's fire-contributed $\PM$}. The posterior mean standard deviation (SD) for additive bias (a,b) and multiplicative bias (c, d).}\label{fig:bias1_maps}
	\begin{subfigure}[b]{.49\textwidth}
		\includegraphics[width=\textwidth]{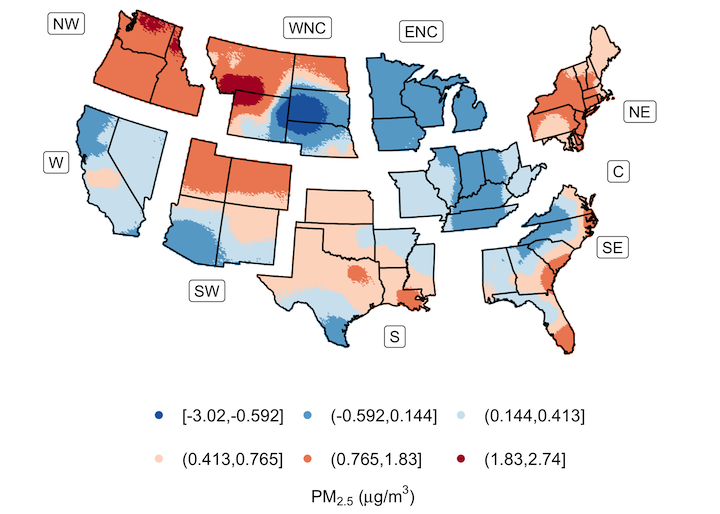}
		\caption{Posterior Mean of $\alpha_1(\bs)$ (\unit)}
	\end{subfigure}
	\begin{subfigure}[b]{.49\textwidth}
		\includegraphics[width=\textwidth]{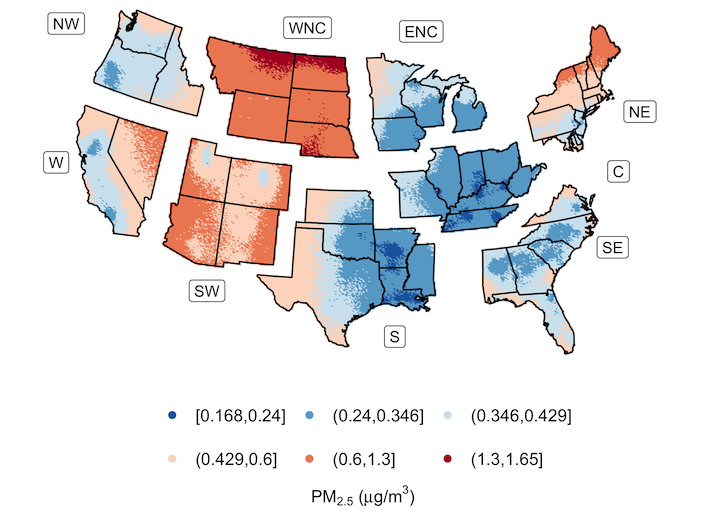}
		\caption{Posterior SD of $\alpha_1(\bs)$ (\unit)}
	\end{subfigure}
	\vskip\baselineskip    
	\begin{subfigure}[b]{.49\textwidth}
		\includegraphics[width=\textwidth]{Plots_Figures/b1}
		\caption{Posterior Mean of $\beta_1(\bs)$ (\unit)}\label{fig:bias1_mean}
	\end{subfigure}    
	\begin{subfigure}[b]{.49\textwidth}
		\includegraphics[width=\textwidth]{Plots_Figures/b1_se}
		\caption{Posterior SD of $\beta_1(\bs)$ (\unit)}\label{fig:bias1_se}
	\end{subfigure}
\end{figure}

\begin{figure}[H]
	\centering
	\caption{\textbf{Estimates from the Bayesian model of background $\PM$}. The posterior means (a) and associated standard deviation (b), as well as estimated total $\PM$ (c).}\label{fig:theta_maps}
	\begin{subfigure}[b]{.49\textwidth}
		\includegraphics[width=\textwidth]{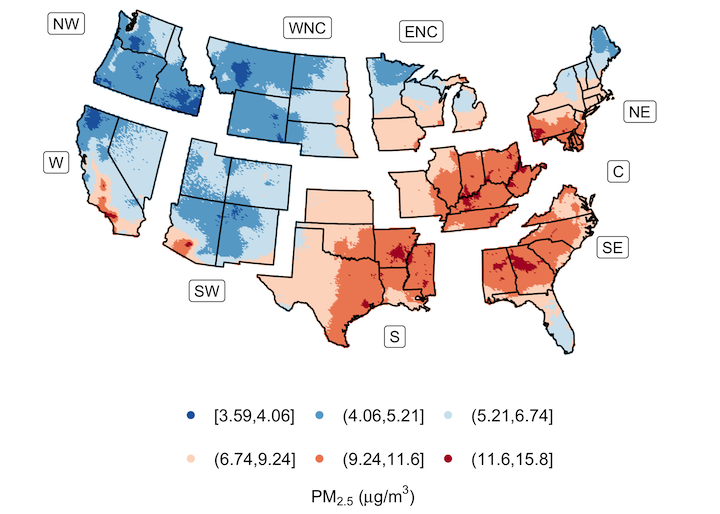}
		\caption{Posterior Mean of $\theta_t(\bs)$ (\unit)}
	\end{subfigure}
	\begin{subfigure}[b]{.49\textwidth}
		\includegraphics[width=\textwidth]{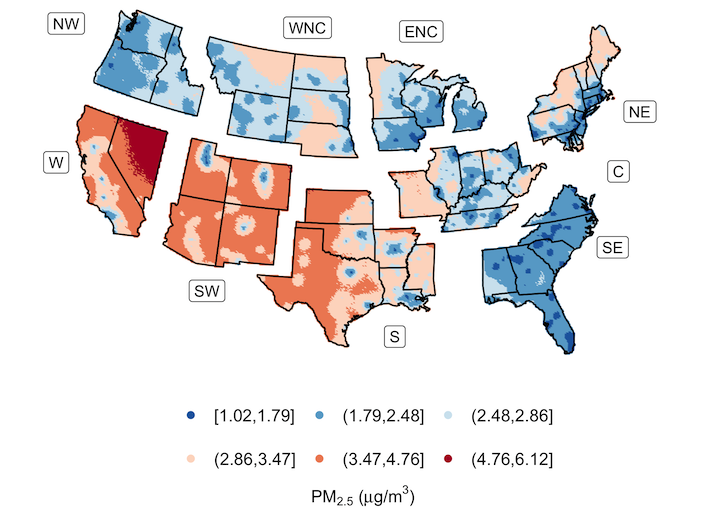}
		\caption{Posterior SD of $\theta_t(\bs)$ (\unit)}
	\end{subfigure}
	\vskip\baselineskip
	\begin{subfigure}[b]{\textwidth}
		\centering
		\includegraphics[width=\textwidth]{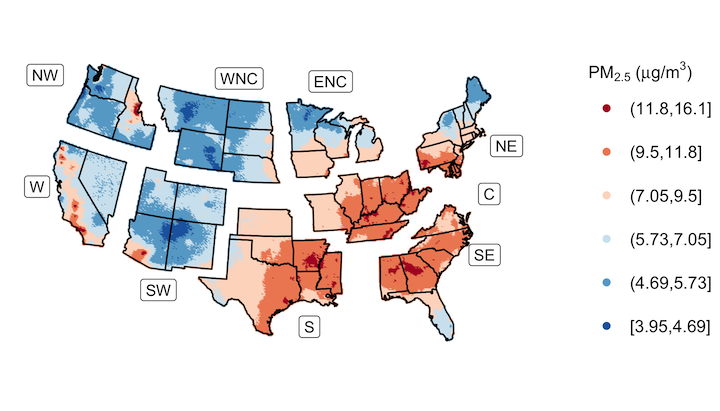}
		\caption{Total $\PM$ (\unit)}
	\end{subfigure}
\end{figure}

\subsection{Health Burden Analysis}\label{app:health-burden}

\begin{table}[H]
\centering
\caption{Age groups and relative rates ($r_a$) used to calculate health burden impact.}
\label{tab:ages_relative_risk}
\begin{tabular}{cccc}
\multicolumn{3}{c}{Age Group (years)} & \multirow{2}{*}{\begin{tabular}[c]{@{}c@{}}Wildfire Period Relative Rate of \\ Respiratory Hospitalization*\end{tabular}} \\
U.S. Census July 2010 & Delfino et al. 2009           & BenMaps/Our Analysis &                         \\
0-4                   & 0-4                           & 0-1                  & 1.045                   \\
5-9, 10-14, 15-19     & 5-19                          & 2-17                 & 1.027                   \\
20-24                 & \multirow{5}{*}{20-64}        & 18-24                & \multirow{5}{*}{1.024}  \\
25-29, 30-34          &                                                      & 25-34                &  \\
35-39, 40-44          &                                                      & 35-44                &  \\
45-49, 50-54          &                                                      & 45-54                &  \\
55-59, 60-64          & \multirow{4}{*}{65 and older}  & 55-64               &  \multirow{3}{*}{1.030} \\
65-69, 70-74          &                                                      & 65-74                &  \\
75-79, 80-84          &                                                      & 75-84                &  \\
85 and older          &                                                      & 85-99                &  \\
\multicolumn{4}{l}{*Delfino et. al 2009}                                  
\end{tabular}
\end{table}

\end{appendices}
\end{document}

%\linenumbers

% Title Page -----------------------------------------------
\begin{center}
{\Large {\bf A spatial causal analysis of wildland fire-contributed $\PM$ using numerical model output}}\\ \vspace{12pt}
{\large Supplemental Materials}
\end{center}

\section{MCMC Convergence Diagnostics}
\begin{figure}[H]
    \centering
    \caption{\textbf{Trace-plots and effective sample size for the causal effect estimates.} To monitor MCMC convergence, we visually examined trace-plots and computed the effective sample size (ESS). These plots show the effective sample sizes and trace-plots for the causal effect estimates at ten randomly selected monitoring sites (s) out of the 96 total sites the West Region.}
    \includegraphics[width=\textwidth]{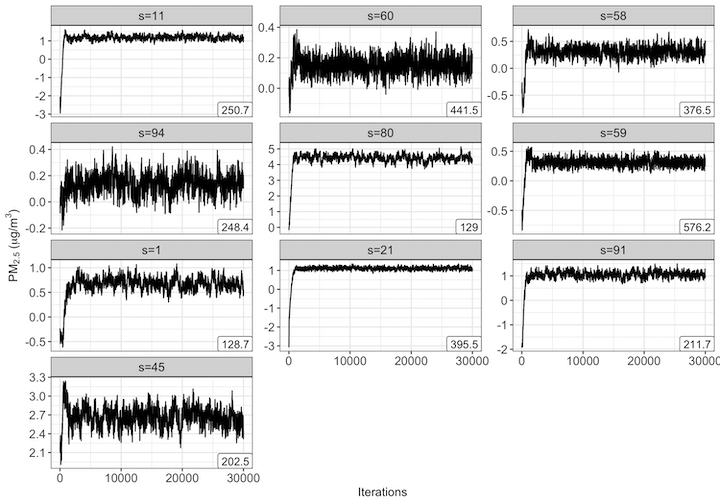}
    \label{fig:trace_ess_ce}
\end{figure}

\begin{figure}[H]
    \centering
    \caption{\textbf{Effective sample sizes for the causal effect estimate at the CMAQ centroids.} To monitor MCMC convergence, we computed the effective sample size (ESS) for the causal effect estimates in the West Region. This map shows the effective sample sizes at each CMAQ centroid.}
    \includegraphics[width=.5\textwidth]{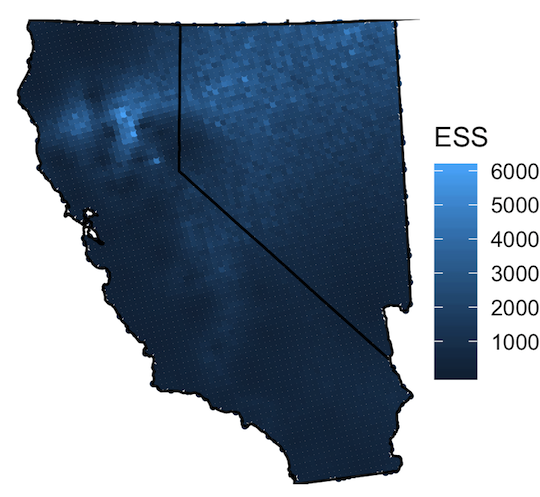}
    \label{fig:map_ess}
\end{figure}

%Each parameter in the model estimated via MCMC converged satisfactorily. 
MCMC convergence was assessed by visual examination of trace-plots and by calculating the effective sample sizes (ESS), displayed in this section for the most fire-prone and therefore most representative area, the West region. In Figure \ref{fig:trace_ess_ce}, we show the ESS and trace-plots for the causal effect estimate at ten randomly selected monitoring sites. The average ESS for the causal effect over all 96 of the monitoring sites was 281.09 (SD=139.649). This was calculated after a burn-in period of 5,000 iterations from 30,000 total iterations; trimming was not included. We also calculated the ESS of the causal effect at the CMAQ centroids (i.e. the Kriging points) was 1,089.7 (SD=950.5). This was computed after a burn-in period of 5,000 iterations; trimming was not included. %Due to the computational burden of each Kriging step, we calculated ESS for only 20,000 MCMC iterations.
The ESS for the causal effect estimate at each CMAQ centroid is displayed in Figure \ref{fig:map_ess}. 

\section{Sensitivity to Regional Blocking}
We conducted separate analyses for each region of the contiguous U.S. included in our study in order to run our analysis in parallel, thereby speeding up computation. A potential downside of this approach is that correlation between neighboring regions is ignored. Here, we demonstrate that ignoring this correlation has only a small impact. 

\begin{figure}
    \centering
    \caption{\textbf{Causal effect estimate for the  West and Northwest regions separately versus combined.} Posterior means (top) and standard deviations (bottom) of the causal effect for the West and Northwest regions as one and separately. We present the $2^{\text{nd}}$, $25^{\text{th}}$, $50^{\text{th}}$, $75^{\text{th}}$, and $98^{\text{th}}$ percentiles of the means and standard deviations.}\label{fig:blocked_w_nw}
    \begin{subfigure}[b]{\textwidth}
        \includegraphics[width=\textwidth]{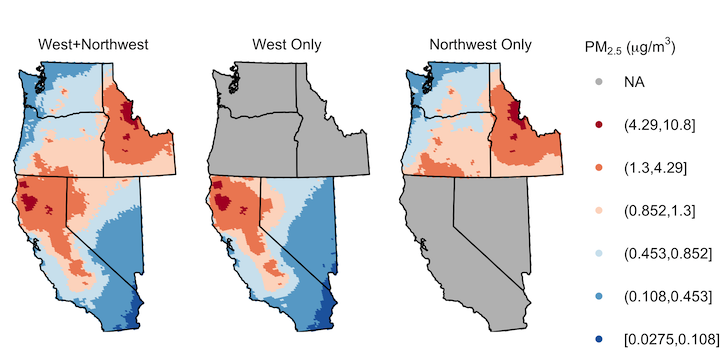}
        \caption{Posterior Mean}
    \end{subfigure}
    \vskip\baselineskip
    \begin{subfigure}[b]{\textwidth}
        \includegraphics[width=\textwidth]{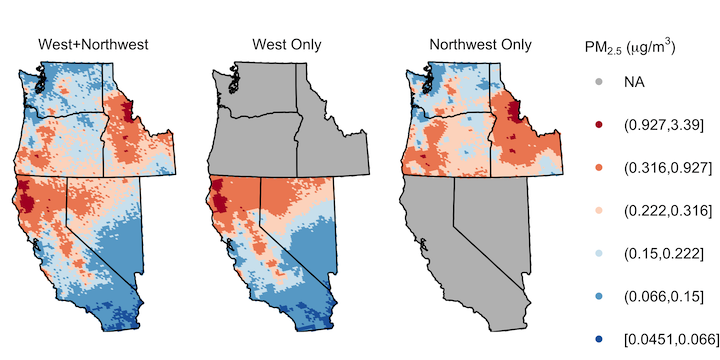}
        \caption{Posterior Standard Deviation}
    \end{subfigure}
%     \begin{subfigure}[b]{0.4\textwidth}
%         \includegraphics[width=\textwidth]{Plots_Figures/map_w_nw_ce.png}
%     \end{subfigure}
%     \quad
%     \begin{subfigure}[b]{0.4\textwidth}
%         \includegraphics[width=\textwidth]{Plots_Figures/map_w_nw_ce_sd.png}
%     \end{subfigure}
%     \vskip\baselineskip
%     \begin{subfigure}[b]{0.4\textwidth}
%         \includegraphics[width=\textwidth]{Plots_Figures/map_nw_ce.png}
%     \end{subfigure}
%   \quad
%   \begin{subfigure}[b]{0.4\textwidth}
%         \includegraphics[width=\textwidth]{Plots_Figures/map_nw_ce_sd.png}
%     \end{subfigure}
%     \label{fig:blocked_w_nw}
%     \vskip\baselineskip
%         \begin{subfigure}[b]{0.4\textwidth}
%         \includegraphics[width=\textwidth]{Plots_Figures/map_w_ce.png}
%     \end{subfigure}
%     \quad
%     \begin{subfigure}[b]{0.4\textwidth}
%         \includegraphics[width=\textwidth]{Plots_Figures/map_w_ce_sd.png}
%     \end{subfigure}    
\end{figure}

We conducted a sensitivity test to determine the effects of running separate regional analyses that entailed running the analysis for the Northwest and Western regions as one and compared the resulting causal effect estimates to those from each region run with separate analyses. We chose to focus on the West and Northwest regions since the western US is the most prone to fires. In Figure \ref{fig:blocked_w_nw}, we display the estimated posterior means and standard deviations of the causal effects for the Northwest and Western regions as one and separately. The resulting spatial patterns are similar, leading us to conclude that our analysis is robust to regional blocking. %Comparing these values to those displayed in Figure 6 of the main document, the results are similar. The highest causal effect estimates (1.31-10.6 $\unit$) from the blocked analysis coincide with those in the unblocked analysis (1.91-10.8 $\unit$) and exist in the same areas of the Northwest and Western regions. The same is true of the lowest estimates from the blocked analysis ($0.033-0.12 \unit$) and the unblocked analysis ($-0.024-0.1 \unit$), as well as the middle estimates (blocked: $0.12-1.31\unit$; unblocked: $0.1-1.91\unit$) . Hence, we conclude that our analysis is robust to regional blocking.

\section{Spatial Covariance Function}
\begin{figure}
	\centering
	\caption{\textbf{Spatial covariance functions}. The covariance functions from Equation (5) in the text evaluated at the posterior means of the model parameters for each combination of $(C_t(\bs), C_t(\bs'))$ at each region.}\label{fig:covs}
	\begin{subfigure}[b]{0.3\textwidth}
		\includegraphics[width=\textwidth]{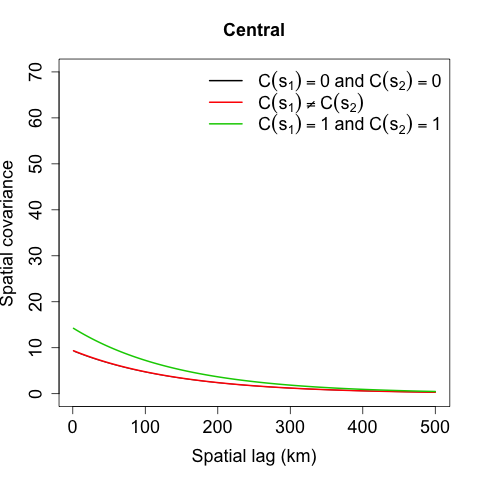}
		\caption{Central}
	\end{subfigure}	
	\quad
	\begin{subfigure}[b]{0.3\textwidth}
		\includegraphics[width=\textwidth]{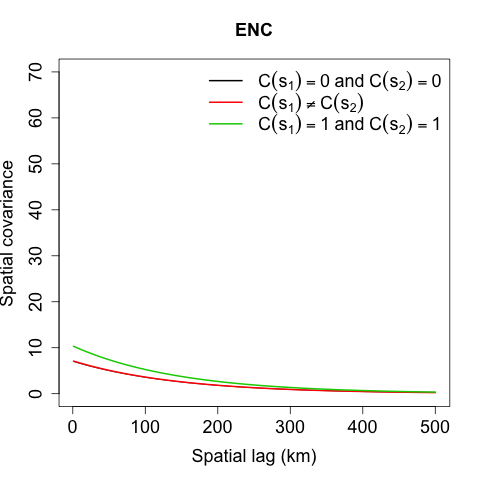}
		\caption{East North Central}
	\end{subfigure}
	\quad
	\begin{subfigure}[b]{0.3\textwidth}
		\includegraphics[width=\textwidth]{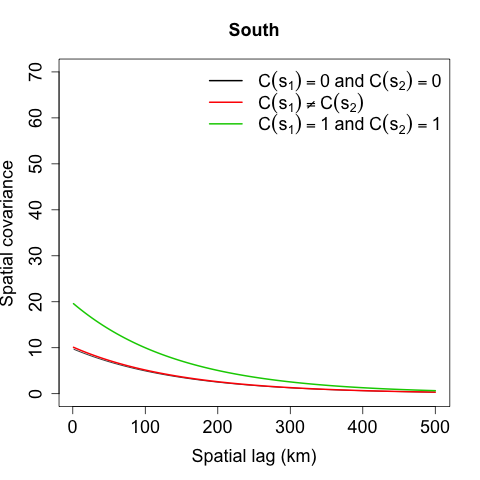}
		\caption{South}
	\end{subfigure}
	\vskip\baselineskip
	\begin{subfigure}[b]{0.3\textwidth}
		\includegraphics[width=\textwidth]{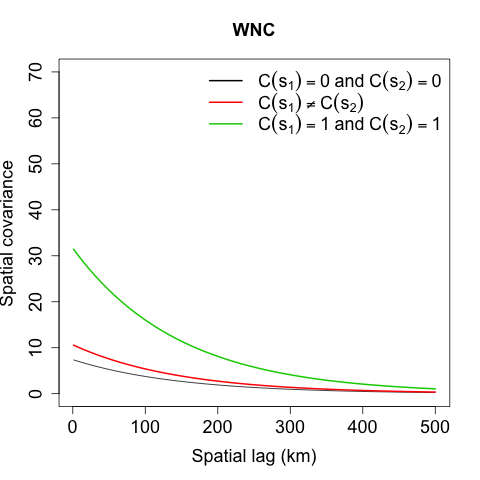}
		\caption{West North Central}		
	\end{subfigure}
	\quad
	\begin{subfigure}[b]{0.3\textwidth}
		\includegraphics[width=\textwidth]{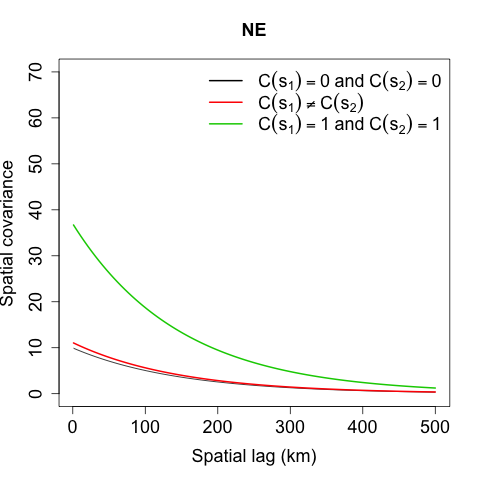}
		\caption{Northeast}
	\end{subfigure}
	\quad
	\begin{subfigure}[b]{0.3\textwidth}
		\includegraphics[width=\textwidth]{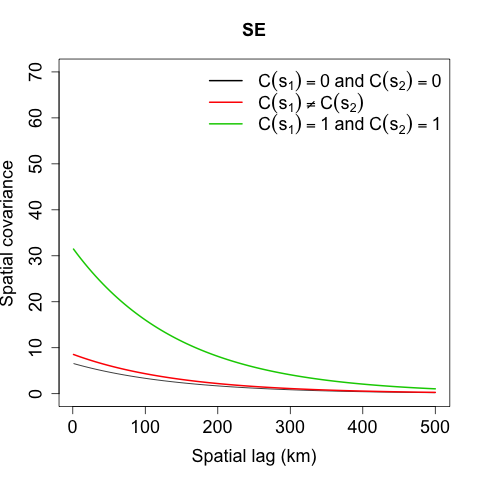}
		\caption{Southeast}
	\end{subfigure}
	\vskip\baselineskip
	\begin{subfigure}[b]{0.3\textwidth}
		\includegraphics[width=\textwidth]{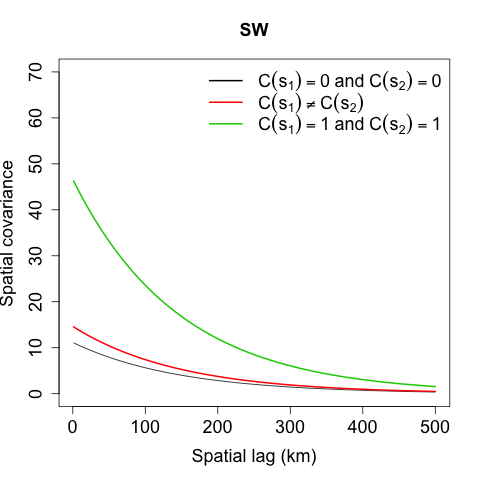}
		\caption{Southwest}
	\end{subfigure}
	\quad
	\begin{subfigure}[b]{0.3\textwidth}
		\includegraphics[width=\textwidth]{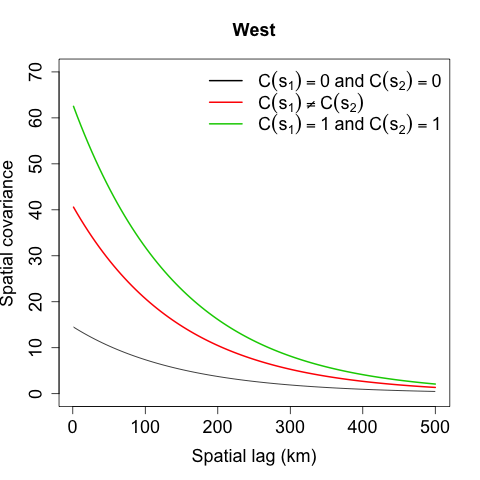}
		\caption{West}
	\end{subfigure}
	\quad
	\begin{subfigure}[b]{0.3\textwidth}
		\includegraphics[width=\textwidth]{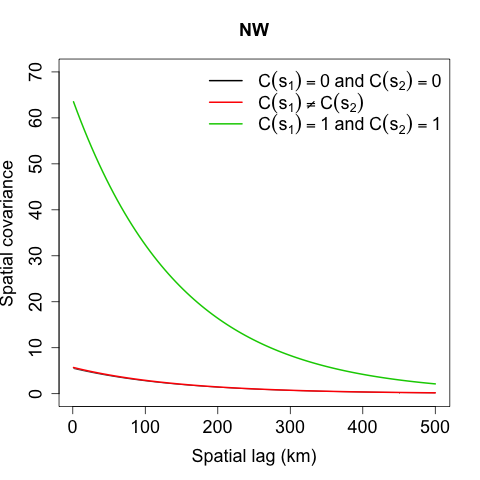}
		\caption{Northwest}
	\end{subfigure}
\end{figure}

Here, we demonstrate the strength of the correlation between sites. We evaluate $$\mbox{Cov}[Y_t(\bs), Y_t(\bs')|{\hat \theta}_t(\bs),{\hat \delta}_t(\bs)]$$ as defined in Equation (5) in the text, at the posterior mean of the model parameters for each combination of $(C_t(\bs), C_t(\bs'))$. Figure \ref{fig:covs} displays the plotted covariance functions for each region. The black curve denotes the covariance for observations where neither site is co-located with smoke, the red curve is the covariance for sites where only one is co-located with smoke, and the green curves denote the covariance for two sites co-located with smoke. The strength of the covariance between observations varies between all of the regions, and is strongest in those regularly impacted with wildfire smoke (e.g. West, West North Central).

\section{Sensitivity to $\tau$}
\begin{figure}[H]
    \centering
    \caption{\textbf{The causal effect given different smoke thresholds.} The causal effect estimate is robust to the choice of smoke threshold, $\tau$. This is demonstrated in the maps below, which show the causal effect estimate in the Northwest Region for $\tau = 0, 0.1, 1, 5, 10$.}
    \label{fig:tau_maps}
    \includegraphics[width=\textwidth]{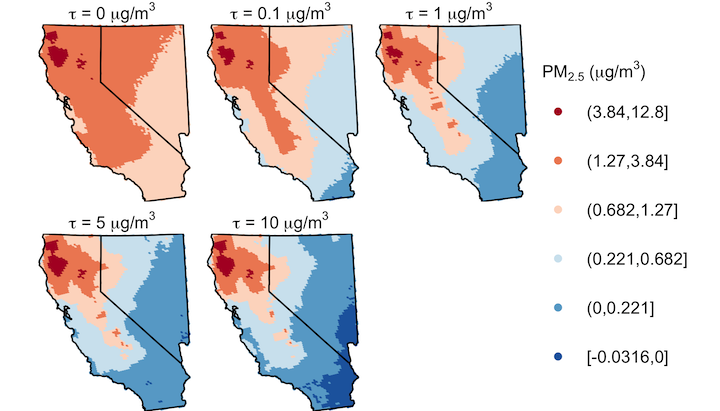}
%     \begin{subfigure}[b]{0.35\textwidth}
% 		\includegraphics[width=\textwidth]{Plots_Figures/map_w_ce_tau=0.png}
% 		\caption{$\tau=0$}
% 	\end{subfigure}	
% 	\quad
% 	\begin{subfigure}[b]{0.35\textwidth}
% 		\includegraphics[width=\textwidth]{Plots_Figures/map_w_ce_tau=01.png}
% 		\caption{$\tau=0.1$}
% 	\end{subfigure}
% 	\vskip\baselineskip
% 	\begin{subfigure}[b]{0.35\textwidth}
% 		\includegraphics[width=\textwidth]{Plots_Figures/map_w_ce_tau=1.png}
% 		\caption{$\tau=1$}
% 	\end{subfigure}
% 	\quad
% 	\begin{subfigure}[b]{0.35\textwidth}
% 		\includegraphics[width=\textwidth]{Plots_Figures/map_w_ce_tau=5.png}
% 		\caption{$\tau=5$}
% 	\end{subfigure}
% 	\vskip\baselineskip
% 	\begin{subfigure}[b]{0.35\textwidth}
% 		\includegraphics[width=\textwidth]{Plots_Figures/map_w_ce_tau=10.png}
% 		\caption{$\tau=10$}
% 	\end{subfigure}
\end{figure}

In this section, we provide sensitivity analysis to the choice of $\tau$, the smoke presence threshold. In Table \ref{tab:tau_cv}, we provide results from a cross-validation analysis and in Figure \ref{fig:tau_maps} we provide plots of the causal effect estimated under different values of $\tau$ to demonstrate robustness to choice of $\tau$.

\begin{table}[H]
\centering
\caption{\textbf{Five-fold cross-validation results for threshold selection, $\tau$ (\unit)}, using data from California. We report average (over space and time) mean-squared error (MSE), root mean-squared error (RMSE), mean absolute difference (MAD), standard deviation of the predicted values (SD) and coverage of 95\% prediction intervals to asses each model's ability to predict total $\PM$.}\label{tab:tau_cv}
\begin{tabular}{rrrrrr}
 \multicolumn{6}{c}{Threshold, $\tau$ (\unit)} \\
  \hline
 & $\tau = 0$ & $\tau=0.1$ & $\tau=1$ & $\tau=5$ & $\tau=10$ \\ 
  \hline
  MSE & 12.59 & 12.63 & 12.58 & 12.71 & 12.61 \\ 
  RMSE & 3.55 & 3.55 & 3.55 & 3.57 & 3.55 \\ 
  MAD & 1.78 & 1.77 & 1.78 & 1.78 & 1.77 \\ 
  SD & 11.84 & 11.68 & 11.58 & 11.47 & 11.48 \\ 
  Coverage & 0.98 & 0.98 & 0.98 & 0.98 & 0.98 \\ 
   \hline
\end{tabular}
\label{tab:cv}
\end{table}

%\begin{figure}[H]
%    \centering
%    \caption{\textbf{Time Series of the Causal Effect for Different Smoke Thresholds ($\tau$).}}
%    \label{fig:tau_ce}
%    \includegraphics[width=.85\textwidth]{Plots_Figures/sensitivity_to_tau.png}
%\end{figure}

\section{Residual Autocorrelation}
\begin{figure}[H]
    \centering
    \caption{Residual Autocorrelation Function (ACF).}
    \includegraphics[width=.85\textwidth]{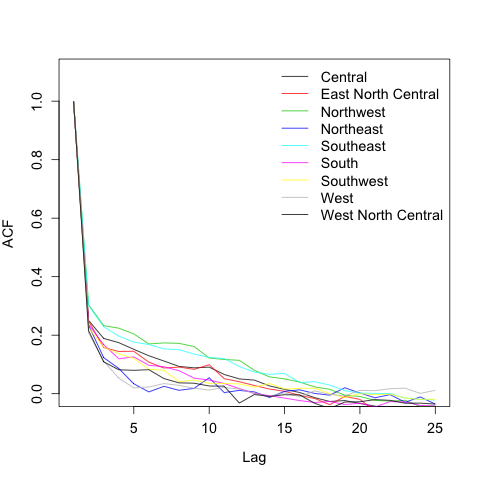}
    \label{fig:autocorr}
\end{figure}

In this section, we present diagnostics to support our assumption of temporal independence between observations. In Figure \ref{fig:autocorr}, we plotted residual autocorrelation functions for each region. The residuals are from the linear regression (separate at each site) of the observations $Y_t(\bs)$ onto the CMAQ covariates $C_t(\bs)$, ${\hat \theta}_t(\bs)$ and $C_t(\bs){\hat \delta}_t(\bs)$. We observe that for all of the regions, the lag-one autocorrelation is around 0.2 for most regions.

\section{Spatial Dependence Model Goodness-of-Fit}
\begin{figure}[H]
	\centering
	\caption{\textbf{Spatial variograms}.}\label{fig:variograms}
	\begin{subfigure}[b]{0.3\textwidth}
		\includegraphics[width=\textwidth]{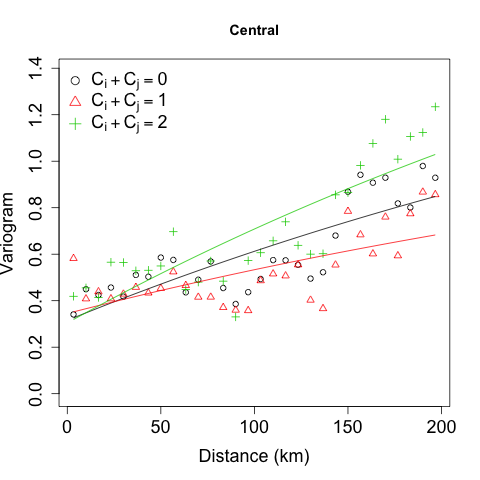}
		\caption{Central}
	\end{subfigure}	
	\quad
	\begin{subfigure}[b]{0.3\textwidth}
		\includegraphics[width=\textwidth]{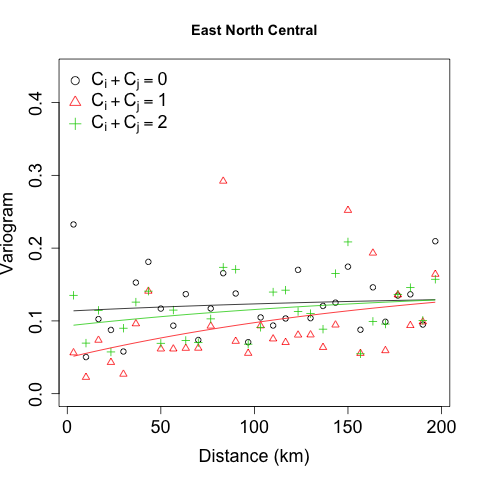}
		\caption{East North Central}
	\end{subfigure}
	\quad
	\begin{subfigure}[b]{0.3\textwidth}
		\includegraphics[width=\textwidth]{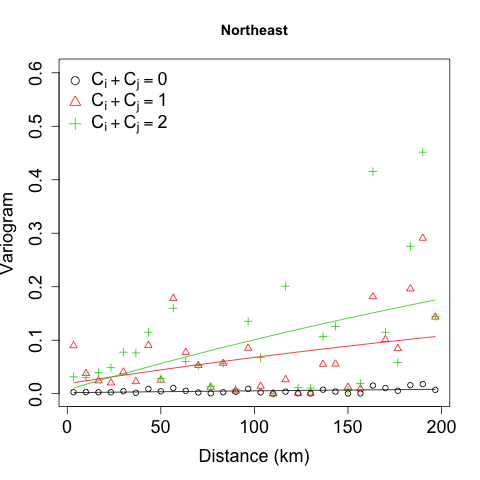}
		\caption{Northeast}
	\end{subfigure}
	\vskip\baselineskip
	\begin{subfigure}[b]{0.3\textwidth}
		\includegraphics[width=\textwidth]{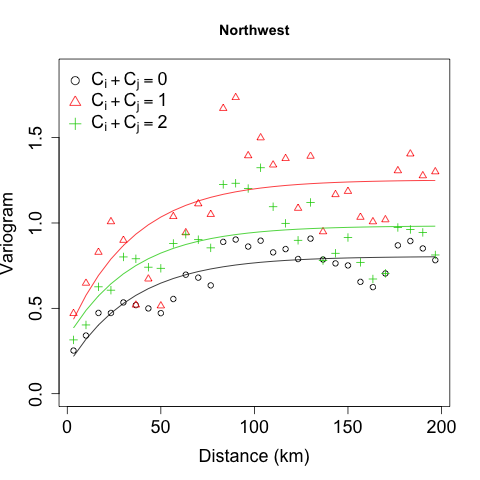}
		\caption{Northwest}
	\end{subfigure}
	\quad
	\begin{subfigure}[b]{0.3\textwidth}
		\includegraphics[width=\textwidth]{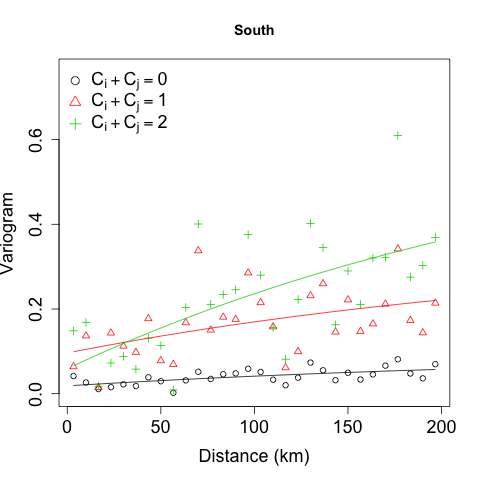}
		\caption{South}
	\end{subfigure}
	\quad
	\begin{subfigure}[b]{0.3\textwidth}
		\includegraphics[width=\textwidth]{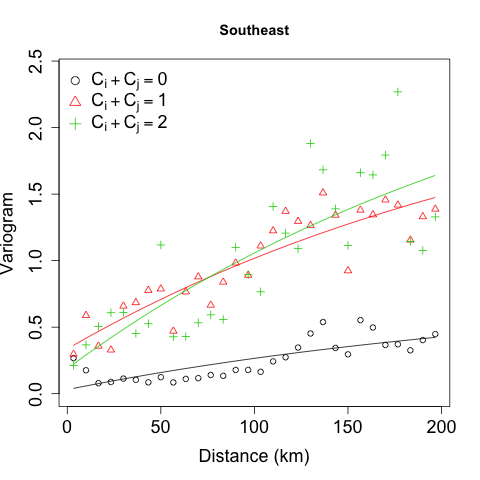}
		\caption{Southeast}
	\end{subfigure}
	\vskip\baselineskip
	\begin{subfigure}[b]{0.3\textwidth}
		\includegraphics[width=\textwidth]{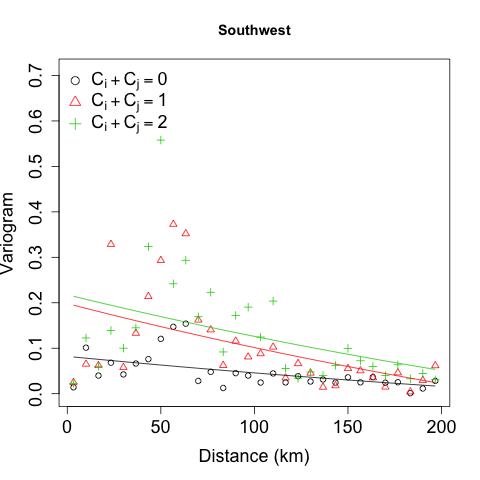}
		\caption{Southwest}
	\end{subfigure}
	\quad
	\begin{subfigure}[b]{0.3\textwidth}
		\includegraphics[width=\textwidth]{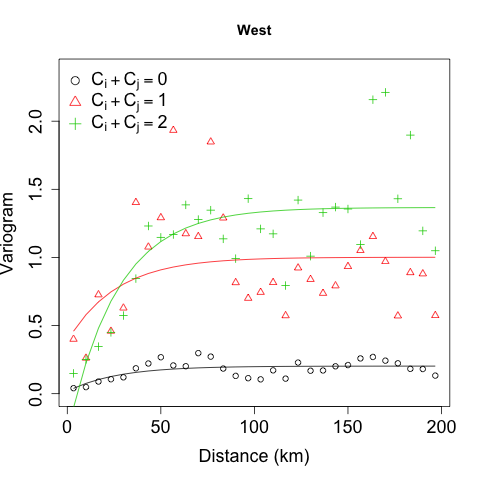}
		\caption{West}
	\end{subfigure}
	\quad
	\begin{subfigure}[b]{0.3\textwidth}
		\includegraphics[width=\textwidth]{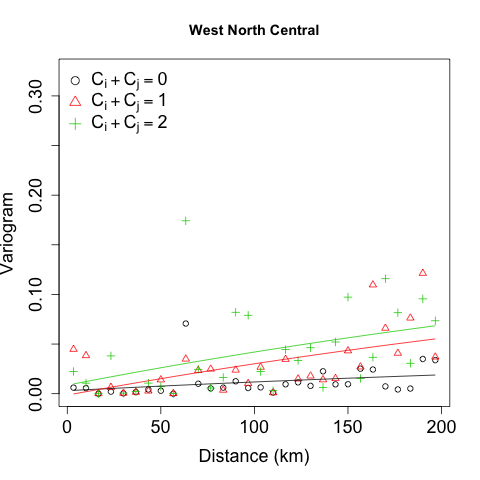}
		\caption{West North Central}		
	\end{subfigure}
\end{figure}

We examined variograms of the residuals for each region to determine goodness-of-fit for our spatial dependence model. Figure \ref{fig:variograms} shows the variograms for all nine regions. We computed the empirical variogram for each combination of $C_t(\bs)$, shown in different colors, as well as the variogrom curves evaluated at the posterior means of the covariance parameters, shown as lines. 

